\documentclass[a4paper,fleqn,usenatbib]{mnras}

\usepackage[T1]{fontenc}
\usepackage{ae,aecompl}

\usepackage{graphicx}	
\usepackage{amsmath}	
\usepackage{amssymb}	
\usepackage{array}
\usepackage{booktabs}
\usepackage{pdflscape}
\newcommand\bvijh{\mbox{$BV\!I\!J\!H$}}
\newcommand\bvijhk{\mbox{$BV\!I\!J\!H\!K$}}
\newcommand\vijh{\mbox{$V\!I\!J\!H$}}
\newcommand\bvi{\mbox{$BV\!I$}}
\newcommand\bvih{\mbox{$BV\!I\!H$}}
\newcommand\vri{\mbox{$V\!RI$}}
\newcommand\jhk{\mbox{$J\!H\!K$}}
\newcommand\vi{\mbox{$V\!-\!I$}}
\newcommand\bv{\mbox{$B\!-\!V$}}
\newcommand\jh{\mbox{$J\!-\!H$}}
\newcommand\hk{\mbox{$H\!-\!K$}}
\newcommand\phn{\phantom{0}} 
\newcommand\phnn{\phantom{00}} 
\newcommand\phs{\phantom{$-$}} 
\newcommand{\kms}{\,km\,s$^{-1}$}
\newcommand{\hnotunits}{\,km\,s$^{-1}$\,Mpc$^{-1}$} 

\title[SNe II As Distance Indicators At Near-IR]{Type II Supernovae as Distance Indicators at Near-IR Wavelengths\thanks{This paper includes data gathered with the 6.5 meter Magellan Telescopes located at Las Campanas Observatory, Chile.}}
\author[\'{O}. Rodr\'{i}guez et al.]{\'{O}. Rodr\'{i}guez,$^{1,2}$\thanks{E-mail: olrodrig@gmail.com}
G. Pignata,$^{1,2}$
M. Hamuy,$^{3,2}$
A. Clocchiatti,$^{4,2}$
M. M. Phillips,$^{5}$
\newauthor
K. Krisciunas,$^{6}$
N. I. Morrell,$^{5}$
G. Folatelli,$^{7,8}$
M. Roth,$^{5,9}$
S. Castell\'{o}n,$^{5}$
I. S. Jang,$^{10}$
\newauthor
Y. Apostolovski,$^{1,2}$
P. L\'{o}pez,$^{3}$
S. Marchi,$^{3}$
R. Ram\'{i}rez,$^{3}$
and
P. S\'{a}nchez$^{3,11}$
\\
$^{1}$Departamento de Ciencias Fisicas, Universidad Andres Bello, Avda. Republica 252, Santiago, Chile\\
$^{2}$Millennium Institute of Astrophysics, Casilla 36-D, Santiago, Chile\\
$^{3}$Departamento de Astronom\'ia, Universidad de Chile, Casilla 36-D, Santiago, Chile\\
$^{4}$Instituto de Astrof\'isica, Pontificia Universidad Cat\'olica de Chile, Casilla 306, Santiago 22, Chile\\
$^{5}$Carnegie Observatories, Las Campanas Observatory, Casilla 60, La Serena, Chile\\
$^{6}$Department of Physics and Astronomy, Texas A\&M University, College Station, TX 77843, USA\\
$^{7}$Facultad de Ciencias Astron\'omicas y Geof\'isicas, UNLP, IALP, CONICET, Paseo del Bosque S/N, B1900FWA La Plata, Argentina\\
$^{8}$Kavli Institute for the Physics and Mathematics of the Universe (WPI), The University of Tokyo, 5-1-5 Kashiwanoha, Kashiwa, Chiba 277-8583, Japan\\
$^{9}$GMTO Corporation, Avenida Presidente Riesco 5335, Suite 501, Las Condes, Santiago, Chile\\
$^{10}$Leibniz-Institut f\"ur Astrophysik Potsdam, D-14482 Potsdam, Germany\\
$^{11}$European Southern Observatory, Casilla 19001, Santiago 19, Chile
}

\date{Accepted XXX. Received YYY; in original form ZZZ}

\pubyear{2018}

\begin{document}
\label{firstpage}
\pagerange{\pageref{firstpage}--\pageref{lastpage}}
\maketitle

\begin{abstract} 
Motivated by the advantages of observing at near-IR wavelengths, we investigate Type II supernovae (SNe II) as distance indicators at those wavelengths through the Photospheric Magnitude Method (PMM). For the analysis, we use \bvijh\ photometry and optical spectroscopy of 24 SNe II during the photospheric phase. To correct photometry for extinction and redshift effects, we compute total-to-selective broadband extinction ratios and $K$-corrections up to $z=0.032$. To estimate host galaxy colour excesses, we use the colour-colour curve method with the \vi\ versus \bv\ as colour combination. We calibrate the PMM using four SNe II in galaxies having Tip of the Red Giant Branch distances. Among our 24 SNe II, nine are at $cz>2000$ \kms, which we use to construct Hubble diagrams (HDs). To further explore the PMM distance precision, we include into HDs the four SNe used for calibration and other two in galaxies with Cepheid and SN Ia distances. With a set of 15 SNe II we obtain a HD rms of 0.13 mag for the $J$-band, which compares to the rms of 0.15--0.26 mag for optical bands. This reflects the benefits of measuring PMM distances with near-IR instead of optical photometry. With the evidence we have, we can set the PMM distance precision with $J$-band below 10 per cent with a confidence level of 99 per cent.
\end{abstract}




\section{Introduction}\label{sec:introduction}

Type II supernovae (SNe II) are the explosive end of massive stars ($M_\text{ZAMS}\!>8\!\text{ M}_{\sun}$) that retain an important amount of hydrogen in their envelopes at the moment of the explosion. These events, consequence of the gravitational collapse of their iron cores, are characterized by a luminosity comparable to the total luminosity of their host galaxies, which make them interesting objects for distance measurements.

The pioneering work of \citet{Kirshner_Kwan1974} marks the beginning of the use of SNe II as distance indicators. In their work they applied the Expanding Photosphere Method (EPM, a variant of the Baade-Wesselink method) to two SNe II, using optical photometry and spectroscopy during the photospheric phase (the phase between the maximum light and the transition to the radioactive tail) to estimate angular and physical sizes, respectively. For the first implementation of the EPM, \citet{Kirshner_Kwan1974} assumed SNe II emit like blackbodies. Years after, \citet{Wagoner1981} demonstrated that the flux of SNe II is diluted as a consequence of their scattering-dominated atmospheres, making necessary SN II atmosphere models to quantify that effect and thus to correct derived distances. Since then, the EPM has been applied using different theoretical atmosphere models \citep[e.g.,][]{Eastman_etal1996,Dessart_Hillier2005} to an ample number of SNe II \citep[e.g.,][]{Schmidt_etal1992,Schmidt_etal1994,Hamuy2001,Dessart_etal2008,Jones_etal2009,Bose_Kumar2014,Gall_etal2016,Gall_etal2018}, where the typical EPM distance precision is found to be about 15 per cent.

Empirically, \citet{Hamuy_Pinto2002} found a correlation between the bolometric luminosity at 50 d since explosion and the expansion velocity of the photosphere at the same epoch. This is due to the fact that a more energetic explosion corresponds to a more luminous SN with higher envelope expansion velocities. The latter correlation is the basis of the Standardized Candle Method (SCM), which allows to estimate distances using photometry and expansion velocities inter- or extrapolated at 50 d since explosion. The SCM has been applied to several SN II sets \citep[e.g.,][]{Nugent_etal2006,Poznanski_etal2009,Olivares_etal2010,DAndrea_etal2010,deJaeger_etal2015,deJaeger_etal2017,Gall_etal2018}, yielding a distance precision about 12--14 per cent.

Despite apparent differences between the EPM and the SCM, \citet{Kasen_Woosley2009} showed that the SCM is a recasting of the EPM at 50 d since explosion. Additionally, by means of SN II models, they proposed a generalization of the SCM, which can be applied in any epoch during the photospheric phase. The same idea was investigated empirically by \citeauthor{Rodriguez_etal2014} (\citeyear{Rodriguez_etal2014}, hereafter R14), who called it the Photospheric Magnitude Method (PMM) to the SCM generalization. Measuring distances with all expansion velocities available during the photospheric phase, decreases observational errors and reduces uncertainties introduced by the interpolation/extrapolation at a certain fiducial epoch. The PMM distance precision is around 6--11 per cent (R14).

For the EPM, SCM, and PMM, optical spectroscopy is necessary in order to estimate expansion velocities. Since the spectroscopy is more time consuming than photometry, expansion velocities are not always available. For this reason, \citet{deJaeger_etal2015} proposed a method based solely on photometry to standardize SNe II, known as the Photometric Colour Method (PCM). \citet{deJaeger_etal2017} applied the PCM to a SN II sample with redshift up to 0.5, finding that the PCM distance precision is around 17 per cent.

Most of distances measurements with the latter methods have been performed with optical photometry. However, observing at near-IR wavelengths has two clear benefits that in principle can improve their use as distance indicators:

1. \textit{Near-IR light is less affected by dust}. Methods to measure colour excess due to SNe II host galaxies (e.g., \citealt{Schmidt_etal1992,Krisciunas_etal2009,Olivares_etal2010,Poznanski_etal2012}; R14; \citealt{Pejcha_Prieto2015}) are still not well established. Therefore, it is propitious to observe at near-IR wavelengths in order to reduce the effect of miscalculation of the colour excess. Moreover, the estimation of a representative extinction curve along the SN II line of sight is still controversial. Assuming the family of extinction curves of \citet{Cardelli_etal1989}, some studies are in favour of a Galactic $R_V\!=\!3.1$ \citep[e.g.,][]{Pejcha_Prieto2015}, while other authors found results in favour of lower values \citep{Poznanski_etal2009,Olivares_etal2010,deJaeger_etal2015}. Since the choice of a certain extinction curve has more impact at optical than at near-IR wavelengths \citep[e.g.,][]{Schlafly_etal2016}, it is preferable to perform photometric observations at those wavelengths in order to diminish systematics induced by the assumption of an incorrect extinction curve.

2. \textit{Contamination by metal lines is less severe at near-IR wavelengths}. Among the few metal lines identified in the near-IR, we remind: in the $J$-band range there is a feature at $\lambda\!=\!1.2$ $\mu$m possibly due to a \ion{Si}{I} multiplet \citep{Valenti_etal2015}, \ion{Mg}{I} $\lambda$1.53 $\mu$m is detected in the $H$-band range \citep{Maguire_etal2010_04et,Valenti_etal2015,Yuan_etal2016}, while in the $K$-band range the Brackett $\gamma$ is possibly blended with \ion{Na}{I} \citep{DallOra_etal2014}. The low number and weakness of metal lines reduce the risk of systematics effects produced by differences in progenitor metallicity \citep[e,g.,][]{Dessart_etal2014,Anderson_etal2016}.

\citet{Schmidt_etal1992} had already pointed out the benefits of measuring distances to SNe II using near-IR photometry. However, at present, there have been very few systematic studies (e.g., \citealt{Schmidt_etal1992}, \citealt{Hamuy_etal2001} for the EPM; \citealt{Maguire_etal2010_NIR_SCM}, \citealt{deJaeger_etal2015} for the SCM). In particular, \citet{Maguire_etal2010_NIR_SCM} suggested that it may be possible to reduce the scatter in the Hubble Diagram (HD) to 0.1--0.15 mag (distance precision of 5--7 per cent) using near-IR instead of optical photometry. However, this result is based on the analysis of a set of 12 SNe II, 11 of them at $z<0.01$, so being highly affected by peculiar velocities. To test this promising result, \citet{deJaeger_etal2015} applied the SCM to a set of 24 SNe II at $0.01<z<0.04$, obtaining a HD rms of 0.28 mag (distance precision of 13 per cent) for the $J$-band and therefore questioning the improvements of the SCM distance precision using near-IR photometry. 

The goal of this study is to investigate the PMM distance precision using near-IR photometry.

We organize our work as follows. In Section \ref{sec:observational_material} we describe the photometric and spectroscopic data. In Section \ref{sec:pmm_distances} we present the PMM developed in R14. In Section \ref{sec:light_curve_fits} we develop an algorithm to achieve nonparametric light curve fitting. In section \ref{sec:galactic_broadband_extinction} and \ref{sec:K_correction} we compute Galactic total-to-selective broadband extinction ratios and $K$-corrections for \bvijhk\ bands, respectively. In Section \ref{sec:host-galaxy_reddening} we compute host galaxy total-to-selective broadband extinction ratios and host galaxy colour excesses through the analysis of colour-colour curves. In Section \ref{sec:vph_texp} we estimate expansion velocities and explosion epochs. In Section \ref{sec:PMM} we apply the PMM to our SN II sample, constructing HDs for \bvijh\ bands. Discussion about the PMM distance precision and systematics are in Section \ref{sec:discussion}. In Section \ref{sec:conclusions} we present our conclusions.

\section{OBSERVATIONAL MATERIAL}\label{sec:observational_material}

We base our work on data obtained over the course of the Carnegie Type II Supernova Survey (CATS; PI: Hamuy, 2002--2003), a program whose main objective was to study nearby ($z<0.05$) SNe II. Optical photometry and spectroscopy, along with some near-IR photometry, were obtained with the 1-m Swope, 2.5-m du Pont, and 6.5-m Magellan Baade and Clay telescopes at Las Campanas Observatory. A few additional optical images were obtained with the 0.9-m and 1.5-m telescopes at Cerro Tololo Inter-American Observatory. During the CATS survey, 34 SNe II were observed. Optical photometry and spectroscopy of these SNe II, along with the description of the data reduction, is presented in \citet{Galbany_etal2016} and \citet{Gutierrez_etal2017_I}, respectively. Next, we briefly summarize the general techniques used to obtain the near-IR photometric data, which will be released in a forthcoming publication.

\subsection{Near-IR Photometric Data}

The near-IR photometric observations were obtained with the $\jhk$ bands mounted in the Swope Telescope IR camera and the Wide Field IR Camera on the du Pont Telescope. Images where processed with a collection of IRAF\footnote{IRAF is distributed by the National Optical Astronomy Observatory, which is operated by the Association of Universities for Research in Astronomy (AURA) under cooperative agreement with the National Science Foundation.} tasks. These include dark subtraction, flat field correction, sky subtraction, image registration and stacking. Instrumental magnitudes were obtained using the point spread function (PSF) technique, implemented in the SNOoPY\footnote{SNOoPy is a package for SN photometry using PSF fitting and/or template subtraction developed by E. Cappellaro. A package description can be found at \url{http://sngroup.oapd.inaf.it/snoopy.html}} package. The near-IR magnitudes of the reference stars were calibrated using standard star fields obtained soon before or after the target field with an airmass similar to the target field.

\subsection{Sample of Supernovae}

Among the 34 SNe II observed over the course of the CATS survey, we select a subset of 10 SNe II which comply with the following requirements: (1) having at least two photometric measurements in the \bvijh\ bands at 35--75 d since explosion (see Section~\ref{sec:PMM}), and (2) having at least one measurement of the expansion velocity at an epoch covered by the photometry mentioned in point 1. To this sample, we add 14 SNe II from the literature. Table~\ref{table:SN_sample} lists our final sample of 24 SNe II, which includes the SN name (Column 1), the name of the host galaxy and its type (Column 2 and 3), the heliocentric SN redshift and its source (Column 4 and 5), host galaxy distance measured with Cepheids, the Tip of the Red Giant Branch (TRGB), or SN Ia (Column 6), Galactic colour excess (Column 7), and references for the data (Column 8). We also use optical and near-IR spectra of SNe II with the purpose of computing total-to-selective broadband extinction ratios (Section~\ref{sec:galactic_broadband_extinction} and \ref{sec:host-galaxy_reddening}) and $K$-corrections (Section~\ref{sec:K_correction}), and to estimate explosion epochs (Section~\ref{sec:explosion_epochs}).

\begin{table*}
\caption{SN II sample.}
\label{table:SN_sample}
\begin{tabular}{lllclccl}
\hline
SN     & Host Galaxy    &Host Type$^\dagger$ &$cz_{\text{helio}}$ &Source$^\ddagger$ &$\mu_{\text{host}}^\star$ &$E_{\text{G}}(\bv)^\diamond$ &References$^\ast$\\
       &                &           & (\kms)            &      &  (mag)             &  (mag)              &          \\
\hline
1999em & NGC 1637       & SAB(rs)c  &\phn800$\pm$50\phn & L02  &$30.21\pm0.15^{\oslash,\triangleleft}$ &0.035$\pm$0.006 & a, b, c, d, e\\ 
2002gd & NGC 7537       & SAbc?     &   2536$\pm$59\phn & here & ...                       &0.058$\pm$0.009 & a, f, g, h  \\ 
2002gw & NGC  922       & SB(s)cd   &   3143$\pm$31\phn & here & ...                       &0.016$\pm$0.003 & a, f         \\ 
2002hj & NPM1G +04.0097 & ...       &   7079$\pm$20\phn & here & ...                       &0.101$\pm$0.016 & a, f         \\ 
2003B  & NGC 1097       & SB(s)b    &   1141$\pm$52\phn & here & ...                       &0.023$\pm$0.004 & a, f         \\ 
2003E  & MCG --4--12--4 & Sc?       &   4484$\pm$21\phn & here & ...                       &0.041$\pm$0.007 & a, f         \\ 
2003T  & UGC 4864       & SA(r)ab   &   8368$\pm$6\phnn & NED  & ...                       &0.027$\pm$0.004 & a, f         \\ 
2003bl & NGC 5374       & SB(r)bc?  &   4295$\pm$41\phn & here & ...                       &0.023$\pm$0.004 & a, f         \\ 
2003bn & LEDA 831618    & ...       &   3897$\pm$25\phn & here & ...                       &0.056$\pm$0.009 & a, f         \\ 
2003ci & UGC 6212       & Sb        &   9052$\pm$21\phn & here & ...                       &0.051$\pm$0.008 & a, f         \\ 
2003cn & IC 849         & SAB(rs)cd &   5430$\pm$162    & NED  & ...                       &0.018$\pm$0.003 & a, f         \\ 
2003hn & NGC 1448       & SAcd?     &   1305$\pm$35\phn & S05  & $31.25\pm0.07^{\otimes}$  &0.012$\pm$0.002 & a, d, f, i  \\ 
2004et & NGC 6946       & SAB(rs)cd &\phnn40$\pm$2\phnn & NED  & $29.39\pm0.14^{\otimes}$  &0.293$\pm$0.047 & h, j, k, l, m, n\\
2005ay & NGC 3938       & SA(s)c    &\phn850$\pm$26\phn & here & ...                       &0.018$\pm$0.003 & h, o        \\
2005cs & M51a           & SA(s)bc   &\phn463$\pm$3\phnn & NED  & $29.66\pm0.06^{\otimes}$  &0.032$\pm$0.005 & h, p, q, r\\
2008in & M61            & SAB(rs)bc &   1566$\pm$2\phnn & NED  & ...                       &0.019$\pm$0.003 & f, o, s     \\
2009N  & NGC 4487       & SAB(rs)cd &    905$\pm$21\phn & here & ...                       &0.018$\pm$0.003 & t           \\
2009ib & NGC 1559       & SB(s)cd   &   1304$\pm$162    & NED  & $31.72\pm0.20^{\triangle}$ &0.026$\pm$0.004 & u, v      \\
2009md & NGC 3389       & SA(s)c    &   1308$\pm$162    & NED  & ...                       &0.023$\pm$0.004 & w       \\
2012A  & NGC 3239       & IB(s)m    &\phn753$\pm$162    & NED  & ...                       &0.027$\pm$0.004 & x          \\
2012aw & M95            & SB(r)b    &\phn778$\pm$4\phnn & NED  & $29.89\pm0.07^{\otimes}$  &0.024$\pm$0.004 & y, z, aa   \\
2012ec & NGC 1084       & SA(s)c    &   1407$\pm$162    & NED  & ...                       &0.023$\pm$0.004 & bb          \\
2013ej & M74            & SA(s)c    &\phn657$\pm$1\phnn & NED  & $29.95\pm0.06^{\otimes}$  &0.060$\pm$0.010 & r, cc, dd, ee  \\
2014G  & NGC 3448       & I0        &   1160$\pm$84\phn & here & ...                       &0.010$\pm$0.002 & ff         \\
\hline
\multicolumn{8}{m{0.98\linewidth}}{$^\dagger$ From NASA/IPAC Extragalactic Database (NED).}\\
\multicolumn{8}{m{0.98\linewidth}}{$^\ddagger$ Source of the heliocentric SN redshift. L02: \citet{Leonard_etal2002_99em}; S05: \citet{Sollerman_etal2005}; here: this work.}\\
\multicolumn{8}{m{0.98\linewidth}}{$^\star$ Host galaxy distance moduli measured with Cepheids ($\oslash$), TRGB ($\otimes$) with the \citet{Jang_Lee2017} calibration, or with SN Ia ($\triangle$).}\\
\multicolumn{8}{m{0.98\linewidth}}{$^\diamond$ Galactic colour excesses from \citet{Schlafly_Finkbeiner2011}, with an error of 16 per cent \citep{Schlegel_etal1998}.}\\
\multicolumn{8}{m{0.98\linewidth}}{$^\triangleleft$ \citet{Saha_etal2006} distance was shifted by $-0.19\pm0.13$ mag to be consistent with the \citet{Riess_etal2016} calibration (Section \ref{sec:PMM_calibration}).}\\
\multicolumn{8}{m{0.98\linewidth}}{$^\ast$ (a)  \citet{Galbany_etal2016};
                                                               (b)  \citet{Hamuy_etal2001};
                                                               (c)  L02
                                                               (d)  \citet{Krisciunas_etal2009};
                                                               (e)  \citet{Saha_etal2006};
                                                               (f)  \citet{Gutierrez_etal2017_I};
                                                               (g)  \citet{Spiro_etal2014};
                                                               (h) \citet{Faran_etal2014_IIP};           
                                                               (i) \citet{Hatt_etal2018};
                                                               (j) \citet{Sahu_etal2006};
                                                               (k) \citet{Maguire_etal2010_04et};
                                                               (l) \citet{Tikhonov2014};
                                                               (m) \citet{Murphy_etal2018};
                                                               (n) \citet{Anand_etal2018_NGC6946_TRGB_distance};
                                                               (o) \citet{Hicken_etal2017};
                                                               (p) \citet{Pastorello_etal2006};
                                                               (q) \citet{Pastorello_etal2009};
                                                               (r) \citet{McQuinn_etal2017};
                                                               (s) \citet{Roy_etal2011};
                                                               (t) \citet{Takats_etal2014};
                                                               (u) \citet{Takats_etal2015};
                                                               (v) \citet{Brown_etal2010};
                                                               (w) \citet{Fraser_etal2011};
                                                               (x) \citet{Tomasella_etal2013};
                                                               (y) \citet{Bose_etal2013};
                                                               (z) \citet{DallOra_etal2014};
                                                               (aa) \citet{Rizzi_etal2007};
                                                               (bb) \citet{Barbarino_etal2015};
                                                               (cc) \citet{Yuan_etal2016};
                                                               (dd) \citet{Dhungana_etal2016};
                                                               (ee) \citet{Bose_etal2015};
                                                               (ff) \citet{Terreran_etal2016}.
           }\\
\end{tabular}
\end{table*}

\section{PHOTOSPHERIC MAGNITUDE METHOD}\label{sec:pmm_distances}

The absolute magnitude of a SN II during the photospheric phase depends strongly on the temperature and the size of the photosphere (e.g., \citealt{Kasen_Woosley2009}; R14; \citealt{Pejcha_Prieto2015}). The latter can be estimated from the velocity of the material instantaneously at the photosphere (hereafter, photospheric velocity, $v_\text{ph}$) and the time since the SN explosion epoch $t_0$, under the assumption of homologous expansion \citep[e.g.,][]{Kirshner_Kwan1974}. R14 found that the time since explosion works better than the $\vi$ colour (used as a proxy for temperature) to standardize the brightness of SNe II (see Fig. 9 in R14), showing that for a given band $x$ the absolute magnitude in any moment $t_i$ during the photospheric phase, $M_{x,\Delta t_i,v_{\text{ph},i}}$, can be parametrized as
\begin{equation}\label{eq:Mabs_SN}
M_{x,\Delta t_i,v_{\text{ph},i}}= a_{x,\Delta t_i}-5\log\left(\frac{v_{\text{ph},i}}{5000\text{ km s}^{-1}}\right).
\end{equation}
Here, $\Delta t_i\equiv(t_i-t_0)/(1+z)$ is the elapsed time since the explosion in the SN rest frame at redshift $z$, and $a_{x,\Delta t_i}$ is a function that can be calibrated empirically. Previously, \citet{Kasen_Woosley2009} found similar results for the SN II brightness standardization, but using SN II models.

With the knowledge of $t_0$ and a measurement of $v_\text{ph}$ in any stage of the photospheric phase, we can compute the absolute magnitude at $t_i$ (equation~\ref{eq:Mabs_SN}) and, therefore, compute the SN distance modulus given by
\begin{align}
\mu_{x,i} &= m_{x,i}^\text{corr}-a_{x,\Delta t_i}+5\log\left(\frac{v_{\text{ph},i}}{5000\text{ km s}^{-1}}\right),\label{eq:mu}\\
m_{x,i}^\text{corr} &= m_{x,i} -A^\text{G}_{x,i}- K_{x,i} -A^\text{h}_{x,i}.\label{eq:m_corr}
\end{align}
Here, $m_{x,i}$ is the apparent magnitude, $A^\text{G}_{x,i}$ and $A^\text{h}_{x,i}$ are the Galactic and host galaxy broadband extinction, respectively, and $K_{x,i}$ is the $K$-correction. If more than one measurement of $v_\text{ph}$ is available, then we compute the distance modulus through a likelihood maximization (see Section~\ref{sec:PMM_calibration}).

\section{LIGHT CURVE FITS}\label{sec:light_curve_fits}

In equation~(\ref{eq:mu}) we need all quantities at the same epoch. Being more time consuming, spectroscopy is in general less abundant than photometry, so performing photometric interpolations is a reasonable choice. Previous efforts to fit SN II light curves use both parametric and nonparametric methods. Parametric methods assume parametric functions that capture the behaviour of the light curve from early to late stages, where parameters are obtained through least-square minimization \citep[e.g.,][]{Olivares_etal2010} or through Bayesian methodologies \citep[e.g.,][]{Sanders_etal2015}. Nonparametric methods are based on nonparametric regressions (NPR) like local regressions \citep[e.g.,][]{Olivares2008} and Gaussian processes \citep[e.g.,][]{deJaeger_etal2017_SN2016jhj}. Since the light curves of the SNe in our sample are in general well sampled, we prefer to use NPR methods for the light curve fitting, thus avoiding the use of heuristic models.

In this work we make use of \texttt{loess}, a NPR method that performs polynomial fits over local intervals along the domain \citep{Cleveland_etal1992}. To perform a \texttt{loess} fit, we have to specify: (1) the class of the local polynomial, which can be linear or quadratic, (2) the smoothing parameter, which defines the neighbourhood size around each element of the independent variable, where data can be well approximated by the aforementioned local polynomial, and (3) the distribution of random errors, which can be normal or symmetric \citep[for more details, see][]{Cleveland_etal1992}. We assume the null hypothesis that residuals are normally distributed, which can be checked with a normality test. An optimal value for the smoothing parameter can be obtained from data using the ``an'' information criterion \citep[AIC,][see Appendix \ref{model_selection}]{Akaike1974}. Therefore, to perform a \texttt{loess} fit, we only have to decide the local polynomial. We choose a quadratic polynomial in order to give more freedom to the \texttt{loess} fitting procedure. When the \texttt{loess} routine cannot perform a fit (e.g., for light curves with less than six points), we perform a low-order (linear or quadratic) polynomial fit.

To test whether photometry errors can account for the observed dispersion around the light curve fit, $f_{x,t}^\text{fit}$ (in flux units), we compute its log-likelihood given by
\begin{equation}
\ln\mathcal{L}=-0.5\sum_i\left(\ln(\sigma_{f_{x,i}}^2+\sigma_{x,\text{int}}^2)+\frac{(f_{x,i}-f_{x,t_i}^\text{fit})^2}{\sigma_{f_{x,i}}^2+\sigma_{x,\text{int}}^2}\right),
\end{equation}
where $f_{x,i}$ and $\sigma_{f_{x,i}}$ are the apparent magnitude and its error in flux units, respectively, and $\sigma_{x,\text{int}}$ is the intrinsic error. If an intrinsic error is necessary to maximize the log-likelihood, then we add it in quadrature to the photometry errors and perform again the light curve fitting. We repeat this process until an intrinsic error is not necessary.
\begin{figure*}
\includegraphics[width=\textwidth]{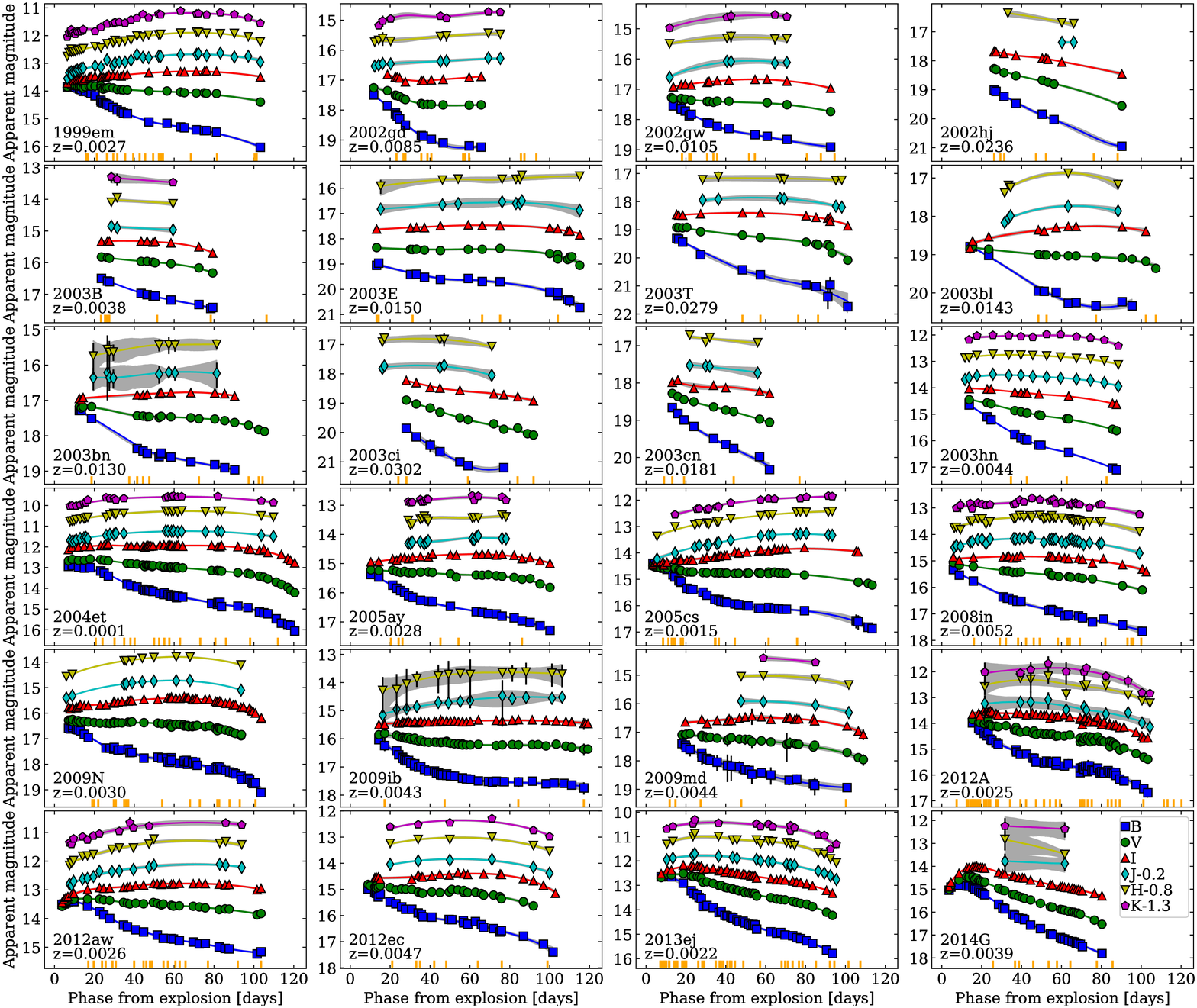}
\caption{Light curves of SNe II used in this work. The estimated explosion epoch of each SN is used as reference time. Orange ticks mark the epochs of the spectroscopy. Solid lines and shaded regions correspond to the light curve fits and its 80 per cent CIs, respectively.}
\label{fig:CATS}
\end{figure*}

To test the normality of the residuals, we use the Rescaled Moment \citep[RM;][]{Imon2003} test (see Appendix \ref{model_selection}). Among all light curves fits, 80 per cent have residuals with RM $p$-values $\geq0.05$, for which the hypothesis that residual are normally distributed cannot be rejected within a confidence level (CL) of 95 per cent. For the remaining 20 per cent, light curve fits are still unbiased and consistent, but the confidence interval (CI) of the parameters may be untrustworthy \citep{Doane_Seward2016}. Anyway, in this work, to prevent any shortcoming related to the non-normality of the residuals, we perform simulations to compute CIs.

To compute the CI around a light curve fit, we perform $10^4$ simulations varying randomly the photometry according to its error. For each realization we perform a \texttt{loess} (or a low-order polynomial) fit, thus obtaining $10^4$ simulated light curves per band. These simulations will allow us to compute its probability density function (pdf) at different epochs.

Fig.~\ref{fig:CATS} shows results of the aforementioned fitting procedure applied to the SN II light curves used in this work, where solid lines are the \texttt{loess} (or low-order polynomial) fits, while shaded regions indicate values between the 10th and the 90th percentile, i.e., the 80 per cent CI.

\section{GALACTIC BROADBAND EXTINCTION}\label{sec:galactic_broadband_extinction}

In equation~(\ref{eq:m_corr}), the Galactic broadband extinction in a photometric band $x$ is given by
\begin{equation}\label{eq:AG}
 A^\text{G}_{x,i}\equiv-2.5\log\left(\frac{\int{d\lambda_\text{o} S_{x,\lambda_\text{o} }F_{\lambda_\text{r} ,i}\lambda_\text{r} 10^{-0.4(A^\text{h}_{\lambda_\text{r}}+A^\text{G}_{\lambda_\text{o}})}}}{\int{d\lambda S_{x,\lambda_\text{o}}F_{\lambda_\text{r},i}\lambda_\text{r} 10^{-0.4A^\text{h}_{\lambda_\text{r}}}}}\right)
\end{equation}
\citep{Olivares2008}. Here, $\lambda_\text{o}$ is the wavelength in the observer's frame, $\lambda_\text{r}\!=\!\lambda_\text{o}/(1+z)$ is the wavelength in the SN rest frame at redshift $z$, $S_{x,\lambda_\text{o}}$ is the $x$-band transmission function, $F_{\lambda_\text{r},i}$ is the spectral energy distribution (SED) of the SN at epoch $t_i$. $A^\text{G}_{\lambda_\text{o}}$ and $A^\text{h}_{\lambda_\text{r}}$ are the Galactic and host galaxy monochromatic extinctions, respectively, given by
\begin{align}
 A^\text{G}_{\lambda_\text{o}} &= R_{\lambda_\text{o}}^\text{G} \cdot E_\text{G}(\bv),\\
 A^\text{h}_{\lambda_\text{r}} &= R_{\lambda_\text{r}}^\text{h} \cdot E_\text{h}(\bv),
\end{align}
where $R_\lambda \equiv A_{\lambda}/E(\bv)$ is the extinction curve for our Galaxy ($R_{\lambda}^\text{G}$) and hosts ($R_{\lambda}^\text{h}$), and $E_\text{G}(\bv)$ and $E_\text{h}(\bv)$ are the Galactic and host galaxy colour excess, respectively. 

Since the SED of SNe II evolve with time, we expect that the broadband extinction $A^\text{G}_{x,i}$ also evolve with time\footnote{We remark the difference between a monochromatic extinction $A_{\lambda}$, which is constant for a fixed wavelength $\lambda$, and a broadband extinction $A_{x,i}$, which depends on the SED and the $x$-band transmission function.}. As the SED of SNe II has a blackbody nature, hereafter we use the intrinsic $\bv$ colour (a proxy for temperature) to represent its evolution.

In a previous work, \citet{Olivares2008} computed the dependence of $A^\text{G}_{V}$ on $\bv$. In this work, in order to convert Galactic colour excesses directly into Galactic broadband extinctions suitable for SNe II, we compute the Galactic total-to-selective broadband extinction ratios $R_{x,i}^{\text{G}}$, such that
\begin{equation}\label{eq:EG_to_AG}
A^\text{G}_{x,i}\equiv R_{x,i}^\text{G}\cdot E_\text{G}(\bv).
\end{equation}

\begin{figure*}
\includegraphics[width=\columnwidth]{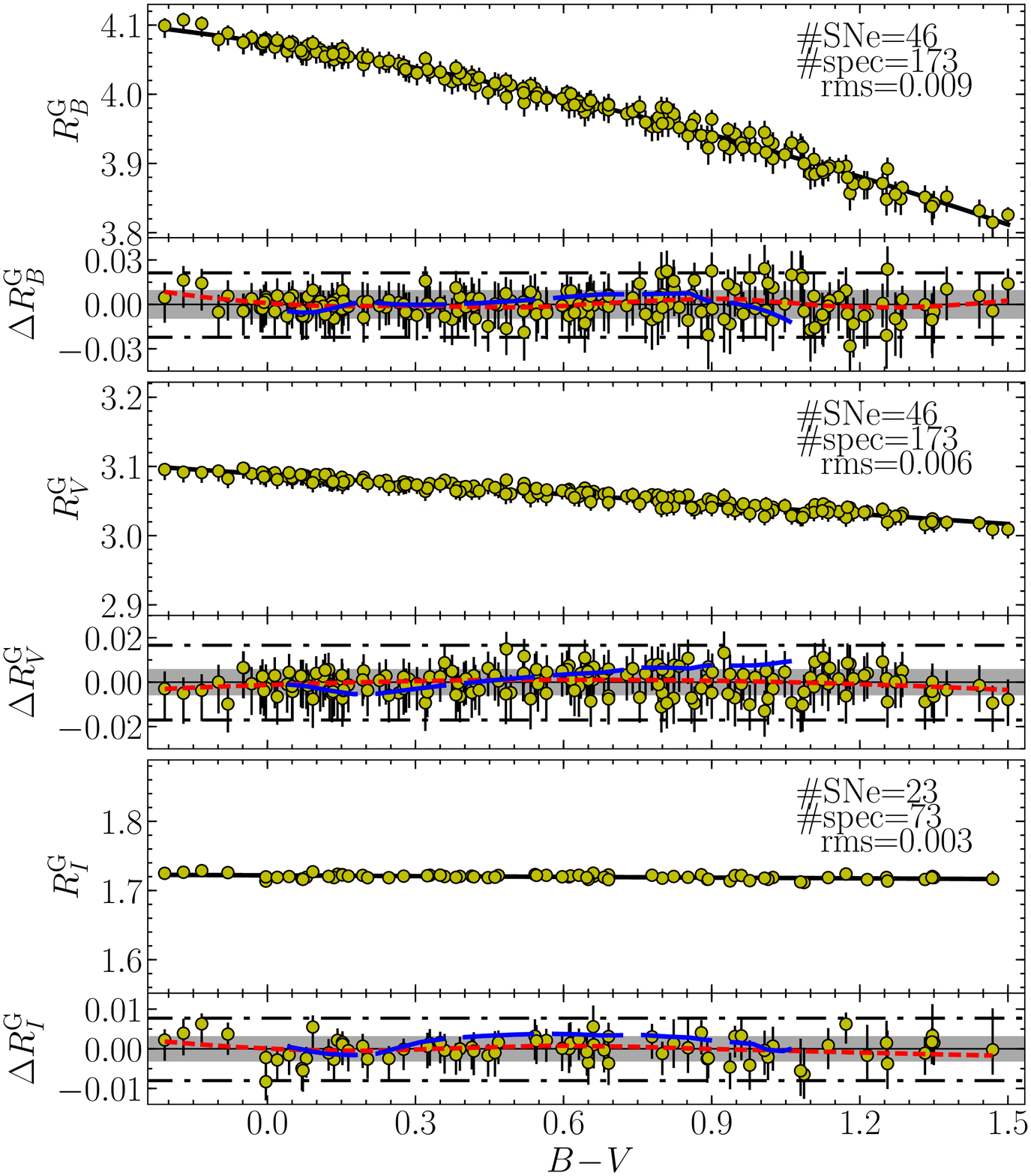}
\includegraphics[width=\columnwidth]{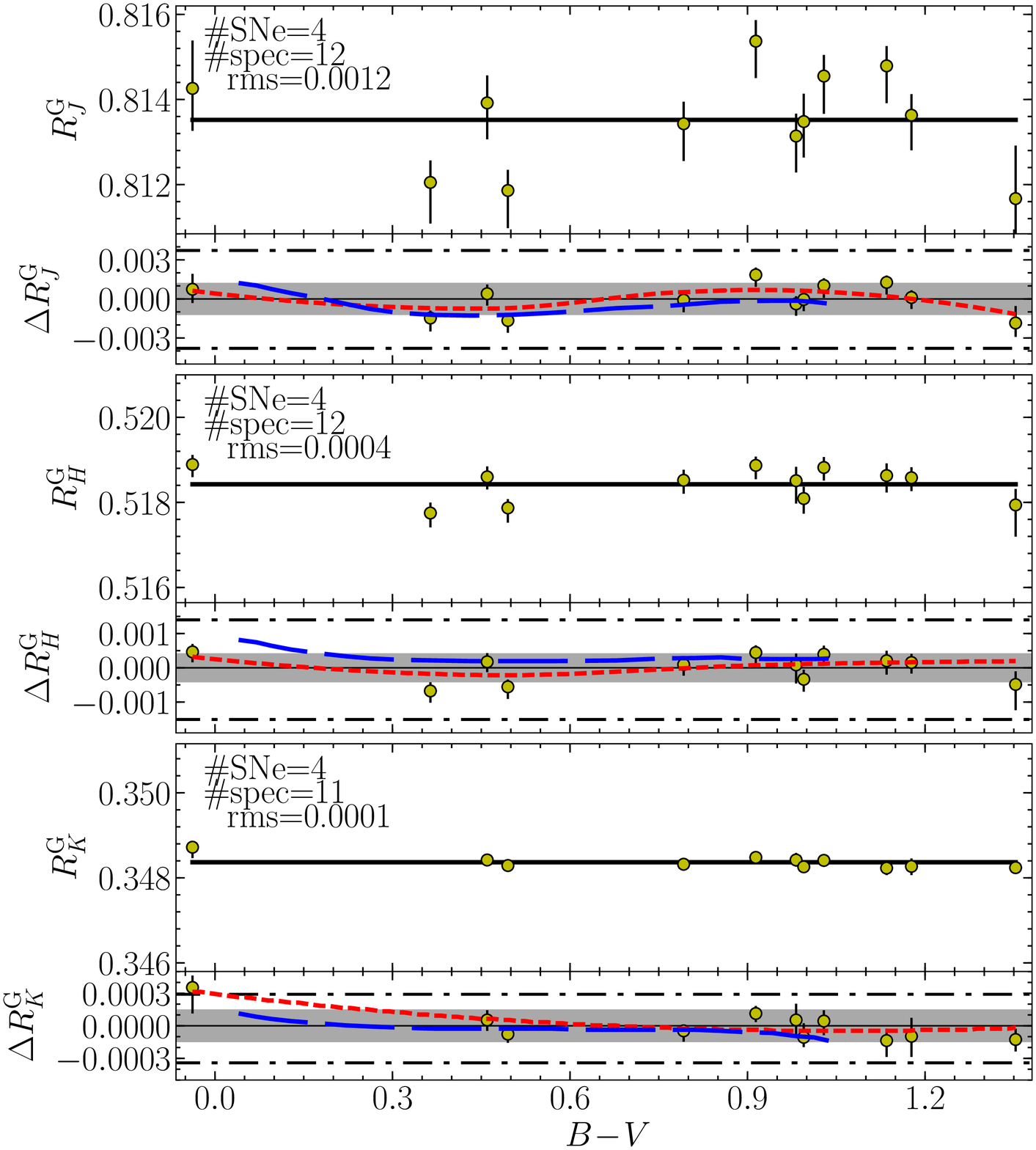}
\caption{Galactic total-to-selective broadband extinction ratios for $\bvi$ (left) and $\jhk$ (right) for SNe II as a function of the intrinsic $\bv$ colour, along with residuals. Solid black lines correspond to the polynomial fits, red short-dashed lines correspond to \texttt{loess} regressions to the residuals, while blue long-dashed lines correspond to residuals between the m15mlt3 model \citep{Dessart_etal2013} and polynomial fits from observations. Gray regions indicate values within one rms, while black dash-dotted lines are the inner fences.}
\label{fig:R_BVIJHK}
\end{figure*}

With the purpose of obtaining representative $R_{x,i}^\text{G}$ values for a local SN II sample through equation~(\ref{eq:EG_to_AG}) and (\ref{eq:AG}), we use: (1) a library of dereddened and deredshifted SN II spectra (see Appendix \ref{SNII_spectra_library}), (2) colour excesses and redshifts from the following representative ranges: $E_{\text{G}}(\bv)\!=\!0.0$--0.36 mag, $E_{\text{h}}(\bv)\!=\!0.0$--0.83 mag, which were taken from the SN sample reported in R14, and $z\!=\!0.0$--0.032,  and (3) an extinction curve to redden the spectra for both our Galaxy and hosts. For the latter, since a representative extinction curve along the line of sight of SNe II is still controversial, we adopt the \citet{Fitzpatrick1999} extinction curve with $R_V\!=\!3.1$. For each spectrum, we perform $10^4$ simulations picking randomly values of $E_{\text{G}}(\bv)$, $E_{\text{h}}(\bv)$, and $z$ from the aforementioned ranges, adopting the median as the $R_{x,i}^\text{G}$ representative value and the 80 per cent CI as its error.

The left side of Fig.~\ref{fig:R_BVIJHK} shows the $R_{x,i}^\text{G}$ values as a function of $\bv$ for $\bvi$ bands. There is a clear dependence of $R_{B,i}^\text{G}$ and $R_{V,i}^\text{G}$ on $\bv$. The $y$-axis scale at the left of Fig.~\ref{fig:R_BVIJHK} is the same in the three panels, so we can see that the redder the band the less the dependence on $\bv$, with $R_{I,i}^\text{G}$ being nearly constant. This behaviour is due to the blackbody nature of the SN II SED, where for the longer wavelengths the less the dependence of the SED slope on temperature.

To express the dependence of $R_{x,i}^\text{G}$ on $\bv$ we perform polynomial fits. The latter, unlike NPR methods like \texttt{loess}, allow to perform corrections in an easy and less time-consuming way (see Appendix~\ref{sec:C3_linearity}).

To determine the optimal degree for the polynomial fit, we consider two criteria: the AIC and the Bayesian information criterion \citep[BIC,][]{Schwarz1978}. For more details, see Appendix \ref{model_selection}. Based on evidence ratios (Table~\ref{table:RKRf_statistics}), for the $B$-, $V$-, and $I$-band, the AIC favours degrees $\geq$ 3, $\geq$ 1, and $\geq$ 1, respectively, while the BIC favours degrees between 2 and 6, 1 and 3, and 1 and 4, respectively. Results for both criteria are consistent. By the principle of parsimony (a.k.a. Occam's razor), we adopt the lowest degrees, i.e., 2, 1, and 1 for the $B$-, $V$-, and $I$-band, respectively. For \jhk\ bands (right of Fig.~\ref{fig:R_BVIJHK}) we adopt constant values. Although the small number of near-IR spectra means that the results are not fully statistically robust, we are confident about the negligible dependence of $R_{x,i}^\text{G}$ on $\bv$ for \jhk\ bands based on the small rms values we obtained ($\lesssim0.001$).

Once the optimal polynomial degrees for $R_{x}^\text{G}$ versus $\bv$ are determined, we perform $10^4$ bootstrap resampling of the data in order to compute the polynomial fit parameters and their errors, adopting the median as the representative value. Results are summarized in Table~\ref{table:AKAf_parameters}.

The bottom of each panel in Fig.~\ref{fig:R_BVIJHK} shows the residuals of the polynomial fit. To identify possible outliers we use the \citet{Tukey1977} rule, where values below $\text{Q1}\!-\!1.5(\text{Q3}\!-\!\text{Q1})$ or above $\text{Q3}\!+\!1.5(\text{Q3}\!-\!\text{Q1})$ (known as inner fences, where Q1 and Q3 are the first and third quartile, respectively) are considered outliers. The few points detected as outliers are consistent with being within inner fences considering their errors, so we do not discard them from the analysis. To analyse possible trends not captured by the polynomial fit, we perform a \texttt{loess} regression (red short-dashed line) to the residuals. Variations in the \texttt{loess} fits are mostly within one rms, meaning that the evolution of $R_{x}^\text{G}$ on $\bv$ can be well represented by a polynomial fit of degree determined with the AIC/BIC. For all bands we obtain RM $p$-values $>0.05$, which means that we cannot reject the null hypothesis that residual are normally distributed (95 per cent CL). Based on this, we can treat the $R_{x}^\text{G}$ rms error as a normal one.

For comparison, we compute $R_{x}^\text{G}$ for \bvijhk\ bands using synthetic spectra of the m15mlt3 model of \citet{Dessart_etal2013}. Residuals between the m15mlt3 model and polynomial fits from observations (blue long-dashed lines in Fig.~\ref{fig:R_BVIJHK}) are mostly contained within one rms. 

\begin{figure*}
\includegraphics[width=\columnwidth]{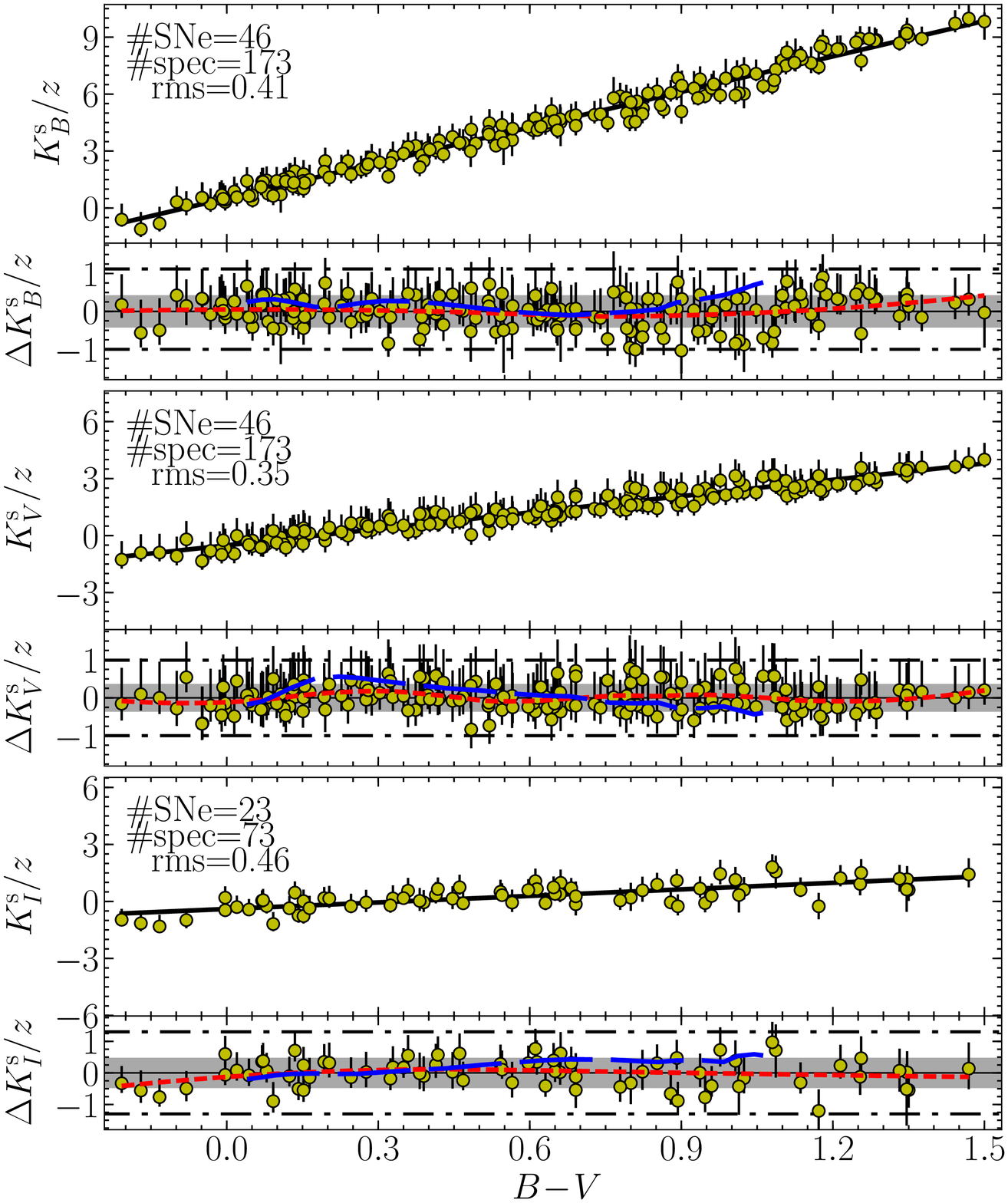}
\includegraphics[width=\columnwidth]{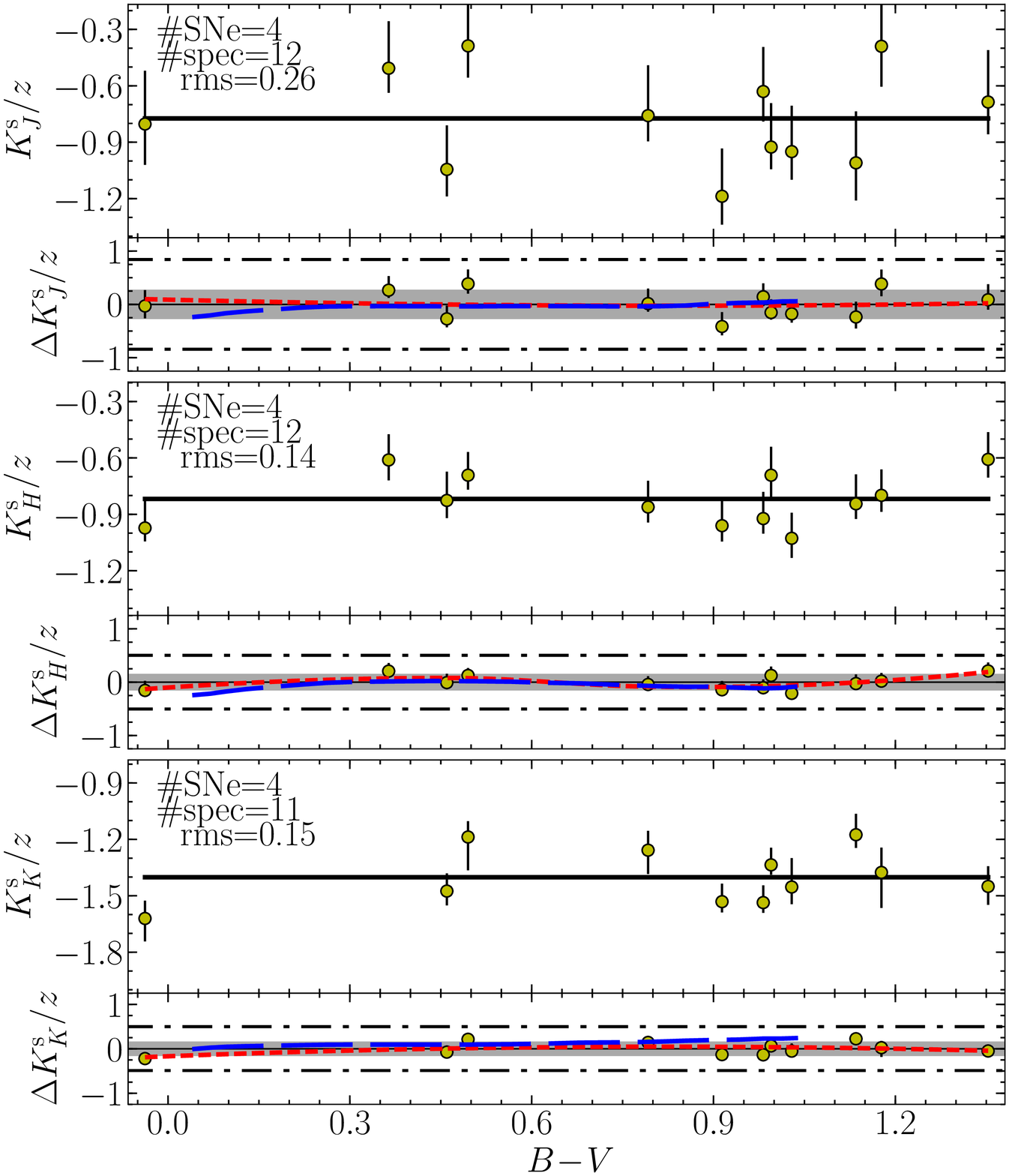}
\caption{Selective term of the $K$-correction over redshift for $\bvi$ (left) and $\jhk$ (right) for SNe II as a function of the intrinsic $\bv$ colour, along with residuals. Solid black lines correspond to the polynomial fits. Red short-dashed lines correspond to \texttt{loess} regressions to the residuals, while blue long-dashed lines correspond to residuals between the m15mlt3 model \citep{Dessart_etal2013} and polynomial fits from observations. Gray regions indicate values within one rms, while black dash-dotted lines are the inner fences.}
\label{fig:Ks_BVIJHK}
\end{figure*}

\section{$K$-CORRECTION}\label{sec:K_correction}

The $K$-correction in a photometric band $x$ is given by
\begin{align}
 K_{x,i}&\equiv-2.5\log(1+z)+K^s_{x,i},\label{eq:Ki}\\
 K^\text{s}_{x,i}&=2.5\log\left(\frac{\int{d\lambda_\text{o} S_{x,\lambda_\text{o}}F_{\lambda_\text{o},i}\lambda 10^{-0.4A^\text{h}_{\lambda_\text{o}}}}}{\int{d\lambda_\text{r} S_{x,\lambda_\text{o}}F_{\lambda_\text{r},i}\lambda_\text{r} 10^{-0.4A^\text{h}_{\lambda_\text{r}}}}}\right)\label{eq:Ki_selective}
\end{align}
\citep{Olivares2008}, being $K^\text{s}_{x,i}$ the selective term. We proceed in the same way than in Section~\ref{sec:galactic_broadband_extinction}, but now the evolving SED is modified by SN host galaxy colour excess and redshift. 

As in Section~\ref{sec:galactic_broadband_extinction} we aim for an analytical expression for $K^\text{s}_x$, for which we perform polynomial surface fits as a function of $\bv$ and $z$. Since $K_x=0$ for $z=0$, any $z$-independent term on the $K^\text{s}_x$ polynomial fit is zero. Dividing by $z$, the polynomial surface to adjust will be of the form
\begin{equation}\label{eq:Ki_selective_fit}
K^\text{s}_x/z=\sum_{j_1=0}^{\mathcal{O}_{B\!-\!V}}\sum_{j_2=0}^{\mathcal{O}_{z,j_1}}a_{j_1,j_2}z^{j_2}(\bv)^{j_1},
\end{equation}
being $\mathcal{O}_{B\!-\!V}$ and $\mathcal{O}_{z,j_1}$ the orders in $\bv$ and $z$, respectively, and $a_{j_1,j_2}$ the fit parameters.

To determine the orders $\mathcal{O}_{B\!-\!V}$ and $\mathcal{O}_{z,i}$, we generate $10^5$ spectral samples, where for each sample we assign to each spectrum a random redshift up to 0.032. For each realization, we obtain optimal order values using the AIC/BIC and the principle of parsimony. In all cases we obtain that $K_x^\text{s}/z$ depends only on $\bv$, i.e., it is z-independent for $z\leq0.032$.

Fig.~\ref{fig:Ks_BVIJHK} shows the $K_x^\text{s}/z$ values as a function of $\bv$ for \bvi\ (left) and \jhk\ (right). We perform the same analysis than in Section~\ref{sec:galactic_broadband_extinction}. For \bvi\ bands we adopt straight lines, while for \jhk\ bands we fit constant values (see Table~\ref{table:RKRf_statistics}). Results are summarized in Table~\ref{table:AKAf_parameters}. Variations of the \texttt{loess} fits to the residuals are within one rms, meaning that the dependence of $K_x^\text{s}/z$ on $\bv$ can be well represented by the polynomial fit of degree determined with the AIC/BIC. For all bands we obtain RM $p$-values $> 0.05$, which means that we cannot reject the null hypothesis that residual are normally distributed (95 per cent CL). Based on the latter, we can treat the $K_x^\text{s}/z$ rms error as a normal one.

\section{HOST GALAXY BROADBAND EXTINCTION}\label{sec:host-galaxy_reddening}

In equation~(\ref{eq:m_corr}), the host galaxy broadband extinction in a photometric band $x$ is given by
\begin{equation}\label{eq:Ah}
 A^\text{h}_{x,i}\equiv-2.5\log\left(\frac{\int{d\lambda S_{x,\lambda_\text{o}}F_{\lambda_\text{o},t}\lambda_\text{o} 10^{-0.4 A^\text{h}_{\lambda_\text{o}}}}}{\int{d\lambda_\text{o} S_{x,\lambda_\text{o}}F_{\lambda_\text{o},t}\lambda_\text{o}}}\right)
\end{equation}
\citep{Olivares2008}. We proceed in the same way than in Section~\ref{sec:galactic_broadband_extinction} and \ref{sec:K_correction}, but now the evolving SED is modified only by the SN host galaxy colour excess.

Similar to Section~\ref{sec:galactic_broadband_extinction}, we define the host galaxy total-to-selective broadband extinction ratios $R_{x,i}^\text{h}$, such that
\begin{equation}\label{eq:Eh_to_Ah}
A^\text{h}_{x,i}\equiv R_{x,i}^\text{h}\cdot E_\text{h}(\bv).
\end{equation}

The optimum $R_{x}^\text{h}$ versus $\bv$ polynomials and their parameters are summarized in Table~\ref{table:AKAf_parameters}.

R14 showed that for SNe II the $\bv$ versus $\vi$ colour-colour curve (C3) can be used to estimate $E_\text{h}(\bv)$ through the method proposed by \citet{Natali_etal1994}, which was originally developed to estimate interstellar colour excess for open clusters. 

The C3 method states that, under the assumptions that (1) the C3 can be well-represented by a straight line, and (2) all SNe II have the same C3 (which means the same slope and intercept), the host galaxy colour excess can be estimated with the formula
\begin{equation}\label{eq:C3_eq}
E_\text{h,i}(\bv) = \frac{1}{R_{c_x,i}^\text{h}}\frac{n_{S,i}-n_{S,\text{zp}}}{\gamma_{S,i}-\text{M}\{m_S\}}
\end{equation}
(e.g., \citealt{Munari_Carraro1996}; R14). Here, $S\equiv\{c_x,c_y\}$ indicates the colours used as $x$- and $y$-axis in the colour-colour diagram, corrected for Galactic colour excess and $K$-correction. $\text{M}\{m_S\}$ is the median of a set of SN II C3 slopes $\{m_S\}$, $n_{S,i}=c_{y,i}-\text{M}\{m_S\}\times c_{x,i}$ and $n_{S,\text{zp}}$ are the $y$-intercept of the C3 linear fit using a fixed slope $\text{M}\{m_S\}$ and that of the SN II less affected by colour excess, respectively. $R_{c,i}^\text{h}=R_{x_1,i}^\text{h}-R_{x_2,i}^\text{h}$ for a colour $c=x_1\!-\!x_2$, and $\gamma_{S,i} = R_{c_y,i}^\text{h}/R_{c_x,i}^\text{h}$ is the slope of the reddening vector. The subindex $i$ in equation~(\ref{eq:C3_eq}) denotes the dependence of $R_{x,i}^\text{h}$ on the intrinsic $\bv$, so equation~(\ref{eq:C3_eq}) must be evaluated separately at each point of the C3. In principle, one colour-colour observation is enough to estimate the colour excess, however more observations allow to check internal consistency and reduce observational errors. 

The C3 method relies strongly on the aforementioned two assumptions. In a previous work, R14 assumed the linearity of C3s based on the blackbody nature of the SED of SNe II during the photospheric phase, while assumption 2 was adopted based on the dependence of the emergent flux mainly on temperature displayed by SN II atmosphere models \citep[e.g.,][]{Eastman_etal1996, Jones_etal2009}. In this work we show that C3s can indeed be expressed as straight lines for several colour combinations (see Appendix~\ref{sec:C3_linearity}). Therefore, the major source of systematics comes from assumption 2. There are indeed some effects, like the line blanketing, that modify the SED continuum shape. In addition, differences in photometric systems \citep[$S$-correction;][]{Stritzinger_etal2002} are expected to produce further changes on C3 parameters. Therefore it is propitious to search for a colour combination where the effect of a colour excess on a C3 is greater than the effect of systematics.

An analysis of the effect of systematics on the C3 $y$-intercept is beyond the scope of this work because it requires an ample set of unreddened SNe II. However, the effect of systematics on C3 slopes and its consequent effects on the $E_\text{h}(\bv)$ estimation through the C3 method can be quantified in a simple way.

The presence of dust along the line of sight produces a vertical displacement of the C3 (for a graphical representation, see R14) where, following equation~(\ref{eq:C3_eq}), the magnitude of the displacement and its rms error are
\begin{align}
\left|n_{S,i}-n_{S,\text{zp}}\right|=E_\text{h}(\bv)\cdot R_{c_x,i}^{\text{h}}\cdot \left|\gamma_{S,i}-\text{M}\{m_S\}\right|,\label{eq:n-nzp}\\
\text{rms}({\left|n_{S,i}-n_{S,\text{zp}}\right|})\approx E_\text{h}(\bv)\cdot R_{c_x,i}^{\text{h}}\cdot \text{rms}\{m_S\},\label{eq:rms_n-nzp}
\end{align}
respectively. In equation~(\ref{eq:rms_n-nzp}), we do not include the error induced by errors in $\gamma_{S,i}$, which is lower that 17 per cent of the uncertainty induced by the error in $\text{M}\{m_S\}$. In order to find the colour combination that maximizes the dust effect (equation~\ref{eq:n-nzp}) and minimizes its error (equation~\ref{eq:rms_n-nzp}), we define the quantity (a signal-to-noise ratio)
\begin{equation}\label{eq:chi_s}
\xi_{S,i}\equiv\frac{\left|n_{S,i}-n_{S,\text{zp}}\right|}{\text{rms}(\left|n_{S,i}-n_{S,\text{zp}}\right|)}\approx\frac{\left|\gamma_{S,i}-\text{M}\{m_S\}\right|}{\text{rms}\{m_S\}}.
\end{equation}
Therefore, the most appropriate colour set $S$ to compute $E_\text{h}(\bv)$ with the C3 method is one that maximizes $\xi_{S,i}$. 

\begin{figure}
\centering
\includegraphics[width=0.98\columnwidth]{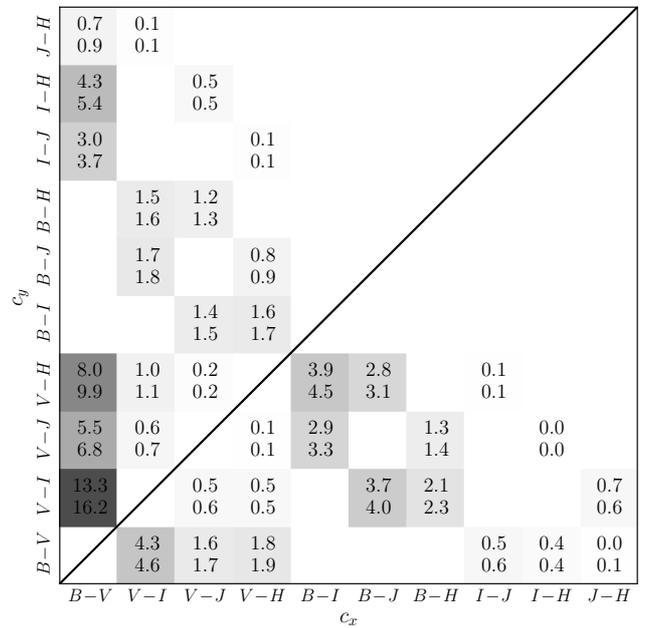}
\caption{Values of $\xi_{S,i}$ for different combinations of $c_x$ and $c_y$, using $\bv=0.0$ (top values) and $1.4$ (bottom values). Empty spaces indicate superfluous colour combinations.}
\label{fig:BVIJH}
\end{figure}

Fig.~\ref{fig:BVIJH} shows the $\xi_{S,t}$ values for all possible independent colours combinations with the \bvijh\ bands, using the $\text{M}\{m_S\}$ and $\text{rms}\{m_S\}$ values computed with our SN II sample (see Appendix~\ref{sec:C3_linearity}), and using $\bv=0.0$ and $1.4$, which are typical colours at the start and end of the photospheric phase, respectively. We do not include the $K$-band in this analysis because of the scarcity of SNe II with photometry in that band. The best colours combinations, independent of the intrinsic $\bv$, are those involving $\bv$, with $\vi$ versus $\bv$ the best. For this combination we obtain $\text{M}\{m_S\}=0.45\pm0.07$. We remark that colours combinations that do not include the $B$-band have $\xi_{S,t}\lesssim1.0$, which indicates that the noise induced by intrinsic differences of C3 slopes is greater than the effect of host galaxy dust, and therefore those combinations are not suitable for $E_\text{h}(\bv)$ measurement through the C3 method. We point out that colours combinations under the diagonal correspond to those above the diagonal but with the axes exchange. In principle they give the same information. However, by construction, they maximize displacement in $x$-axis instead of $y$-axis.

To compute the pdf of $n_{S,\text{zp}}$ for $S\!=\!\{\bv$, $\vi\}$, we use the data of SN 2003bn and SN 2013ej, which are affected by a negligible host galaxy colour excess (R14), maximizing the likelihood of a straight line with slope $0.45\pm0.07$. With this process, we obtain a pdf with median of 0.107 mag and $\text{rms}=0.053$ mag. Since the RM $p$-value for the latter distribution is $>\!0.05$, we treat it as a normal distribution.

To estimate $E_\text{h}(\bv)$, we use equation~\ref{eq:C3_eq} and each point in the $\vi$ versus $\bv$ colour-colour plot. The pdf of $E_\text{h}(\bv)$ is obtained in a similar way than the pdf of $n_{S,\text{zp}}$, but maximizing the likelihood of a constant-only model. We include in the final pdf of $E_\text{h}(\bv)$ the error induced by errors on $n_{S,\text{zp}}$ and $\text{M}\{m_S\}$. Median values and errors of $E_\text{h}(\bv)$ are listed in Column 2 of Table~\ref{table:SN_EBV_t0}. For our SN set we obtain $E_\text{h}(\bv)$ rms errors between 0.082 and 0.128 mag, with a median of 0.097 mag.

\begin{figure}
\centering
\includegraphics[width=0.90\columnwidth]{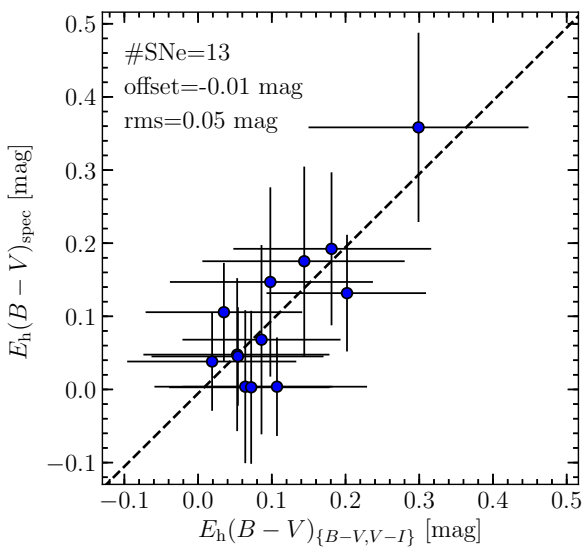}
\caption{Comparison between $E_\text{h}(\bv)$ obtained with the spectrum-fitting method ($y$-axis) and those obtained with the \vi\ versus \bv\ C3 method ($x$-axis). Dashed line represents the one-to-one correlation, with a median offset of $-0.01$ mag.}
\label{fig:SPEC_vs_BVI}
\end{figure}

Fig.~\ref{fig:SPEC_vs_BVI} shows the comparison between host galaxy colour excesses computed with $S=\{\bv,\vi\}$ with those obtained by \citet{Olivares_etal2010}, $E_\text{h}(\bv)_\text{spec}$ (Column 5 of Table~\ref{table:SN_EBV_t0}), which are based on the fit between observed spectra and SN II models. We measure a median offset of $-0.01$ mag, meaning that our estimations of $E_\text{h}(\bv)$ are slightly lower than those estimated by \citet{Olivares_etal2010}. Both methods are consistent within $\pm$0.05 mag.

\section{EXPLOSION EPOCH AND PHOTOSPHERIC VELOCITY}\label{sec:vph_texp}

The explosion epoch and the photospheric velocity are, under the assumption of homologous expansion, the unique parameters determining the actual size of the photosphere \citep{Kirshner_Kwan1974}.

\subsection{Photospheric Velocities}\label{sec:photospheric_velocity}

The most widely used method to estimate SN photospheric velocities consists of measuring the blueshift of P Cygni absorption minima in SN spectra \citep{Kirshner_Kwan1974, Eastman_Kirshner1989}. Weak lines, like those from \ion{Fe}{II} species, are typically used under the assumption that they are formed near the photosphere \citep[e.g.,][]{Leonard_etal2002_99em}. A more confident method to estimate photospheric velocities is through the cross-correlation technique \citep{Hamuy_etal2001,Takats_Vinko2012}, where observed spectra are compared to those from SN models which have known photospheric velocities. The application of the latter method is beyond the scope of this work, therefore we will use velocities derived from the \ion{Fe}{II} $\lambda$5169 line absorption minima as a proxy for the photospheric velocity.

To estimate \ion{Fe}{II} $\lambda$5169 absorption minima with appropriate errors, we have to consider the uncertainties induced by the noise and spectral resolution ($\Delta\lambda$) of each spectrum, and also by the endpoints we choose for the line profile.

We estimate the noise on the \ion{Fe}{II} $\lambda$5169 line profile of each spectrum performing a \texttt{loess} fit and then removing it to the observed line profile. Then we generate $10^4$ simulated line profiles, varying randomly the noise over the \texttt{loess} fit, wavelengths within $\Delta\lambda$, and endpoints. For each realization we apply a \texttt{loess} fit, registering the minimum value. The output of this process is a distribution of absorption minima, which we convert to velocities using the relativistic Doppler equation. With this process we obtain typical $v_{\text{ph}}$ rms errors between 30 and 230 \kms, with a median of 76 \kms.

Photospheric velocities are estimated from spectroscopic data, corrected for the SN heliocentric redshift. 

In some cases, SN II spectra shows narrow emission lines as result of a superposed \ion{H}{II} region at the SN position. These narrow lines allow a good estimation of the SN heliocentric redshift, under the assumption that the SN is spatially close to the \ion{H}{II} region \citep[e.g.,][]{Anderson_etal2014_blueshifted_emission}. When those lines are not present in the SN spectra, the heliocentric redshift of the host galaxy is used as a proxy for the SN heliocentric redshift. However, since most of the SNe II in our set explode in spiral galaxies, the SN heliocentric redshift has a component due to the galaxy rotation. \citet{Anderson_etal2014_blueshifted_emission} computed heliocentric redshifts of 72 SNe II using \ion{H}{II} region narrow emission lines, and comparing with heliocentric redshifts of the host galaxy nucleus, they obtained a zero-centred distribution with a rms of 162 \kms, which is attributed to the galaxy rotation effect. 

In our sample, 11 SNe II (SN 2002gd, SN 2002gw, SN 2002hj, SN 2003B, SN 2003E, SN 2003bl, SN 2003bn, SN 2003ci, SN 2005ay, SN 2009N, and SN 2014G) show \ion{H}{II} region narrow emission lines in the spectra, which we use to estimate the heliocentric redshift. Another six SNe (SN 2003T, SN 2004et, SN 2005cs, SN 2008in, SN 2012aw, and SN 2013ej) exploded within nearly face-on galaxies, in which case we adopt the redshift of the host galaxy nucleus. For SN 1999em we adopt the value from \citet{Leonard_etal2002_99em}, and for SN 2003hn we use the average of the \ion{Na}{I} D velocities measured by \citet{Sollerman_etal2005}. The remaining five SNe (SN 2003cn, SN 2009ib, SN 2009md, SN 2012A, and SN 2012ec) did not occur within nearly face-on galaxies, and do not show \ion{H}{II} region narrow emission lines in the spectra. For those cases we adopt the redshift of the host galaxy nucleus, with an error of 162 \kms\ (that we assume normal) to take into account the host galaxy rotational velocity. Adopted SN heliocentric redshifts are listed in Table~\ref{table:SN_sample}.

\subsection{Explosion Epoch}\label{sec:explosion_epochs}

The SN explosion epoch can be estimated by means of photometric information; it can be constrained between the last non-detection $t_\text{ln}$ and the first detection $t_\text{fd}$ (e.g., \citealt{Nugent_etal2006,Poznanski_etal2009,DAndrea_etal2010}; R14; \citealt{Valenti_etal2016}), or estimated through a polynomial fit to the rise-time photometry when it is available \citep[e.g.,][]{Gonzalez-Gaitan_etal2015,Gall_etal2015}. The spectroscopy of a SN can also provide information about its explosion epoch by means of the comparison with other spectra of SNe with explosion epoch estimated through photometric information \citep[e.g.,][]{Anderson_etal2014_V_LC,Gutierrez_etal2017_I}.

Column 4 and 5 of Table~\ref{table:SN_EBV_t0} lists the $t_\text{ln}$ and $t_\text{fd}$ values of the SNe in our set, respectively. The explosion epochs for our set are typically constrained within 14 d using photometric information, which is twice the range suggested by R14 (namely, 7 d) to reduce errors induced by $t_0$ errors over PMM distances. We need, therefore, to include spectroscopic information in order to better constrain the explosion epochs.

As was done by \citet{Anderson_etal2014_V_LC} and \citet{Gutierrez_etal2017_I}, to estimate $t_0$ we use optical spectroscopy along with the Supernova Identificator code \citep[SNID;][]{Blondin_Tonry2007}, which finds by cross-correlation the spectra from its SN library that are more similar to the input spectrum. For a good estimation of $t_0$ with SNID, we need a library with spectra of an ample amount of SNe II that sample the high spectral diversity displayed by SNe II \citep[e.g.,][]{Gutierrez_etal2017_I} and with $t_0$ constrained by photometric information. In this work, we compile optical spectroscopy of 59 SNe II with $t_0$ constrained within 10 d (for more details, see Appendix~\ref{sec:phase_estimation_with_SNID}).

To estimate the explosion epoch of a given SN (SN$_\text{input}$) with $N$ spectra ($\lbrace\text{spec}\rbrace$) through SNID and using our SN II templates library, we perform the following procedure:
\begin{enumerate}
\item[1.]We run SNID using as input the $N$ spectra of SN$_\text{input}$ earlier than 40 d since the first detection. The SNID output for each spectrum is a list with the best-matching templates, their phase since explosion, and their $r$lap parameter (which indicates the strength of the correlation).
\item[2.]We convert phases since explosion to explosion epochs (since we know the phase of each SN$_\text{input}$ spectrum). The associated errors are derived from the $r$lap values through a procedure described in Appendix~\ref{sec:phase_estimation_with_SNID}. 
\item[3.]From each of the $N$ lists, we select the first ten best-matching templates with $r$lap $>$ 5.0, compiling them in a unique list. From this list, we extract a sublist for each of the $M$ best-matching SNe (SN$_\text{bm}$). With each of the $M$ sublist, we compute the SN$_\text{in}$ explosion epoch as the average, taking the standard deviation as the associated error, and including the respective explosion epoch error of the SN$_\text{bm}$ through a Monte Carlo error propagation. If a spectrum gives a median $t_0$ greater than 40 d, then we remove it from the analysis.
\item[4.] After that, we compute the likelihood $\mathcal{L}(t_0|\lbrace\text{spec}\rbrace)$ with the $M$ results, including an error of 4.1 d which is the rms obtained from the comparison between explosion epochs constrained with photometric information and those derived with SNID (see Appendix~\ref{sec:phase_estimation_with_SNID}).
\item[5.] Finally, we obtain the posterior pdf of $t_0$, $p(t_0|\lbrace\text{spec}\rbrace)$, combining $\mathcal{L}(t_0|\lbrace\text{spec}\rbrace)$ with the uniform prior on $t_0$, $p(t_0)$, provided by photometric information, i.e., 
\begin{equation}
p(t_0|\lbrace\text{spec}\rbrace)\propto \mathcal{L}(t_0|\lbrace\text{spec}\rbrace)p(t_0).
\end{equation}
\end{enumerate}

Table~\ref{table:SN_EBV_t0} lists the medians and rms values of the explosion epochs obtained with SNID and without any prior on $t_0$ (Column 7), and the median of $p(t_0|\lbrace\text{spec}\rbrace)$ of each SN along with the 80 per cent CI (Column 8).

\section{APPLYING THE PMM}\label{sec:PMM}

Once all observables required for the PMM are available, the next step is to prepare the data before applying the method. As was mentioned in Section~\ref{sec:light_curve_fits}, we interpolate photometry to the epochs of the photospheric velocities. Since we want to study the PMM distance precision at different photometric bands, i.e., changing only the photometry, we use epochs where spectroscopy is covered simultaneously by optical and near-IR photometry. In the case of SN 2002hj, it does not have spectroscopy covered by $J$-band photometry, so we interpolate photospheric velocities (using \texttt{loess}) and $\bvih$ photometry to the epochs of $J$-band photometry.

\subsection{Calibration}\label{sec:PMM_calibration}

For the PMM calibration, we express $a_{x,\Delta t}$ as
\begin{equation}\label{eq:ax_cal}
a_{x,\Delta t}= \mathrm{ZP}_x+a^*_{x,\Delta t},
\end{equation}
where $\mathrm{ZP}_x$ is the zero-point of the PMM in the $x$-band, and $a^*_{x,\Delta t}$ is a function that represents the dependence of $a_{x,\Delta t}$ on $\Delta t$ (without the constant term).

To estimate the evolution of $a^*_{x,\Delta t}$ with $\Delta t$, we use the $a^*_{x,\Delta t_i}$ values of the SNe in our set with two or more $v_\mathrm{ph}$ measurements during the photospheric phase. For each SN, the $a^*_{x,\Delta t_i}$ values are given by
\begin{equation}\label{eq:ax_cal}
a^*_{x,\Delta t_i} = m_{x,i}^\text{corr}+5\log\left(\frac{v_{\text{ph},i}}{5000\text{ km s}^{-1}}\right)+\delta_\mathrm{SN},
\end{equation}
where $\delta_\mathrm{SN}$ is an additive term to normalize the $a^*_{x,\Delta t_i}$ values of each SN to the same scale. Based on the definition given in R14, the dependence of $a^*_{x,\Delta t}$ on $\Delta t$ has the form
\begin{equation}\label{eq:ax_cal}
a^*_{x,\Delta t}= f_{x,\Delta t} -5 \log{\left(\frac{\Delta t}{100\text{ d}}\right)}.
\end{equation}

We express the dependence of $f_{x,\Delta t}$ on $\Delta t$ through polynomials. We use the AIC/BIC to determine the optimum polynomial order for $f_{x,\Delta t}$ and the values of $\delta_\mathrm{SN}$, while to estimate the time range of applicability of the PMM, we group the $f_{x,\Delta t_i}$ values in bins of width 10 d and then we compute the rms of the points in each bin. We found that rms values are lower in a range 35--75 d since the explosion. Among all optimum orders for \bvijh\ bands (see Table~\ref{table:RKRf_statistics}), we select the order that the different bands have in common, i.e., order one. With this, we prevent that differences in the rms($f_{x,\Delta t}$) value for different bands are due to differences in the order of the polynomial fit. To estimate error in the parameters, we perform $10^4$ bootstrap resampling. Table~\ref{table:f_ZP_parameters} lists $f_{x,\Delta t}$ fits parameters for $\bvijh$ bands.
\begin{table}
\caption{Parameters for the PMM calibration.}
\label{table:f_ZP_parameters}
\begin{tabular}{lccccc}
\hline
     &\multicolumn{3}{c}{$f_{x,\Delta t}=c_{x}\cdot\Delta t/(100\text{ d}^{-1})$} &\multicolumn{2}{c}{zero-point}\\
\cmidrule(rr){2-4} \cmidrule(rr){5-6}
$x$ & $c_{x}$                        & rms                      & $p$(RM)                      & $\mathrm{ZP}_x$  & rms \\
\hline
$B$ & $  3.87_{-0.26}^{+0.32}$ & $0.09$ & $0.05$ & $-19.51_{-0.52}^{+0.52}$  & $0.27$ \\
$V$ & $  2.78_{-0.22}^{+0.31}$ & $0.09$ & $0.32$ & $-20.03_{-0.39}^{+0.39}$  & $0.19$ \\
$I$ & $  2.22_{-0.24}^{+0.34}$ & $0.10$ & $0.70$ & $-20.36_{-0.23}^{+0.23}$  & $0.15$ \\
$J$ & $  2.13_{-0.27}^{+0.29}$ & $0.10$ & $0.51$ & $-20.64_{-0.12}^{+0.12}$  & $0.10$ \\
$H$ & $  2.05_{-0.28}^{+0.28}$ & $0.09$ & $0.92$ & $-20.77_{-0.09}^{+0.09}$  & $0.07$ \\
\hline
\multicolumn{6}{m{0.85\linewidth}}{Parameters are valid for $35\text{ d}<\Delta t<75\text{ d}$. Errors are the 99 per cent CI.}
\end{tabular}
\end{table}

\begin{figure*}
\includegraphics[width=\columnwidth]{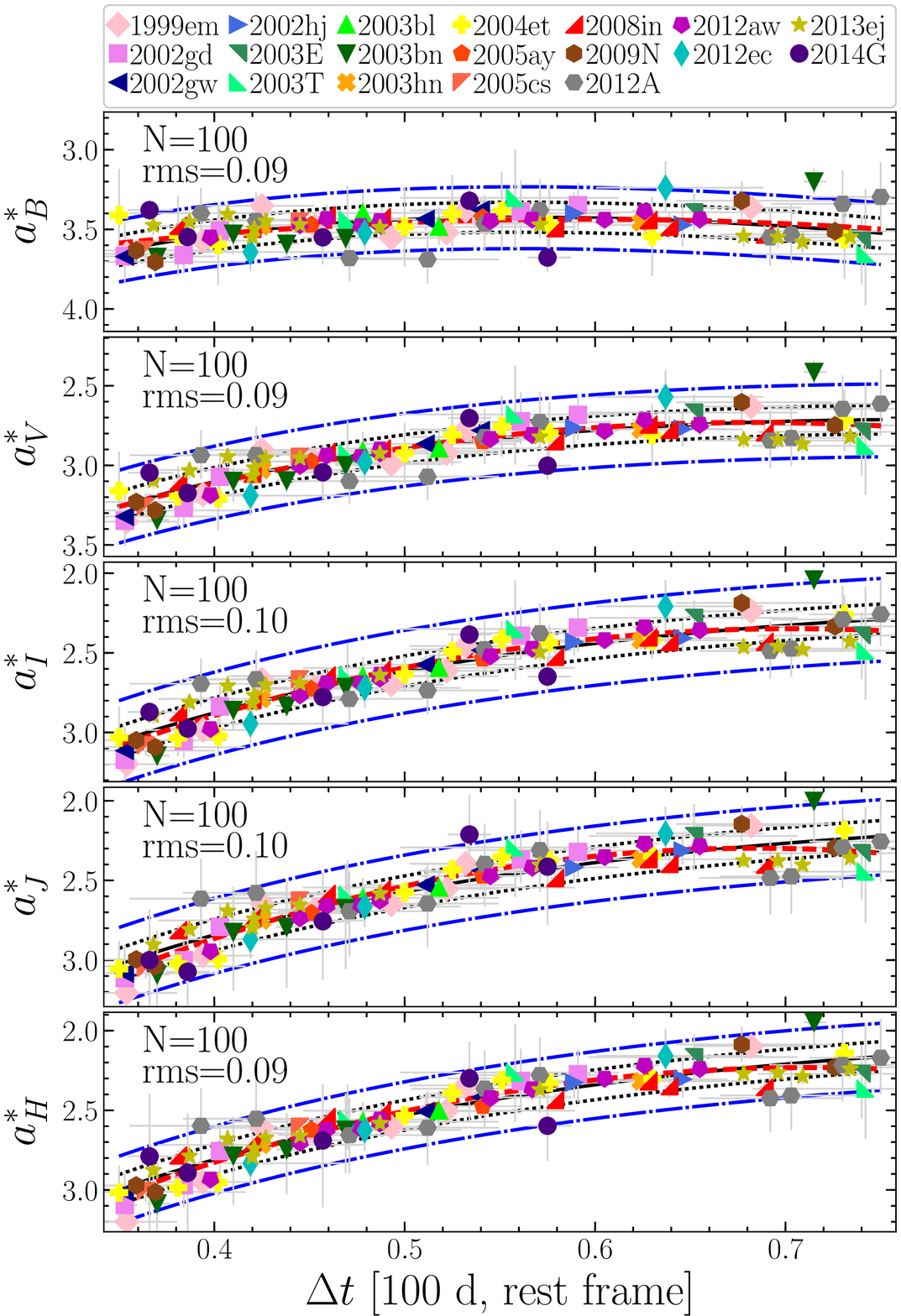}
\includegraphics[width=\columnwidth]{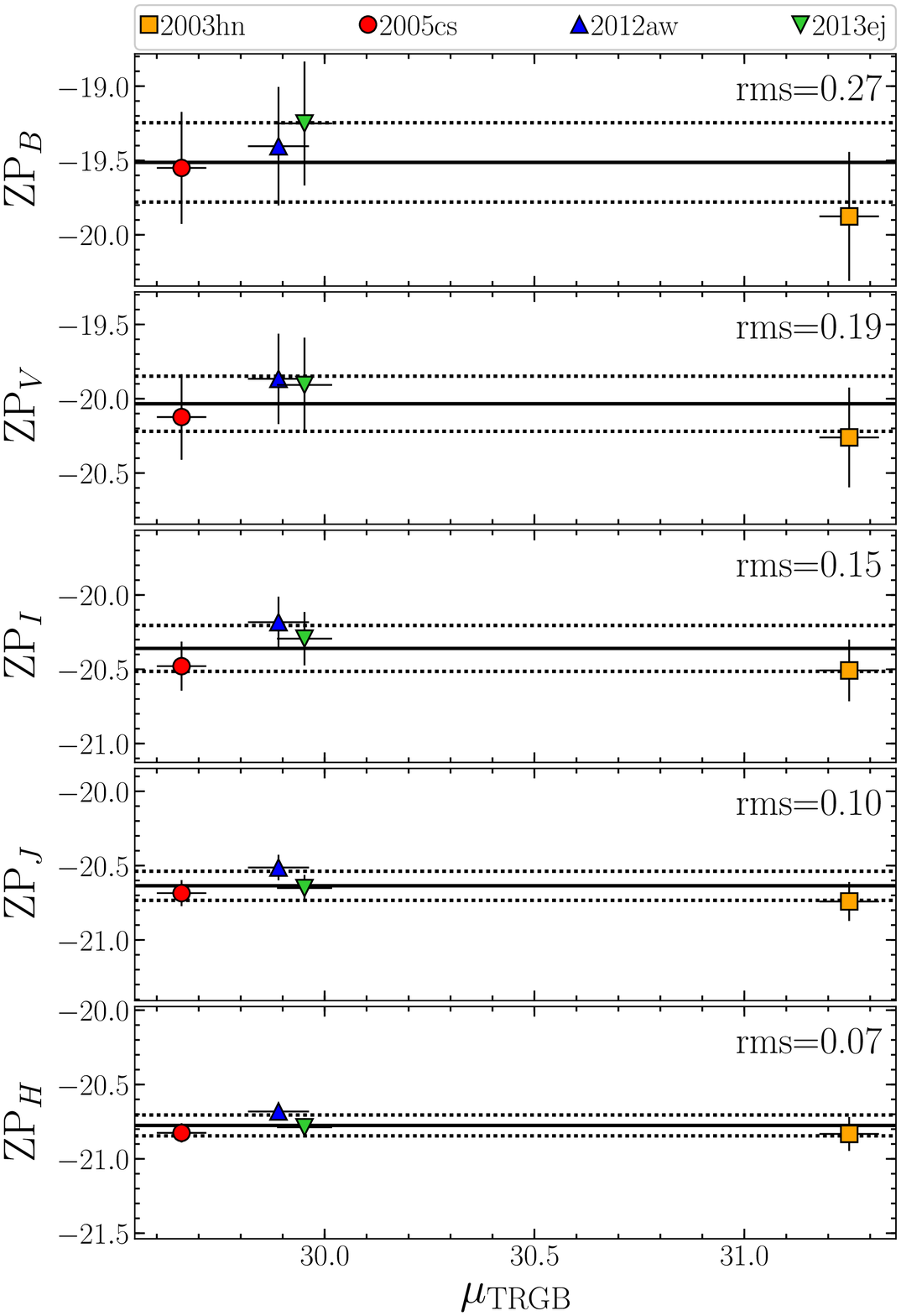}
\caption{Left: values of $a^*_x$ as a function of $\Delta t$ for $\bvijh$. Black solid lines correspond to the parametric fits. Red dashed lines correspond to \texttt{loess} regressions. Dotted lines indicate values within one rms, while blue dash-dotted lines are the inner fences. Right: $\mathrm{ZP}_x$ values for $\bvijh$ derived with four SNe II in galaxies with TRGB distances.}
\label{fig:pmm_BVIJHK}
\end{figure*}

The left half of Fig.~\ref{fig:pmm_BVIJHK} shows the values of $a^*_{x, \Delta t_i}$ as a function of $\Delta t_i$ for $\bvijh$ bands. The variation of the \texttt{loess} fits (red dashed lines) are within one rms (black dotted lines), which means that polynomial fits we adopted capture almost all the dependence on $\Delta t$.

The PMM zero-points can be obtained using a sample of SNe II at known distances where, for each SN, we have
\begin{equation}
\mathrm{ZP}_x^\mathrm{SN}=\mu^*_x-\mu_\mathrm{host}^\mathrm{SN}.
\end{equation}
Here, $\mu_\mathrm{host}^\mathrm{SN}$ is the SN host galaxy distance modulus and $\mu^*_x$ is the SN pseudo-distance modulus. The latter, for each measurement of $v_\mathrm{ph}$ at time $t_i$, is defined similar to equation~\ref{eq:mu} but with $a^*_{x,\Delta t_i}$ instead of $a_{x,\Delta t_i}$, i.e.,
\begin{equation}\label{eq:pseudo_mu}
\mu^*_{x,i} = m_{x,i}^\text{corr}-a^*_{x,\Delta t_i}+5\log\left(\frac{v_{\text{ph},i}}{5000\text{ km s}^{-1}}\right).
\end{equation}
The pdfs of $\mu^*_{x,i}$ are obtained through equation \ref{eq:pseudo_mu} using the pdfs of the observables for each photospheric velocity epoch. Finally, we combine the pdfs of $\mu^*_{x,i}$ in a unique $\mu^*_x$ pdf maximizing the likelihood (equation~\ref{eq:likelihood}) for a constant-only model.

To compute accurate $\mathrm{ZP}_x$ values, we need SNe II in galaxies with distances measured with the best possible precision. Among the SNe that we compiled from the literature, there are only three (SN 1999em, SN 2003hn, and SN 2012aw) in galaxies with distances measured through Cepheid, and five (SN 2003hn, SN 2004et, SN 2005cs, SN 2012aw, and SN 2013ej) in galaxies with distances measured with TRGB. Cepheid distances for the hosts of SN 1999em and SN 2012aw were reported by \citet{Saha_etal2006}, while \citet{Riess_etal2016} reported the Cepheid distance of the host of SN 2003hn. Comparing Cepheid distances of six galaxies in common between the two publications (NGC 1365, NGC 3370, NGC 3982, NGC 4536, NGC 4639, and NGC 5457) we found that Cepheid distances reported by \citet{Saha_etal2006} are, on average, 0.19 mag greater than those reported by \citet{Riess_etal2016}, showing a rms of 0.13 mag. The latter could indicate a systematic difference between the two calibrations, which can introduce an undesirable noise on the $\mathrm{ZP}_x$ estimation if we rescale \citet{Saha_etal2006} distances to the \citet{Riess_etal2016} calibration. For this reason, we decide to use only SNe in galaxies with TRGB distances, that can be homogenized to the \citet{Jang_Lee2017_TRGB_ZP} calibration, which is based on the distance to the Large Magellanic Cloud and NGC 4258. Recalibrated TRGB distances are listed in Column 6 of Table~\ref{table:SN_sample}. From these five SNe II, we discard SN 2004et since the TRGB distance of its host is at least 0.59 mag higher compared to the distances we compute for SN 2004et and two other SNe II that exploded in the same galaxy (see Appendix~\ref{sec:NGC_6946_distance}).

The right half of Fig.~\ref{fig:pmm_BVIJHK} shows the ZP$_x^{\mathrm{SN}}$ values for $\bvijh$. As in the case of $\mu^*_x$, the pdf of ZP$_x$ is obtained combining the pdfs of ZP$_x^\mathrm{SN}$. Median values, 99 per cent CI, and rms values for $\mathrm{ZP}_{x}$ are summarized in Table \ref{table:f_ZP_parameters}.

\begin{figure*}
\includegraphics[width=\columnwidth]{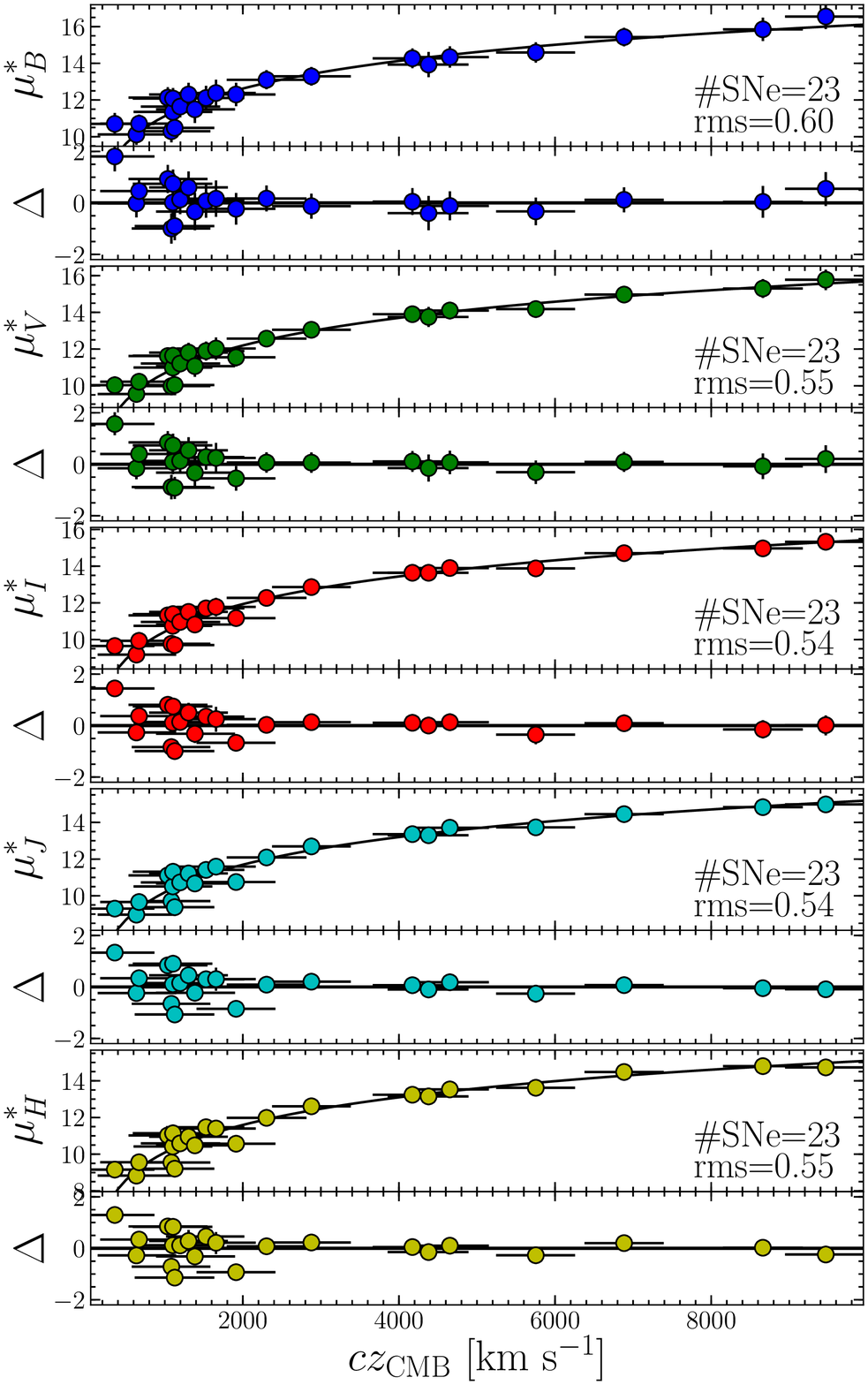}
\includegraphics[width=\columnwidth]{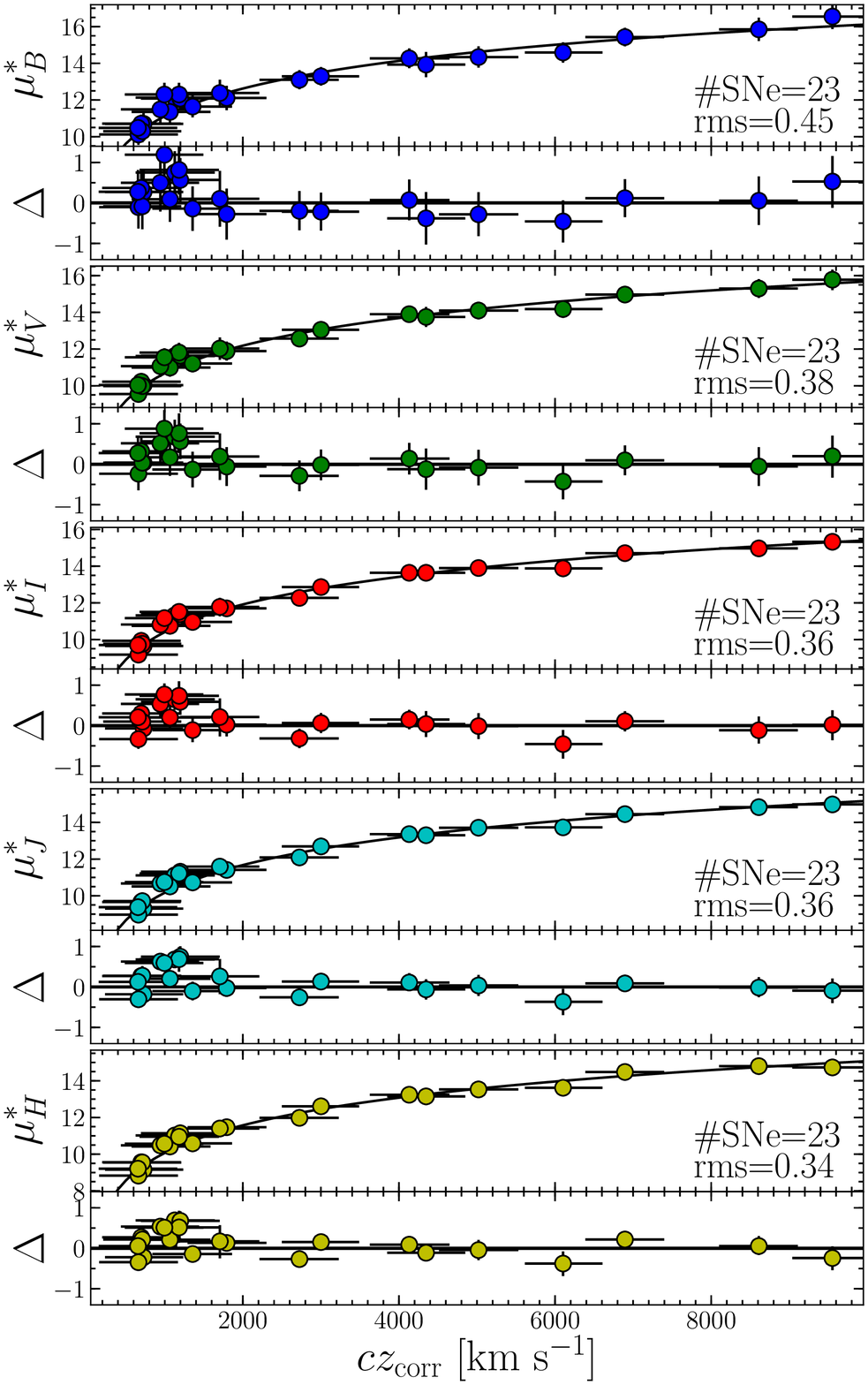}
\caption{Hubble diagrams for SNe II with PMM distances and CMB redshifts (left) and redshifts corrected for the Virgo and Great Attractor infall (right). Solid lines correspond to Hubble law fits. Residuals are plotted at bottom of each panel.}
\label{fig:HD_all}
\end{figure*}

Once the PMM zero-points are computed, we can estimate the distance modulus for each band as $\mu_x = \mu^*_x-\mathrm{ZP}_x$. Median values, 80 per cent CI, and rms values for $\mu_x$ are summarized in Table~\ref{table:SN_distances}, where we include the TRGB zero-point systematic error of 0.058 mag \citep{Jang_Lee2017_TRGB_ZP}.

\subsection{Hubble Diagrams}\label{sec:Hubble_diagrams}

To investigate the PMM distance precision, we construct HDs. We convert heliocentric host galaxy redshifts to cosmological ones using as reference the cosmic microwave background (CMB) dipole \citep{Fixsen_etal1996}. Redshift errors are dominated by peculiar velocities, with a rms of 382 \kms\ for local SN Ia host galaxies \citep[$z<0.08$, ][]{Wang_etal2006}, followed by the error in the determination of the Local Group velocity \citep[rms of 187 \kms,][]{Tonry_etal2000}. CMB redshifts and their rms errors are listed in Table~\ref{table:SN_distances}.

Taking into account the pdfs of the pseudo-distance moduli ($\mu^*_x$) and the pdfs of the CMB redshifts ($cz_\text{CMB}$), we compute the Hubble diagram intercept ($a_{\mathrm{HD},x}$) maximizing the likelihood (equation~\ref{eq:likelihood}), where the model for the pseudo-distance modulus is given by the Hubble law
\begin{equation}\label{eq:Hubble_law}
\mu_x^*=a_{\mathrm{HD},x}+5\log(cz_\text{CMB}).
\end{equation}

\begin{figure*}
\includegraphics[width=\columnwidth]{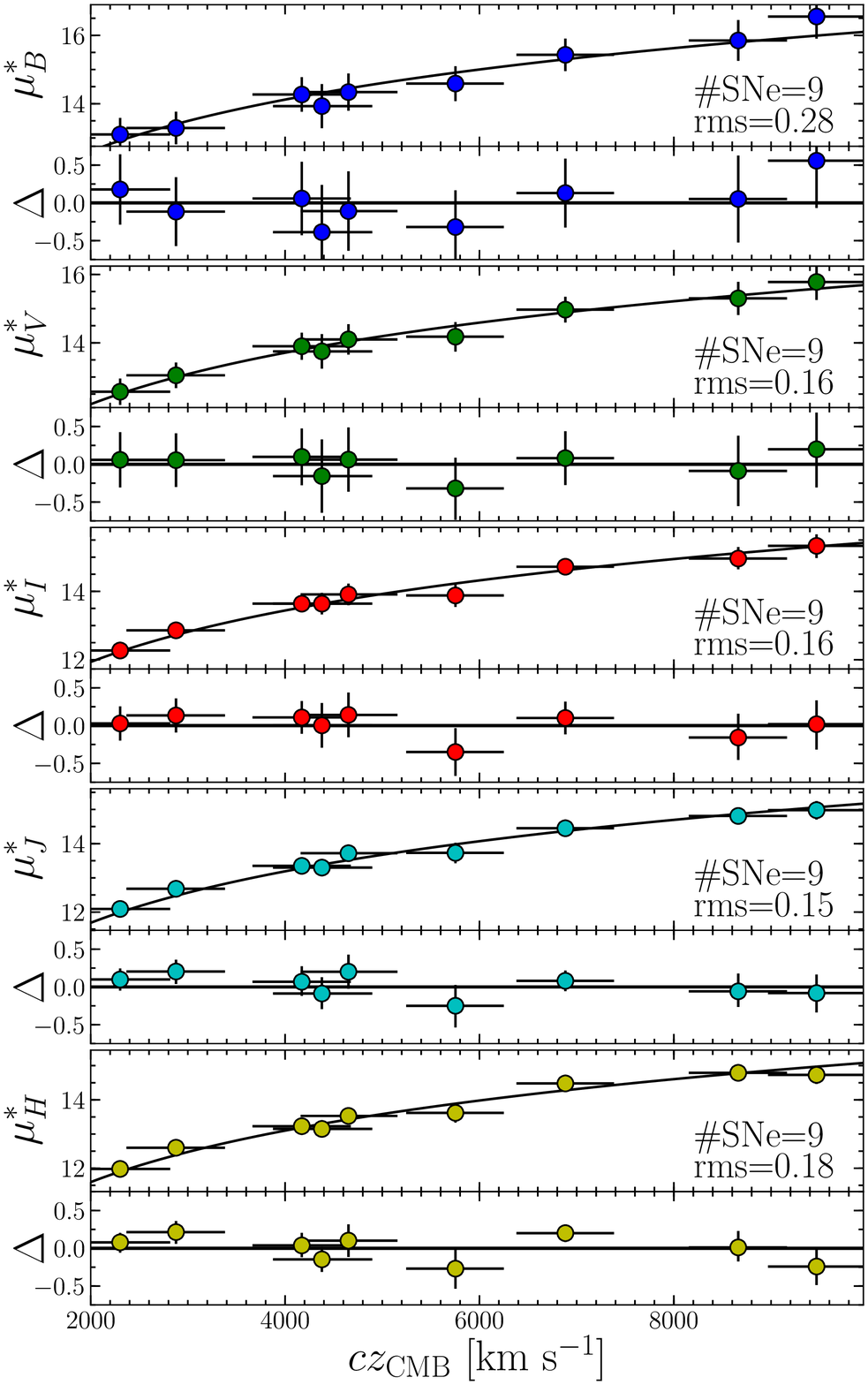}
\includegraphics[width=\columnwidth]{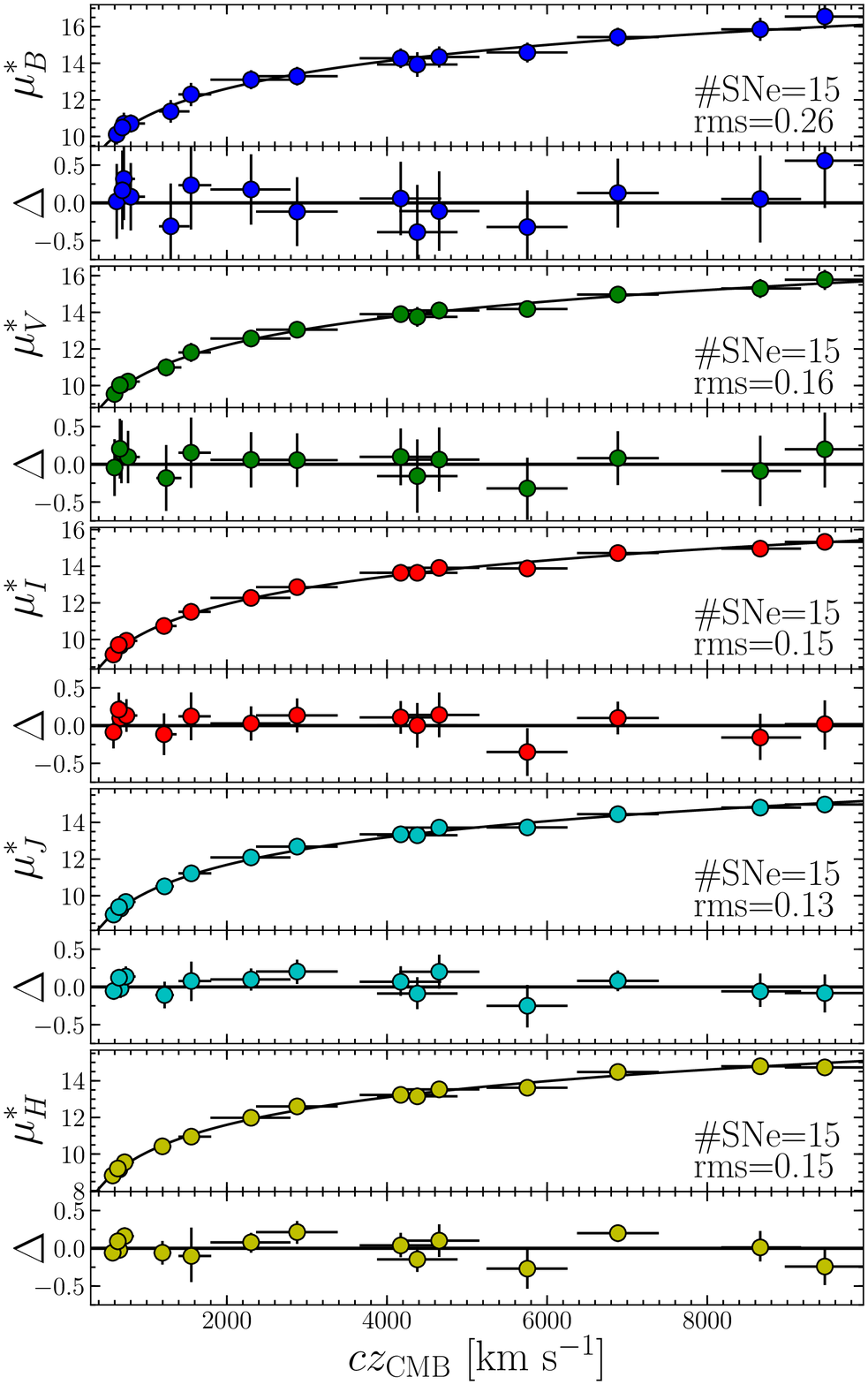}
\caption{Hubble diagrams for SNe II with distances computed with the PMM, using the nine SNe II at $cz>2000$ \kms\ (left) and including the six SNe II in galaxies with distance measured with TRGB, Cepheids, or SN Ia (right), where the $H_0$ values are those computed with the nine SNe II at $cz>2000$ \kms , and used to convert distances of the six nearest SNe II to redshifts. Solid lines correspond to the Hubble law fits. Residuals are plotted at the bottom of each panel.}
\label{fig:HD_final}
\end{figure*}

The left half of Fig.~\ref{fig:HD_all} shows HDs for \bvijh\ bands, using PMM distances for all SNe in our set. The rms, greater than $0.5$ mag for all bands, is mostly produce by peculiar velocities of host galaxies at low redshift. In fact, the median redshift of the host galaxies in the HD is 1528 \kms, where a redshift error of 382 \kms\ translates into a magnitude error of 0.54 mag. Indeed, if we use redshifts corrected for the infall of the Local Group toward the Virgo cluster and the Great Attractor ($cz_\mathrm{corr}$) instead of CMB redshifts, we obtain a HD rms of 0.34--0.38 mag for \vijh\ bands (see the right half of Fig.~\ref{fig:HD_all}). We note that even after infall corrections the scatter in the HDs is still mostly due to SNe in galaxies with $cz<2000$ \kms . Therefore, to estimate the PMM distance precision and the Hubble constant ($H_0$), given by
\begin{equation}
\log{H_0}=(25-a_{\mathrm{HD},x}+\mathrm{ZP}_x)/5,
\end{equation}
we use only SNe II with $cz>2000$ \kms\ and, as visible in the left half of Fig.~\ref{fig:HD_final}, the HD rms decreases significantly. The corresponding values of $H_0$ and rms are listed in Table~\ref{table:HD_parameters}.

\begin{table}
\caption{Hubble diagram parameters.}
\label{table:HD_parameters}
\begin{tabular}{@{}c@{\,\,\,\,}c@{\,\,}c@{\,\,\,\,}c@{\,\,}c@{\,\,\,\,}c@{\,\,}c@{\,\,\,}c@{}}
\hline
     & \multicolumn{2}{c}{All$^a$} & \multicolumn{4}{c}{$cz_\mathrm{CMB}>2000$ \kms} & Final$^d$ \\
     \cmidrule(rr){2-3} \cmidrule(rr){4-7} \cmidrule(rr){8-8}
     & $cz_\mathrm{CMB}$     &$cz_\mathrm{corr}$ & \multicolumn{2}{c}{$R_V\!=\!3.1^b$} & \multicolumn{2}{c}{$R_V\!=\!2.3^b$} &       \\
     \cmidrule(rr){2-2} \cmidrule(rr){3-3} \cmidrule(rr){4-5}  \cmidrule(rr){6-7}
Band & rms    & rms & $H_0^c$              & rms    & $H_0^c$              & rms    & rms    \\
    
\hline
 $B$ & $0.60$ & $0.45$ & $74.9_{-9.2}^{+10.6}$ & $0.28$ & $75.8_{-9.6}^{+11.0}$ & $0.30$ & $0.26$ \\
 $V$ & $0.55$ & $0.38$ & $71.4_{-7.3}^{+8.2}$  & $0.16$ & $71.5_{-7.1}^{+7.9}$  & $0.16$ & $0.16$ \\
 $I$ & $0.54$ & $0.36$ & $69.5_{-6.1}^{+6.7}$  & $0.16$ & $69.4_{-5.8}^{+6.4}$  & $0.16$ & $0.15$ \\
 $J$ & $0.54$ & $0.36$ & $68.6_{-5.2}^{+5.7}$  & $0.15$ & $68.7_{-5.2}^{+5.7}$  & $0.15$ & $0.13$ \\
 $H$ & $0.55$ & $0.34$ & $67.1_{-5.0}^{+5.4}$  & $0.18$ & $67.2_{-5.0}^{+5.4}$  & $0.18$ & $0.15$ \\
\hline
\multicolumn{8}{m{7.9cm}}{$^a$ We do not report the $H_0$ values for this SN set because they are severely affected by peculiar velocities.}\\
\multicolumn{8}{m{7.9cm}}{$^b$ $R_V$ value adopted for SN II host galaxies.}\\
\multicolumn{8}{m{7.9cm}}{$^c$ In \hnotunits. Errors are the 80 per cent CI, and include the TRGB zero-point systematic error and the error in the determination of the Local Group velocity.}\\
\multicolumn{8}{m{7.9cm}}{$^d$ Set of 15 SNe II: nine SNe at $cz_\text{CMB}>2000$ \kms\ plus six SNe in galaxies with TRGB, Cepheids, or SN Ia distances.}
\end{tabular}
\end{table}

The values of $H_0$ range between 67.1 and 74.9 \hnotunits . Taking into account that the ZP$_x$ values were calibrated using TRGB distances in the scale of \citet{Jang_Lee2017_TRGB_ZP}, our $H_0$ values, as expected, are consistent within the errors with those reported in \citet{Jang_Lee2017}, i.e., $71.17\pm1.66\pm1.87$ \hnotunits, which also use the \citet{Jang_Lee2017_TRGB_ZP} calibration.

As visible in Column 4 of Table~\ref{table:HD_parameters}, all the $H_0$ values are compatible within their errors. However, we note that $H_0$ decreases moving from shorter to longer wavelengths, which could suggest a systematic introduced by: (1) our assumption of the $R_V$ value for the SN host galaxies, or (2) an underestimation and/or an overestimation of the host galaxy colour excesses for the four SNe in the PMM calibration set and the the nine SNe at $cz_\text{CMB}>2000$ \kms, respectively. To test the first possibility, we change the $R_V$ adopted for SN II host galaxies to the lowest $R_V$ value for which the \citet{Fitzpatrick1999} extinction curve is defined ($R_V=2.3$). As is visible in Column 6 of Table~\ref{table:HD_parameters}, there are no significant changes in the $H_0$ values. For the second possibility, we found that an underestimation of $E_\text{h}(\bv)$ for the SNe in the calibration set, or an overestimation of $E_\text{h}(\bv)$ for the SNe at $cz_\text{CMB}>2000$ \kms , of 0.05--0.07 mag can erase the tension between the $H_0$ values for all bands. Given the typical statistical $E_\text{h}(\bv)$ error of 0.097 mag (see Section~\ref{sec:host-galaxy_reddening}), the probability of obtaining an $E_\text{h}(\bv)$ underestimation of 0.05--0.07 mag with four objects is of 8 per cent, while to obtain an overestimation in a same amount for nine objects is of 5 per cent. It is worth mentioning that also the scatter in ZP$_x$ decreases going from shorter to longer wavelengths, suggesting again a trend introduced by the combination of a large uncertainty in the colour excess estimation and the low number statistics.

Regarding the HD scatter, we note that the rms of 0.28 mag obtained with the $B$-band decreases to 0.15--0.18 mag for the $\vijh$ bands. Despite the good results, the low number of SNe II within galaxies at $cz_\text{CMB}>2000$ \kms\ makes the result not statistically robust. Therefore, to check the PMM distance precision, it is necessary to include more SNe II into the analysis. Thus, we include the four SNe II used for the PMM calibration, plus other two in galaxies having Cepheid and SN Ia distances. The latter distances are converted to redshifts through the Hubble law (equation~\ref{eq:Hubble_law}), where for each band we use the $H_0$ value listed in Column 4 of Table~\ref{table:HD_parameters} and the ZP value given in Table~\ref{table:f_ZP_parameters}.

The right half of Fig.~\ref{fig:HD_all} shows the HDs computed with the selected SNe II based on the aforementioned criterion, which correspond to our final sample. For $V\!I$ bands we obtain a HD rms of 0.15--0.16 mag. The lowest HD rms is obtained with the $J$-band, whose rms of 0.13 mag translates into a distance precision of 6 per cent. This value, compared to the rms of 0.15--0.26 mag obtained for optical bands, suggests that using the $J$-band photometry instead of optical one to measure PMM distances can improve the precision by at least 0.07 mag.

For the $H$-band we expected a similar HD rms than for $J$-band since, among \bvijh\ bands, the $H$-band is the least affected by dust extinction. We, however, obtained a HD rms of 0.15 mag. The latter can be partially due to the higher photometry error of the $H$-band (of 0.07 mag) with respect to the error of the $J$-band (of 0.05 mag).

\section{DISCUSSION}\label{sec:discussion}

\subsection{Statistical significance of the result}

Given the small size of our SN sample, the HD rms of 0.13 mag we measured for the $J$-band is not statistically robust enough to be considered as a measure of the PMM distance precision in that band. In particular, we want to know the probability of measuring a rms $\leq0.13$ mag with $N=15$ values drawn from a parent distribution with standard deviation ($\sigma$) $\geq$ 0.13 mag. Assuming a normal parent distribution, the quantity $(N\!-\!1)(\mathrm{rms}/\sigma)^2$ has a chi-square distribution with $N\!-\!1$ degrees of freedom. Using this, we found that there is a 1 per cent chance that the parent distribution has $\sigma=0.23$ mag. Therefore, with the evidence we have, we can set an upper limit on the PMM distance precision with the $J$-band of 10 per cent with a CL of 99 per cent.

\subsection{Comparison with other methods}

For a consistent comparison of our results with those from other SN II distance measurement methods, we select results from works that uses a sample of SNe II at $z\approx0.01$--$0.03$.

Table~\ref{table:other_methods} lists the best distance precisions reached by other SN II distance measurement methods along with results obtained in this work. We note that the precision we report in this work is significantly lower than the best dispersion obtained by other works with SCM and PCM.

We also compare PMM and SCM applied to the same sample for $J$- and $H$-band. For this comparison we discard SN 2002hj because there is not photometry in $J$-band at 50 d since explosion. As visible in Table~\ref{table:other_methods}, the dispersion is lower by $\sim$30--40 per cent in both band.

\begin{table}
\caption{Distance precision of different methods.}
\label{table:other_methods}
\begin{tabular}{l c c c l}
\hline
Method      &HD rms& Band& $\#$SNe & Reference                \\
\hline
PCM         & 0.43 & $Y$ & 30      & \citet{deJaeger_etal2015}\\
SCM         & 0.25 & $B$ & 19      & \citet{Olivares_etal2010}\\
PMM         & 0.13 & $J$ & 15      & This work                \\
\hline
PMM$^{a}$   & 0.14 & $J$ & 14      & This work                \\
PMM$^{a}$   & 0.14 & $H$ & 14      & This work                \\
SCM$^{a,b}$ & 0.21 & $J$ & 14      & This work                \\
SCM$^{a,b}$ & 0.23 & $H$ & 14      & This work                \\
\hline
\multicolumn{5}{l}{$^a$ Excluding SN 2002hj from the SN sample.}\\
\multicolumn{5}{l}{$^b$ Using the PMM parameters evaluated at 50 d.}\\
\end{tabular}
\end{table}

\subsection{Error budget}

Taking into account that the lowest HD rms is obtained with the $J$-band, in Table~\ref{table:rms_budget} we show the statistical error budget for the distances measured for that band. We see that 88.6 per cent of the statistical error is induced by the errors of the first four terms: the host galaxy colour excess, the explosion epoch, the PMM zero-point, and the $J$-band photometry. Therefore, it is possible improve the performance of the PMM in the future developing a better method to estimate $E_\text{h}(\bv)$, selecting SNe II with explosion epoch constrained within a small range of time, including more SNe II in the PMM calibration set, and increasing the quality of the $J$-band photometry.

\begin{table}
\caption{Error budget for the PMM $J$-band distances.}
\label{table:rms_budget}
\begin{tabular}{lccc}
\hline
Error source       &Typical error &Error on $\mu_J$& \% of total error \\
                   &              & (mag)        &              \\
\hline
 $E_\text{h}(\bv)$ & 0.097 mag  & 0.079 & 35.7   \\
             $t_0$ &   2.6 days & 0.066 & 24.9   \\
            ZP$_J$ & 0.050 mag$^a$  & 0.050 & 14.3   \\
             $m_J$ & 0.049 mag  & 0.049 & 13.7   \\
     $v_\text{ph}$ &    60 \kms & 0.040 &  9.2   \\
   $cz_\text{hel}$ &    29 \kms & 0.019 &  2.1   \\
             $K_J$ & 0.003 mag  & 0.003 &  0.05   \\
 $E_\text{G}(\bv)$ & 0.004 mag  & 0.003 &  0.05   \\
\hline
                   & Total      & 0.132      &100.0            \\
\hline
\multicolumn{4}{l}{$^a$ It does not include the TRGB zero-point systematic error.}
\end{tabular}
\end{table}

\subsection{Diminishing systematics}

Our results show that we are reaching a precision in distance modulus of $\pm$0.1 mag with the PMM at near-IR wavelengths. Therefore it is important to control systematics, and push them below 0.1 mag. For the latter in the following, we analyse sources of systematics affecting our results:

1. Explosion epoch: The dependence of the PMM calibration on explosion epoch (left half of Fig.~\ref{fig:pmm_BVIJHK}) is stronger at early times, so $t_0$ errors have a strong effect at those epochs. In order to obtain errors lower than 0.1 mag for observations at $\Delta t\gtrsim35$ d, we need SNe II with explosion epochs constrained within 5 d.

2. SN heliocentric redshift: When the host galaxy heliocentric redshift is used as a proxy of the SN heliocentric redshift, a systematic error of
\begin{equation}
\sigma(M_{x,\Delta t_i,v_{\text{ph},i}})=\frac{5}{\ln{10}}\frac{162\text{ km s}^{-1}}{v_{\text{ph},i}}
\end{equation}
is introduced into the absolute magnitude (equation~\ref{eq:Mabs_SN}). This effect increases when the photospheric velocity decreases, translating into errors $\gtrsim$ 0.1 mag for photospheric velocities $\lesssim$ 3500 \kms. Therefore, if optical spectra of a SN II do not show \ion{H}{II} narrow emission lines due to a nearby \ion{H}{II} region or if the SN is not within a nearly face-on galaxy, then epochs for which photospheric velocities are greater than 3500 \kms\ are preferable.

3. Host galaxy redshift: A galaxy is believed to be within the Hubble flow when its redshift is greater than 0.01. At that redshift, peculiar velocities are thought to be negligible compared with the velocities due to the Universe expansion. However, the typical error of 382 \kms\ translates into a distance modulus error of 0.28 mag for $z=0.01$. Including the error in the determination of the Local Group velocity of 187 \kms, the redshift error increases to 425 \kms. Therefore, in order to reduce the error induced by redshift errors at a level lower than 0.1 mag, it is necessary to observe SNe II within galaxies at $z>0.03$.

\section{CONCLUSIONS}\label{sec:conclusions}

Our main results are the following:
\begin{enumerate}




\item[1.] Using nine SNe II at $cz>2000$ \kms , we obtained $H_0$ ranging between 67.1 and 74.9 \hnotunits , and a HD rms of 0.15--0.28 mag.

\item[2.] Adding six SNe II with host galaxy distances measured with TRGB, Cepheids, or SN Ia (total 15), which distances were converted to redshifts through the Hubble law, we obtain a HD rms of 0.15--0.26 mag in the optical bands, which reduces to 0.13 mag in the $J$-band.

\end{enumerate}

In order to test further the promising results we are obtaining in this work, it is necessary to carry out an optical and near-IR photometric follow-up of SNe II at $z>0.03$ and with explosion epochs constrained within 5 d. For these SNe, it is necessary to take at least one optical spectrum at any epoch between 35--75 d since explosion.

Its evident from Fig.~\ref{fig:CATS} that the quality of the near-IR photometry used in this work is in general lower than the optical one. Therefore, we expect that increasing the quality of the near-IR photometry will further improve our results.

\section*{Acknowledgements}

The authors thank the anonymous referee for the useful comments that helped to improve the original manuscript. O.R, G.P, A.C, and Y.A acknowledge support by the Ministry of Economy, Development, and Tourism's Millennium Science Initiative through grant IC120009, awarded to The Millennium Institute of Astrophysics, MAS. O.R acknowledge support from CONICYT PAI/INDUSTRIA 79090016. This research has made use of the NASA/IPAC Extragalactic Database (NED) which is operated by the Jet Propulsion Laboratory, California Institute of Technology, under contract with the National Aeronautics and Space Administration. This work has made use of the Weizmann Interactive Supernova Data Repository (\url{https://wiserep.weizmann.ac.il}).

\bibliographystyle{mnras}
\bibliography{references} 




\appendix

\section{MODEL SELECTION}\label{model_selection}

For the model selection we consider two criteria: the ``an information criterion'' \citep[AIC,][]{Akaike1974}, which is based on information theory, and the Bayesian information criterion \citep[BIC,][]{Schwarz1978}, which is based on Bayesian inference. From a set of $R$ models, the AIC selects the model that have the least information loss with respect to the unknown true model, while the BIC selects the model with the highest likelihood $\mathcal{L}$, given by
\begin{equation}\label{eq:likelihood}
 \textstyle{\mathcal{L}(\text{model}| \lbrace X_i\rbrace)=\prod_{i=1}^N{p(X_i|\text{model})}}. 
\end{equation}
Here, $X_i$ is the $i$-th observed data, $p(X_i|\text{model})$ is the probability density function (pdf) of $X_i$ given the model, and $N$ is the number of observed data points. Quantitatively, the AIC and BIC search for a balance between overfitting and underfitting penalizing the likelihood. For the AIC and the BIC, the best model is one which minimizes the quantity
\begin{equation}
\text{AIC}\equiv-2\ln{\mathcal{L_\text{max}}}+2Nk/(N-k-1)
\end{equation}
\citep[corrected for small sample sizes,][]{Sugiura1978}, and
\begin{equation}
\text{BIC}\equiv-2\ln{\mathcal{L_\text{max}}}+k\ln{N}
\end{equation} 
\citep{Schwarz1978}, respectively, where $\mathcal{L_\text{max}}$ is the maximum likelihood achievable by the model, and $k$ is the number of parameters of the model.

It is known that a model selection based only on the minimum AIC value reached for a certain model does not provide enough evidence to prefer that model over the other ones \citep[e.g.,][]{Akaike1978,Burnham_anderson2002}. Instead, it is necessary to include into the analysis the strength of evidence in favour of each model. To quantify the latter, it has been proposed to use the likelihood of the model given the data \citep[e.g.,][]{Akaike1978} which, normalized by the sum of likelihoods of all $R$ models, defines the Akaike weights
\begin{equation}\label{eq:Akaike_weights}
w_i=\dfrac{e^{-0.5(\text{AIC}_i-\text{AIC}_\text{min})}}{\sum_{r=1}^{R} e^{-0.5(\text{AIC}_r-\text{AIC}_\text{min})}}
\end{equation}
\citep[e.g.,][]{Burnham_anderson2002}. Here, $\text{AIC}_\text{min}$ is the minimum AIC value reached among the $R$ models used in the analysis. The same idea is applicable for the BIC \citep{Burnham_anderson2002}, which defines the Bayesian weights
\begin{equation}\label{eq:Akaike_weights}
\text{P}_i=\dfrac{e^{-0.5(\text{BIC}_i-\text{BIC}_\text{min})}}{\sum_{r=1}^{R} e^{-0.5(\text{BIC}_r-\text{BIC}_\text{min})}}.
\end{equation}

For the AIC and BIC, the evidence ratios defined as $w_i/w_j$ and $p_i$/$p_j$, respectively, allow comparison of the evidence in favour of the $i$-th model as the best model versus the $j$-th model. As reference, if evidence ratios are greater than 13.0, then there is a strong evidence in favour of the $i$-th model over the $j$-th model \citep[e.g.,][]{Liddle2007}. If several models have substantial support as the best (e.g., evidence ratios $<13.0$), then, by the principle of parsimony, we select the one with less parameters.

In the case of least-square regressions, with random errors normally distributed and with constant variance,
\begin{equation}
-2\ln{\mathcal{L_\text{max}}}=N\ln{2\pi\hat{\sigma}^2}+N
\end{equation}
 \citep[e.g.,][p. 17]{Burnham_anderson2002}, where $\hat{\sigma}^2$ is the average of squared residuals around the model. The AIC and BIC in this case can be expressed as
\begin{align}
\text{AIC}&=\ln{\hat{\sigma}^2}+(N+k-1)/(N-k-1),\label{eq:aic_ls}\\
\text{BIC}&=\ln{\hat{\sigma}^2}+k/N\cdot \ln{N}.\label{eq:bic_ls}
\end{align} 
Since $\hat{\sigma}^2$ is computed from data, it must be considered as a model parameter. 

In the case of nonparametric regressions (NPR), like \texttt{loess}, $k$ is not defined. Instead, it has been proposed to use the trace of the smoother matrix, $\text{tr}(H)$, which is a quantity analogous to the number of parameters in a parametric regression \citep{Cleveland_etal1992,Hurvich_etal1998}. Replacing $k=\text{tr}(H)+1$ in equation~(\ref{eq:aic_ls}) allows us to obtain the AIC version for NPR presented by \citet{Hurvich_etal1998}.

To check the normality of random errors, it is necessary to carry out a normality test. As we do not measure random errors directly, we use residuals instead. However, widely used normality tests like the \citet{Shapiro_Wilk1965} and the \citet{Jarque_Bera1987} test, when applied over residuals, have little power to reject the null hypothesis even when the random errors are not normal \citep{Das_Imon2016}. \citet{Imon2003} proposed a statistic more suitable to verify normality for regressions, which is based on the \citet{Jarque_Bera1987} test and on the use of unbiased moments. The statistic of the test, called Rescaled Moment (RM), is defined as
\begin{equation}
 \text{RM} \equiv Nc^3\left[\hat{\mu}_3^2/\hat{\mu}_2^3/6+c\cdot(\hat{\mu}_4/\hat{\mu}_2^2-3)^2/24\right]
\end{equation}
\citep{Rana_etal2009}, where $\hat{\mu}_i$ is the $i$-th sample moment, and $c\equiv N/(N-k)$. Under the null hypothesis of a normal distribution, the RM statistic is asymptotically distributed as a chi-square distribution with two degrees of freedom. 

\section{SN II SPECTRA LIBRARY}\label{SNII_spectra_library}

In order to compute total-to-selective broadband extinction ratios (Section~\ref{sec:galactic_broadband_extinction} and \ref{sec:host-galaxy_reddening}) and $K$-corrections (Section~\ref{sec:K_correction}) for SNe II, it is necessary to know their SED. The latter can be estimated through theoretical models \citep[e.g.,][]{Sanders_etal2015,deJaeger_etal2015,deJaeger_etal2017} or, as in \citet{Olivares2008} and in this work, through spectroscopic observations. 

In practice, spectra are not always taken with the slit oriented along the parallactic angle (PA), so their shape can be modified due to differential refraction \citep{Filippenko1982}. Even when the slit is oriented along the PA, contamination due to the light from the host galaxy can produce spurious SEDs. Therefore we have to check the flux calibration of spectra before using them as proxies for the SED. To do the latter, we compute colour indices from the spectra and then we compare them with those obtained using the observed photometry. If the spectrum is well flux-calibrated, then colour differences should be small.

Photometric colour indices at the epoch of the spectra can be computed through the light curve fitting procedure presented in Section~\ref{sec:light_curve_fits}, while to compute a $x_1\!-\!x_2$ colour from a spectrum we use
\begin{equation}\label{eq:C_spec}
 x_1\!-\!x_2 = -2.5\log{\left(\frac{\int d\lambda S_{x_1,\lambda} F_\lambda \lambda}{\int d\lambda S_{x_2,\lambda} F_\lambda \lambda}\right)}+\text{ZP}_{x_1\!-\!x_2}.
\end{equation}
Here, $\lambda$ is the wavelength in the observer's frame, $F_{\lambda}$ is the observed SED of the source, $S_{x_1,\lambda}$ and $S_{x_2,\lambda}$ are the transmission functions of the photometric band $x_1$ and $x_2$, respectively, and $\text{ZP}_{x_1\!-\!x_2}$ is the zero point of the colour scale, which can be computed using a star with good spectrophotometric observations. 

We use the Vega SED published by \citet{Bohlin_Gilliland2004}\footnote{\url{ftp://ftp.stsci.edu/cdbs/current_calspec/alpha_lyr_stis_008.fits}} and the magnitudes published by \citet{Fukugita_etal1996}: $B_\text{Vega}=0.03$, $V_\text{Vega}=0.03$, and $I_\text{Vega}=0.024$ mag, and by \citet{Cohen_etal1999}: $J_\text{Vega}=-0.001$, $H_\text{Vega}=0.000$, and $K_\text{Vega}=-0.001$ mag.
We adopt the transmission functions given in \citet{Hamuy_etal2001}. For $\bv$, $\vi$, $\jh$, and $\hk$ we obtained $\text{ZP}_{B\!-\!V}=0.425$, $\text{ZP}_{V\!-\!I}=0.320$, $\text{ZP}_{J\!-\!H}=0.131$, and $\text{ZP}_{H\!-\!K}=-0.077$ mag, respectively.

Among the SN II spectra available from different sources, we select those: (1) observed in the photospheric phase, and (2) covered by $B$- and $V$-band photometry. To check the flux-calibration in the blue and red part of the optical spectra, we compute $\Delta_{x_1\!-\!x_2}\equiv(x_1\!-\!x_2)_\text{phot}\!-\!(x_1\!-\!x_2)_\text{spec}$ using the $\bv$ and $\vi$ colours, respectively, while for the near-IR spectra we use the $\jh$ and $\hk$ colours, respectively. We also compute the intrinsic $\bv$ colour to represent the shape of the SED. For optical spectra we compute this quantity from dereddened and deredshifted spectra, while for near-IR we compute the intrinsic $\bv$ colour from the photometry (see Section~\ref{sec:C3_linearity}).

\begin{figure}
\includegraphics[width=0.98\columnwidth]{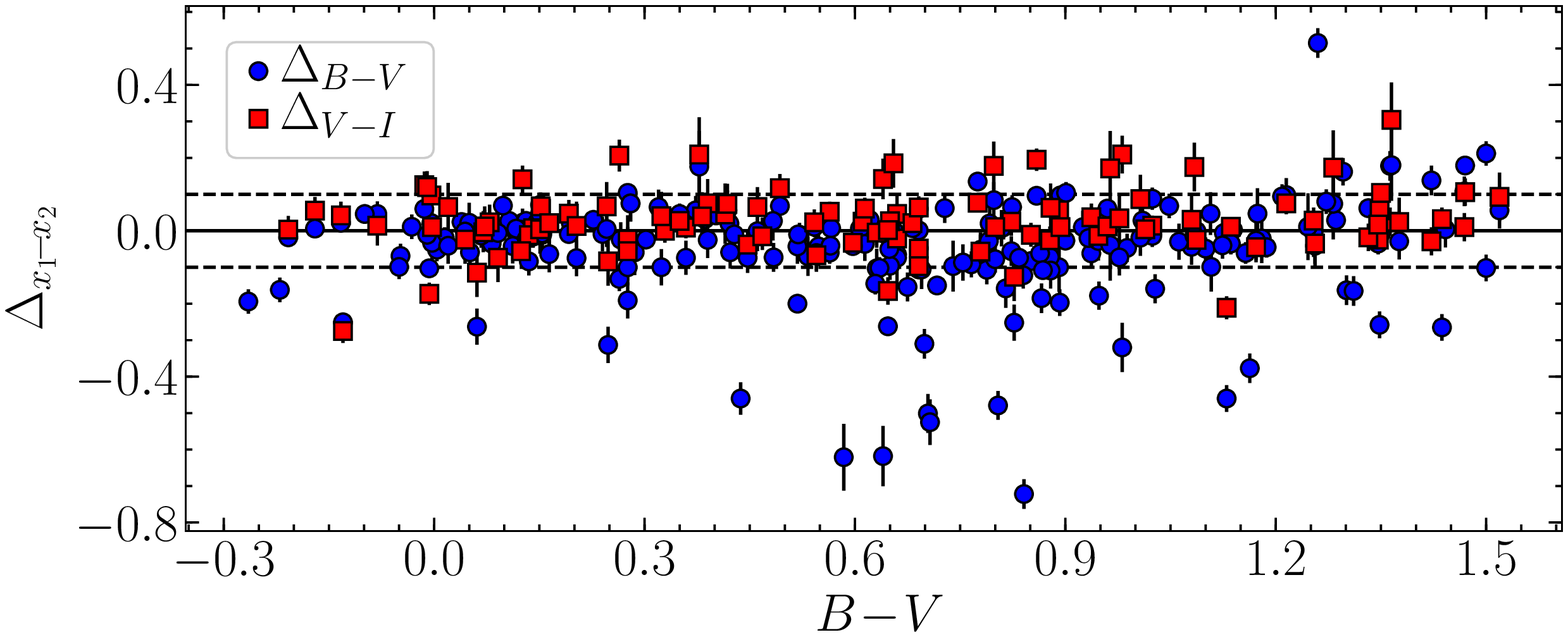}
\includegraphics[width=0.98\columnwidth]{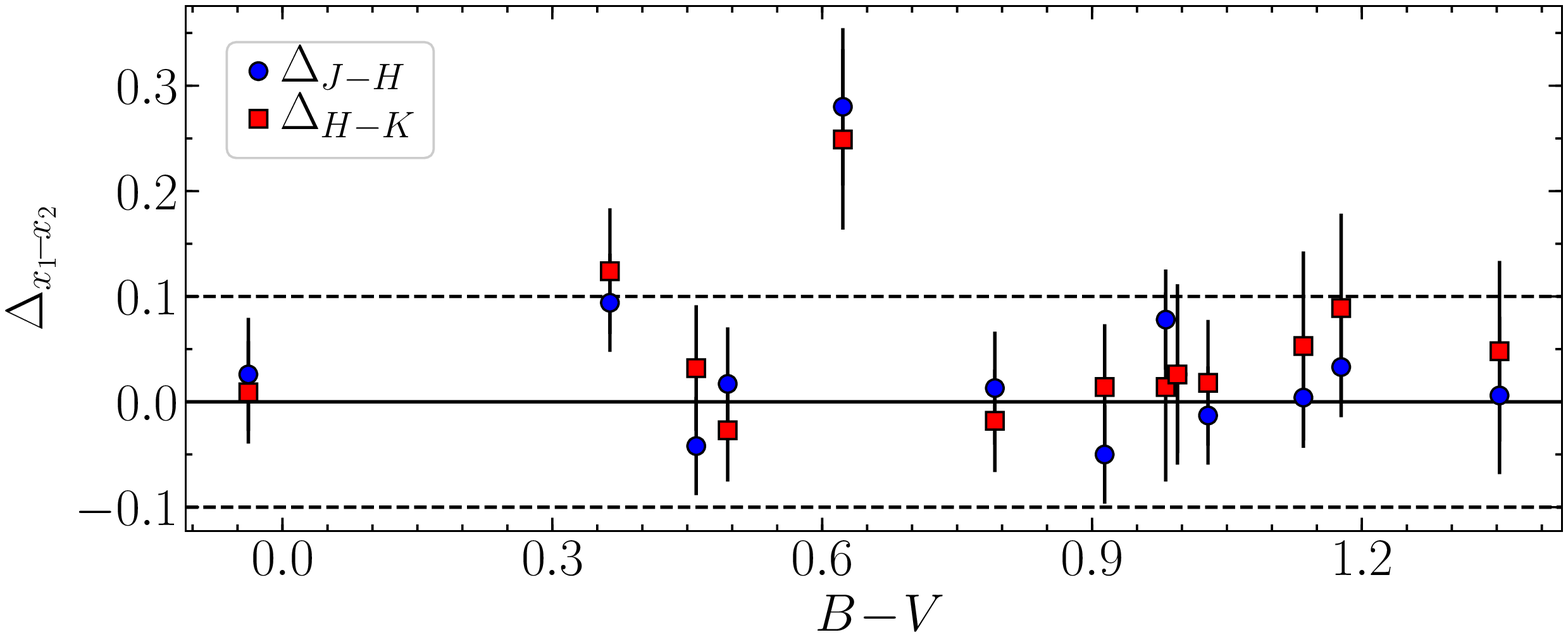}
\caption{Differences between photometric and spectroscopic colours, along with the intrinsic $\bv$ colour. Dashed lines indicate differences within $\pm0.1$ mag.}
\label{fig:color_residuals}
\end{figure}

Fig.~\ref{fig:color_residuals} shows the values of $\Delta_{B\!-\!V}$ and $\Delta_{V\!-\!I}$ (top), and $\Delta_{J\!-\!H}$ and $\Delta_{H\!-\!K}$ (bottom), along with the intrinsic $\bv$ values for the collected spectra. For the SN II spectra library, we select spectra with $|\Delta_{x_1\!-\!x_2}|<0.1$ mag. Finally, we correct spectra for redshift and for Galactic and host galaxy extinction, assuming a \citet{Fitzpatrick1999} extinction curve with $R_V=3.1$  for both our Galaxy and hosts.

Table~\ref{table:SNII_library} summarizes the details of the spectra in the library: SN name (Column 1), Galactic colour excess (Column 2), heliocentric redshift (Columns 3), host galaxy colour excess (Column 4), and references for the data (Column 5).

\section{C3 LINEARITY}\label{sec:C3_linearity}

Assuming that a C3 can be well represented by a polynomial fit, the linearity of a C3 can be demonstrated if there is a high fraction of SNe II showing C3s with straight line as optimal polynomial. Due to the scarcity of $K$-band photometry, we use only \bvijh\ photometry for this analysis.

With \bvijh\ photometry set it is possible to define a total of ten colour indices and, therefore, 90 colour-colour plots (i.e., discarding one-to-one correlations). Among them, only 36 combinations give us non-superfluous information, which we analyse for host galaxy colour excess estimation.

Before computing intrinsic C3 slopes, photometry must be corrected for Galactic and host galaxy extinction, and for $K$-correction. Since we need the prior knowledge of the intrinsic $\bv$, we need to know in advance the value of host galaxy colour excess. For the latter, we apply zero order correction as prior values. The intrinsic $\bv$ can be computed easily from the relation between the observed and the intrinsic $\bv$ colour, i.e., 
\begin{align}
 (\bv)_\text{obs} = \bv+ (R_{B}^\text{G}\!-\!R_{V}^\text{G})\cdot E_\text{G}(\bv)\nonumber \\
 + (K^\text{s}_B/z\!-\!K^\text{s}_V/z)\cdot z + (R_{B}^\text{h}\!-\!R_{V}^\text{h})\cdot E_\text{h}(\bv),
\end{align}
where $(\bv)_\text{obs}$ is the observed $\bv$. 
In Sections~\ref{sec:galactic_broadband_extinction}, \ref{sec:K_correction}, and \ref{sec:host-galaxy_reddening} we found that $R_{V}^{\text{G}}$, $K^\text{s}_B/z$, $K^\text{s}_V/z$ are linear on $\bv$, while $R_{B}^{\text{G}}$, $R_{B}^{\text{h}}$ and $R_{V}^{\text{h}}$ are quadratic on \bv. Therefore, solving a quadratic equation, we can obtain $\bv$ in terms of $(\bv)_\text{obs}$, $E_\text{G}(\bv)$, $z$, $E_\text{h}(\bv)$, and the fit parameters of $R_{B}^{\text{G}}$, $R_{V}^{\text{G}}$, $K^\text{s}_B/z$, $K^\text{s}_V/z$, $R_{B}^{\text{h}}$, and $R_{V}^{\text{h}}$.

For each SN and for each colour combination, we adjust a polynomial fit. The optimum degree is determined using the AIC/BIC and the principle of parsimony. 

\begin{figure}
\includegraphics[width=\columnwidth]{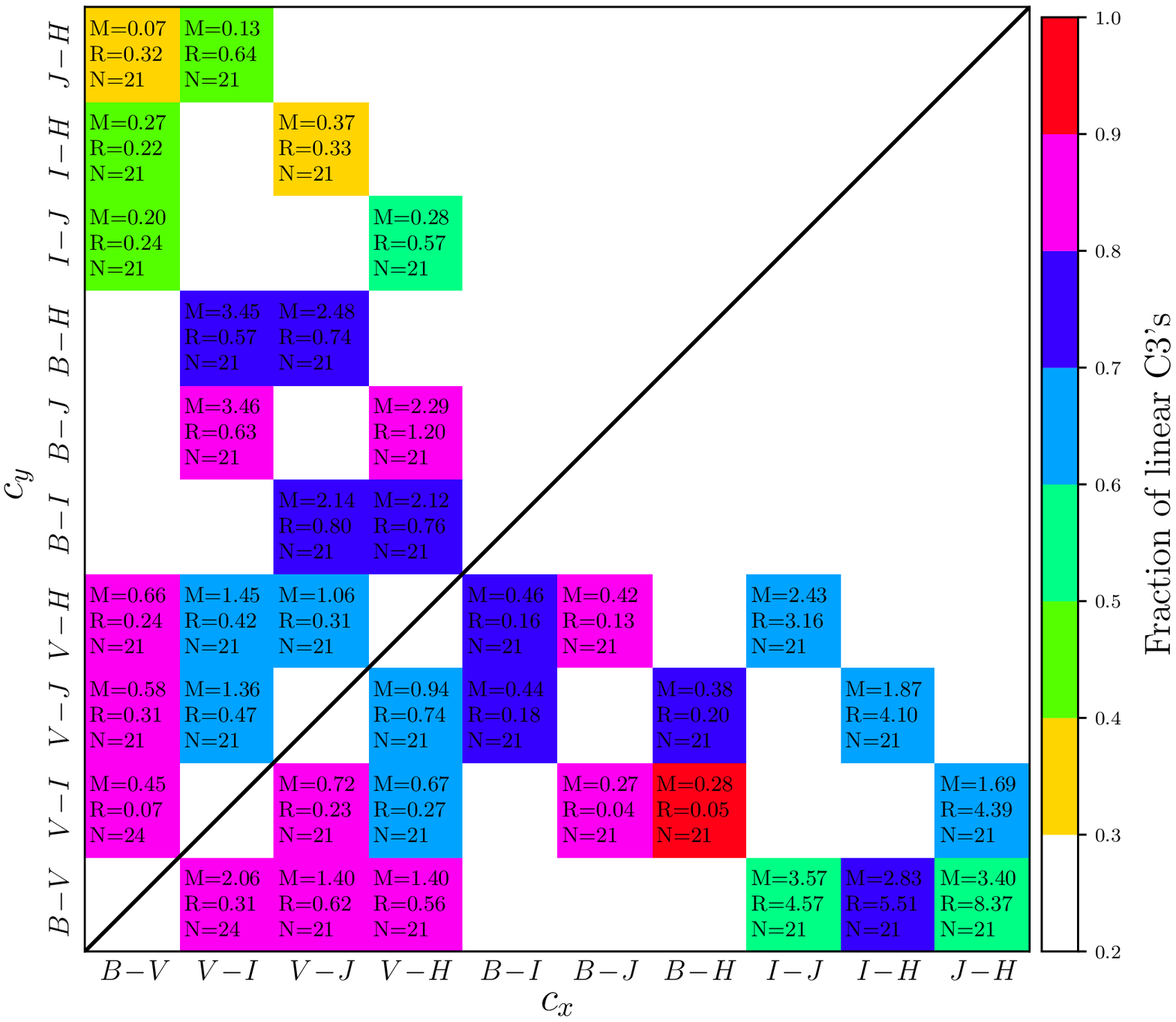}
\caption{Fraction of linear C3s (colourbar) for different colour combinations, along with the median (M) and the rms (R) of the C3 slopes measured from N SNe in our set. Empty spaces correspond to superfluous colour combinations.}
\label{fig:m_BVIJH}
\end{figure}

Fig.~\ref{fig:m_BVIJH} shows the fraction of SNe that are well represented by a straight line. In 20 of the 36 colour combinations, the number of SNe displaying a linear C3 is over 70 per cent. 

\begin{figure}
\includegraphics[width=0.94\columnwidth]{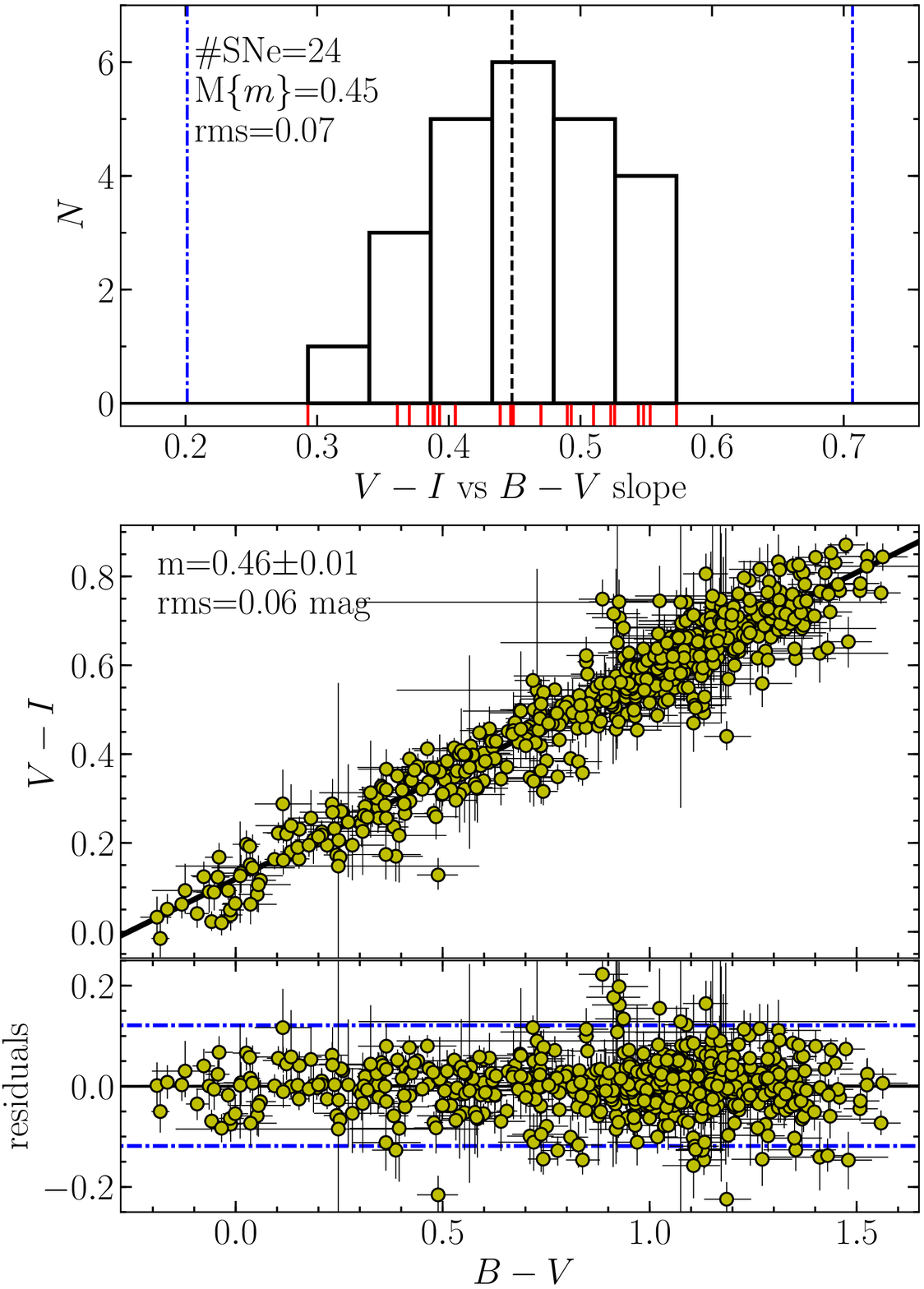}
\caption{Top: $\vi$ versus $\bv$ C3 slopes distribution for the SNe in our set. Red ticks mark the specific slope values, and black dashed line marks the median of the slopes. Bottom: $\vi$ versus $\bv$ diagram showing the 24 SNe II dereddened using host halaxy colour excesses from the C3 method, along with residuals. In both panels, blue dash-dotted lines are the inner fences.}
\label{fig:m_BVI}
\end{figure}

Assuming the C3 linearity for different combinations, we compute slopes of all SNe in our set. For each colour combination we compute the median and rms of the C3 slopes. Fig.~\ref{fig:m_BVI} shows this process for the $\vi$ versus $\bv$ C3s.

\section{SNID EXPLOSION EPOCHS}\label{sec:phase_estimation_with_SNID}

To estimates phases of SNe II with SNID, we follow similar procedures done by \citet{Blondin_Tonry2007}, \citet{Silverman_etal2012}, and \citet{Gutierrez_etal2017_I}.

Among the spectra available from different sources, we select spectra: (1) of SNe II with explosion epoch constrained within 10 d through photometric information, where for these SNe we assume the midpoint between the last non-detection and the first detection as the explosion epoch ($t_{0,\mathrm{ln}}$), (2) spanning a rest-frame wavelength range from 4100 \AA\ to 7000 \AA\ with a S/N $\gtrsim$ 10 \AA$^{-1}$, and (3) within 40 d since the explosion epoch. We do not include spectra at $>40$ d since explosion because in the literature it is less abundant than spectra at earlier epochs \citep[see, for example, Fig. 5 in][]{Gutierrez_etal2017_I} which could bias the age determination toward earlier epochs, and also because at late time the spectral evolution is slower than at early phases, which makes more difficult to accurately determine ages with SNID \citep{Blondin_Tonry2007}. If for a given epoch a SN has several spectra within one day, then we keep that with higher S/N. With the aforementioned constraints, we generate a SNID template library with 242 spectra of 56 SNe II, where each spectrum is corrected by heliocentric redshift. Details of this SN II templates library are summarized in Table~\ref{table:SNID_library}. 

\begin{figure}
\includegraphics[width=0.95\columnwidth]{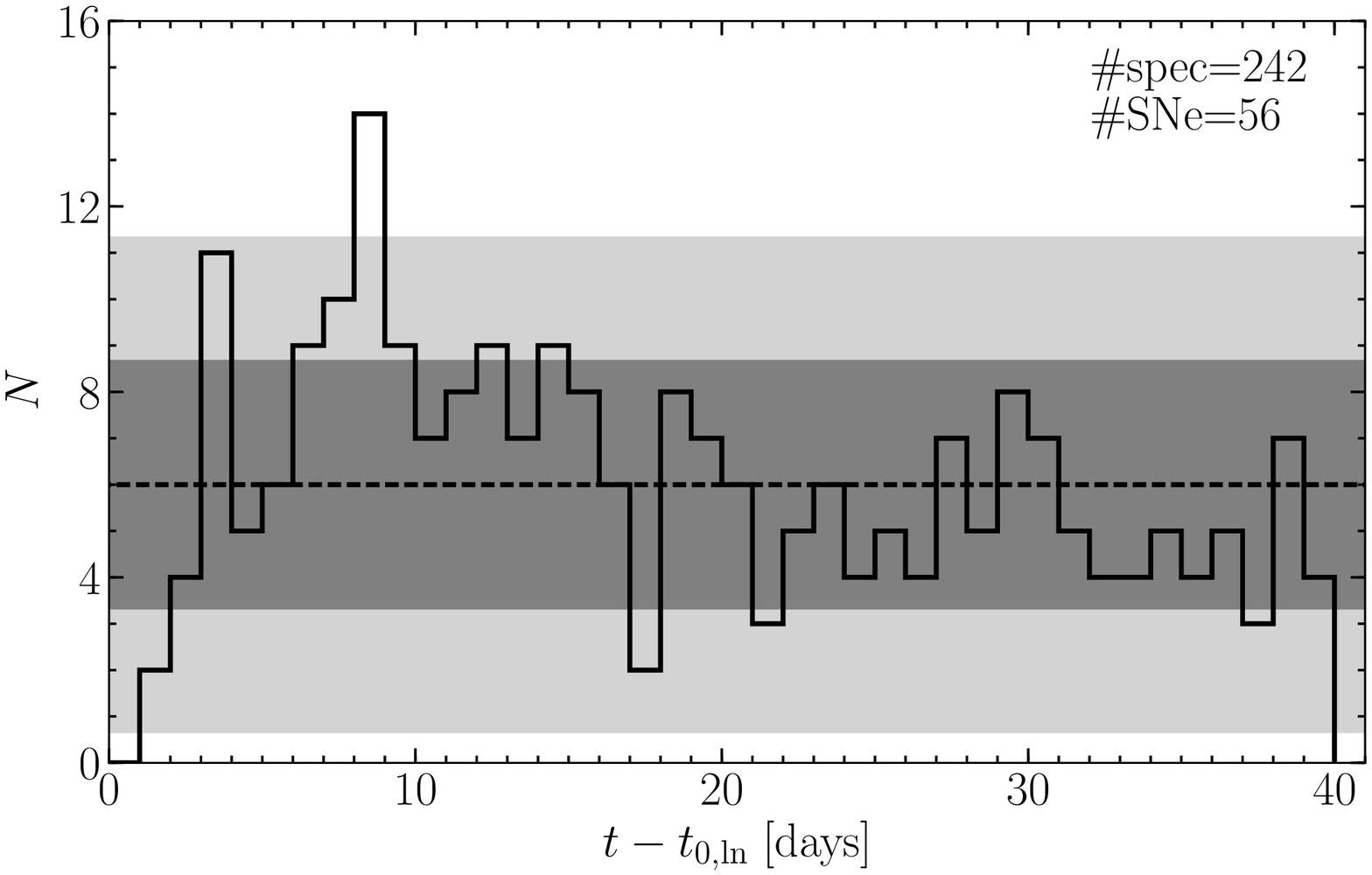}
\caption{Phase distribution of the templates in our SNID SN II library. The dashed line indicates the average of spectra per bin, while dark gray and light gray regions contain values within $\pm1$ rms and $\pm2$ rms around the mean, respectively.}
\label{fig:templates_histogram}
\end{figure}

Fig~\ref{fig:templates_histogram} shows the phase distribution of the templates. The library has, on average, 6 spectra per day, while almost all the variation is within $\pm2$ rms around the mean.

To create the template library, we run the \texttt{logwave} routine (which is part of the SNID program) with the options \texttt{w0=3000 w1=8400 nw=1024}. This generates SNID spectral templates with a bin in the logarithmic wavelength space of ln(8400/3000)/1024$\approx$0.001, equivalent to 300 \kms.

\begin{figure}
\includegraphics[width=0.92\columnwidth]{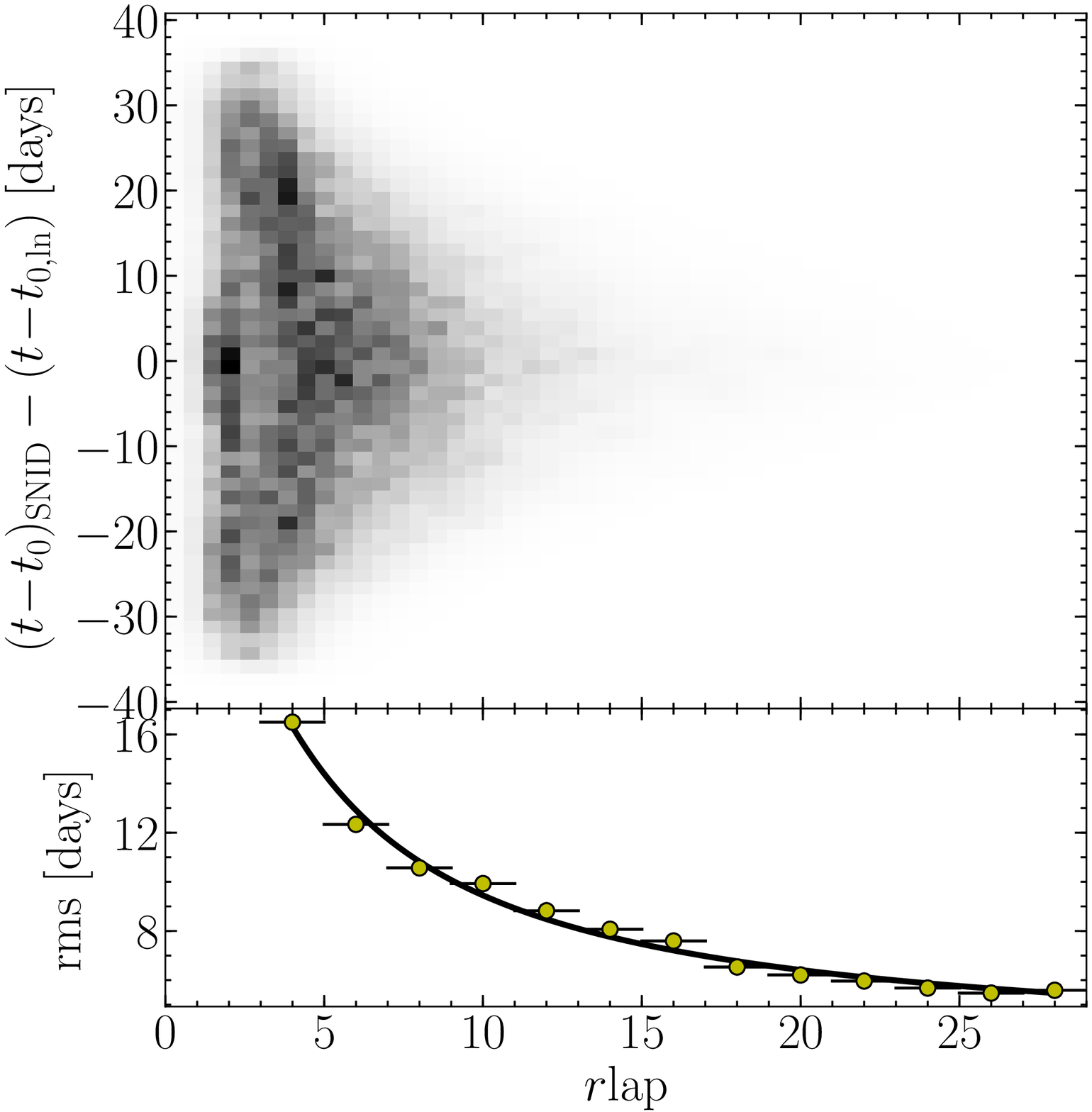}
\caption{2D histogram of differences between phases since explosion estimated from last non-detection $(t\!-\!t_{0,\text{ln}})$ and from SNID $(t\!-\!t_0)_\text{SNID}$ versus the $r$lap parameter (top), and the rms of phase differences versus $r$lap (bottom).}
\label{fig:SNID_delta_t_vs_rlap}
\end{figure}

Once the template library is created, the next step is to test how good are the phases since explosion computed with SNID and our new library. To do this, we run SNID with each library spectrum as input, using the \texttt{avoid} option to avoid templates of the same SN. We record all phases and $r$lap values (which indicate the correlation strength) of the templates found to be similar to the input spectrum and with a redshift within $\pm0.01$. The top panel of Fig.~\ref{fig:SNID_delta_t_vs_rlap} shows the 2D histogram of differences between phases since explosion estimated from last non-detection and from SNID, versus $r$lap. To correlate $r$lap with a rms error in phase, we compute the rms of phase differences in bins of width 2 $r$lap, which is shown at bottom of Fig.~\ref{fig:SNID_delta_t_vs_rlap}. To convert $r$lap to a rms error, we fit a 1/$r$lap polynomial of order 2 (determined by the AIC/BIC), given by
\begin{equation}
\text{rms}=(3.0+72.0/r\mathrm{lap}-75.0/r\mathrm{lap}^2) \text{ d}.
\end{equation}

In general, only one spectrum (e.g., the earliest) is used to estimate explosion epochs with SNID \citep[e.g.,][]{Anderson_etal2014_V_LC,Gutierrez_etal2017_I}. We expect, however, that including all available spectra of a SN in the analysis will result in a best estimation of the explosion epoch. To explore this possibility, we select all SNe in our library with five or more spectra and perform the following procedure:

\begin{enumerate}
\item[1.]For each SN we randomly choose one spectrum, computing the explosion epoch ($t_{0,\mathrm{SNID}}$) and $t_{0,\mathrm{SNID}}-t_{0,\mathrm{ln}}$.
\item[2.]We compute the median (i.e., the offset) and the rms of the $t_{0,\mathrm{SNID}}-t_{0,\mathrm{ln}}$ distribution.
\item[3.]We repeat steps 1 and 2 $10^2$ times, recording the median of the offsets and the rms values.
\item[4.]We repeat steps 1--3, but now randomly choosing two and then three spectra per SN as input.
\end{enumerate}

\begin{table}
\caption{Result of SNID simulations.}
\label{table:rms_vs_n_spectra}
\begin{tabular}{ccc}
\hline
\# of input spectra   & Md($t_{0,\text{SNID}}\!-\!t_{0,\text{ln}}$) & rms($t_{0,\text{SNID}}\!-\!t_{0,\text{ln}}$)\\
  & (d)               & (d)              \\
\hline
1 & $-0.6_{-0.6}^{+1.0}$ & $5.0_{-0.6}^{+0.9}$ \\
2 & $-0.7_{-0.7}^{+0.9}$ & $4.1_{-0.4}^{+0.9}$ \\
3 & $-0.6_{-0.9}^{+0.6}$ & $4.1_{-0.4}^{+0.2}$ \\
\hline
\end{tabular}
\end{table}
Table~\ref{table:rms_vs_n_spectra} shows the result of the aforementioned process, i.e., the offset and the rms as a function of the number of input spectra. Using only one spectrum we obtain a typical rms of 5.0 d, which is similar than the rms of 5.2 d reported by \citet{Gutierrez_etal2017_I}. We see that using more than one spectrum the rms is reduced down to 4.1 d. The median of the offsets is around -0.6 d, independent of the number of input spectra in the analysis. This offset means that explosion epochs computed with SNID are 0.6 d earlier that those estimated with the non-detection. Hereafter, for the explosion epochs derived with SNID we assume an intrinsic error of 5.0 d when only one spectrum is used, or 4.1 d whether more spectra are available.

\begin{table}
\caption{Observed quantities of SN 1980K and SN 2002hh.}
\label{table:SN1980K_SN2002hh_quantitites}
\begin{tabular}{@{}l@{\quad}c@{\,\,}c@{\,\,}c@{\,\,}c@{\,\,\,\,}c@{}}
\hline
SN      & $\Delta t$           & $J$                     & $v_\text{ph}$        & $E_{\text{h}}(\bv)$ & References$^*$ \\
        & (d)               & (mag)                   & (\kms)               & (mag)               & \\
\hline
1980K   & $54.4_{-7.2}^{+7.7}$ & $11.29_{-0.05}^{+0.06}$ & $4013_{-143}^{+165}$ & 0.0                 & 1, 2, 3, 4, 5\\
2002hh  & $44.5_{-2.0}^{+2.0}$ & $12.30_{-0.04}^{+0.04}$ & $4714_{-56}^{+53}$   & 2.74$\pm$0.11$^\dagger$       & 6, 7, 8, 9\\
\hline
\multicolumn{6}{m{0.95\linewidth}}{$^\dagger$ The colour excess was computed adopting the \citet{Cardelli_etal1989} extinction curve and $R_V=1.9$ \citep{Pozzo_etal2006}. For that extinction curve, we obtained $R^\text{h}_J=0.402$.}\\
\multicolumn{6}{m{0.95\linewidth}}{$^*$ (1) IAUC 3532; (2) \citet{Thompson1982}; (3) \citet{Dwek_etal1983}; (4) \citet{Schmidt_etal1992}; (5) WISeREP; (6) IAUC 8005; (7) IAUC 8024; (8) \citet{Pozzo_etal2006}; (9) \citet{Faran_etal2014_IIP}.}
\end{tabular}
\end{table}

\section{The distance to NGC 6946}\label{sec:NGC_6946_distance}

The distance to NGC 6946, host of SN 2004et, was measured with the TRGB method by \citeauthor{Tikhonov2014} (\citeyear{Tikhonov2014}, hereafter T14), \citeauthor{Murphy_etal2018} (\citeyear{Murphy_etal2018}, hereafter M18), and \citeauthor{Anand_etal2018_NGC6946_TRGB_distance} (\citeyear{Anand_etal2018_NGC6946_TRGB_distance}, hereafter A18), and correspond to $\mu=29.39\pm0.14$ mag in the \citet{Jang_Lee2017_TRGB_ZP} calibration. The PMM $J$-band distance for SN 2004et obtained in this work ($\mu_J=28.83\pm0.12$ mag) is in conflict with the TRGB estimation. To investigate the reason for this discrepancy, we compute distances to other two SNe II that exploded in NGC 6946: SN 1980K and SN 2002hh. These SNe have near-IR photometry, but we did not include them into the analysis because they do not have photometry in the five bands we use (i.e., \bvijh).

Using data given in Table~\ref{table:SN1980K_SN2002hh_quantitites}, we obtain $\mu_J=28.73\pm0.18$ and $28.77\pm0.11$ mag for SN 1980K and SN 2002hh, respectively, which are consistent with the distance computed with SN 2004et. There is then a tension of at least $\sim$4 rms between the PMM and the TRGB distance. This discrepancy could be due to: (1) all three SNe II are intrinsically brighter at least 0.56 mag than the SNe II we use for the calibration, or (2) there are issues with the TRGB distances reported by T14, M18, and A18. We noted that the latter two independent studies used almost the same data but obtained significantly different values of the TRGB F814W magnitude (F814W$_\text{TRGB}$): $26.00\pm0.04$ mag in M18 and $25.84\pm0.11$ mag in A18. T14 used another set of image data, which is closer to center of the galaxy than those used in M18 and A18, and obtained a lower F814W$_\text{TRGB}$ value ($25.79\pm0.05$ mag). At this moment, the origin of this large discrepancy is unclear. Taking into account this, we safely remove SN 2004et from the calibration and the final sample.

\section{Tables}

\begin{table}
\caption{Akaike and Bayesian weights, and evidence ratios for $R_x^{\text{G}}$, $K^\text{s}_x/z$, $R_x^{\text{h}}$, and $f_{x,\Delta t}$.}
\label{table:RKRf_statistics}
\begin{tabular}{@{}c@{\,\,\,\,}c@{\,\,}c@{\,\,\,\,}c@{\,\,}c@{\,\,\,\,}c@{\,\,}c@{\,\,\,\,}c@{\,\,}c@{\,\,\,\,}c@{\,\,}c@{\,\,\,\,}c@{\,\,}c@{}}
\hline
    & \multicolumn{4}{c}{$R_B^{\text{G}}$}& \multicolumn{4}{c}{$R_V^{\text{G}}$}& \multicolumn{4}{c}{$R_I^{\text{G}}$} \\
      \cmidrule(rr){2-5} \cmidrule(rr){6-9} \cmidrule(rr){10-13}
$i^a$&$w_i$& P$_i$ &$\frac{w_6}{w_i}$& $\frac{\text{P}_2}{\text{P}_i}$ & $w_i$ & P$_i$ &$\frac{w_2}{w_i}$& $\frac{\text{P}_2}{\text{P}_i}$&$w_i$& P$_i$ &$\frac{w_4}{w_i}$& $\frac{\text{P}_1}{\text{P}_i}$ \\
\hline
0 & 0.00 & 0.00 &  ... &  ... & 0.00 & 0.00 &  ... &  ... & 0.00 & 0.00 &  ... &  ... \\
1 & 0.00 & 0.00 &  ... &  ... & 0.05 & 0.34 &  7.4 &  1.8 & 0.14 & 0.76 &  2.3 &  1.0 \\
2 & 0.02 & 0.50 & 19.5 &  1.0 & 0.37 & 0.60 &  1.0 &  1.0 & 0.05 & 0.09 &  6.4 &  8.4 \\
3 & 0.03 & 0.21 & 13.0 &  2.4 & 0.16 & 0.06 &  2.3 & 10.0 & 0.02 & 0.01 & 16.0 & 76.0 \\
4 & 0.10 & 0.16 &  3.9 &  3.1 & 0.06 & 0.00 &  6.2 &  ... & 0.32 & 0.09 &  1.0 &  8.4 \\
5 & 0.23 & 0.09 &  1.7 &  5.6 & 0.04 & 0.00 &  9.2 &  ... & 0.30 & 0.04 &  1.1 & 19.0 \\
6 & 0.39 & 0.04 &  1.0 & 12.5 & 0.12 & 0.00 &  3.1 &  ... & 0.12 & 0.01 &  2.7 & 76.0 \\
7 & 0.14 & 0.00 &  2.8 &  ... & 0.06 & 0.00 &  6.2 &  ... & 0.03 & 0.00 & 10.7 &  ... \\
8 & 0.09 & 0.00 &  4.3 &  ... & 0.15 & 0.00 &  2.5 &  ... & 0.02 & 0.00 & 16.0 &  ... \\
\hline
  & \multicolumn{4}{c}{$R_J^{\text{G}}$}& \multicolumn{4}{c}{$R_H^{\text{G}}$}& \multicolumn{4}{c}{$R_K^{\text{G}}$}\\
\cmidrule(rr){2-5} \cmidrule(rr){6-9} \cmidrule(rr){10-13}
$i^a$&$w_i$& P$_i$&$\frac{w_0}{w_i}$& $\frac{\text{P}_3}{\text{P}_i}$&$w_i$& P$_i$&$\frac{w_0}{w_i}$& $\frac{\text{P}_3}{\text{P}_i}$&$w_i$& P$_i$&$\frac{w_1}{w_i}$& $\frac{\text{P}_6}{\text{P}_i}$\\
\hline
0 & 0.80 & 0.18 &  1.0 &  2.8 & 0.81 & 0.20 &  1.0 &  2.3 & 0.06 & 0.00 & 13.2 &  ... \\
1 & 0.13 & 0.05 &  6.2 & 10.0 & 0.13 & 0.06 &  6.2 &  7.7 & 0.79 & 0.09 &  1.0 &  2.9 \\
2 & 0.01 & 0.01 & 80.0 & 50.0 & 0.01 & 0.02 & 81.0 & 23.0 & 0.12 & 0.06 &  6.6 &  4.3 \\
3 & 0.06 & 0.50 & 13.3 &  1.0 & 0.05 & 0.46 & 16.2 &  1.0 & 0.03 & 0.16 & 26.3 &  1.6 \\
4 & 0.00 & 0.17 &  ... &  2.9 & 0.00 & 0.14 &  ... &  3.3 & 0.00 & 0.08 &  ... &  3.2 \\
5 & 0.00 & 0.06 &  ... &  8.3 & 0.00 & 0.08 &  ... &  5.8 & 0.00 & 0.20 &  ... &  1.3 \\
6 & 0.00 & 0.02 &  ... & 25.0 & 0.00 & 0.02 &  ... & 23.0 & 0.00 & 0.26 &  ... &  1.0 \\
7 & 0.00 & 0.01 &  ... & 50.0 & 0.00 & 0.02 &  ... & 23.0 & 0.00 & 0.15 &  ... &  1.7 \\
8 & 0.00 & 0.00 &  ... &  ... & 0.00 & 0.00 &  ... &  ... & 0.00 & 0.00 &  ... &  ... \\
\hline
    & \multicolumn{4}{c}{$K^\text{s}_B/z$}& \multicolumn{4}{c}{$K^\text{s}_V/z$}& \multicolumn{4}{c}{$K^\text{s}_I/z$}\\
      \cmidrule(rr){2-5} \cmidrule(rr){6-9} \cmidrule(rr){10-13} 
$i^a$&$w_i$& P$_i$ &$\frac{w_5}{w_i}$& $\frac{\text{P}_2}{\text{P}_i}$ &$w_i$ & P$_i$ &$\frac{w_8}{w_i}$& $\frac{\text{P}_1}{\text{P}_i}$ &$w_i$& P$_i$ &$\frac{w_3}{w_i}$& $\frac{\text{P}_1}{\text{P}_i}$\\
\hline      
0 & 0.00 & 0.00 &  ... &  ... & 0.00 & 0.00 &  ... &  ... & 0.00 & 0.00 &  ... &  ... \\
1 & 0.01 & 0.24 & 24.0 &  2.0 & 0.11 & 0.82 &  4.0 &  1.0 & 0.14 & 0.47 &  2.8 &  1.0 \\
2 & 0.07 & 0.48 &  3.4 &  1.0 & 0.08 & 0.14 &  5.5 &  5.9 & 0.28 & 0.33 &  1.4 &  1.4 \\
3 & 0.16 & 0.21 &  1.5 &  2.3 & 0.09 & 0.03 &  4.9 & 27.3 & 0.39 & 0.18 &  1.0 &  2.6 \\
4 & 0.17 & 0.05 &  1.4 &  9.6 & 0.03 & 0.01 & 14.7 & 82.0 & 0.13 & 0.02 &  3.0 & 23.5 \\
5 & 0.24 & 0.02 &  1.0 & 24.0 & 0.01 & 0.00 & 44.0 &  ... & 0.04 & 0.00 &  9.8 &  ... \\
6 & 0.18 & 0.00 &  1.3 &  ... & 0.12 & 0.00 &  3.7 &  ... & 0.01 & 0.00 & 39.0 &  ... \\
7 & 0.08 & 0.00 &  3.0 &  ... & 0.12 & 0.00 &  3.7 &  ... & 0.01 & 0.00 & 39.0 &  ... \\
8 & 0.09 & 0.00 &  2.7 &  ... & 0.44 & 0.00 &  1.0 &  ... & 0.01 & 0.00 & 39.0 &  ... \\
\hline
& \multicolumn{4}{c}{$K^\text{s}_J/z$}& \multicolumn{4}{c}{$K^\text{s}_H/z$}& \multicolumn{4}{c}{$K^\text{s}_K/z$}\\
\cmidrule(rr){2-5} \cmidrule(rr){6-9} \cmidrule(rr){10-13} 
$i^a$&$w_i$& P$_i$&$\frac{w_0}{w_i}$& $\frac{\text{P}_0}{\text{P}_i}$&$w_i$& P$_i$&$\frac{w_0}{w_i}$& $\frac{\text{P}_3}{\text{P}_i}$&$w_i$& P$_i$&$\frac{w_0}{w_i}$&$\frac{\text{P}_7}{\text{P}_i}$\\
\hline
0 & 0.84 & 0.66 &  1.0 &  1.0 & 0.73 & 0.09 &  1.0 &  6.3 & 0.80 & 0.05 &  1.0 & 13.0 \\
1 & 0.14 & 0.20 &  6.0 &  3.0 & 0.13 & 0.03 &  5.6 & 19.0 & 0.17 & 0.03 &  4.7 & 21.7 \\
2 & 0.02 & 0.06 & 42.0 & 11.0 & 0.01 & 0.01 & 73.0 & 57.0 & 0.03 & 0.02 & 26.7 & 32.5 \\
3 & 0.00 & 0.04 &  ... & 16.5 & 0.13 & 0.57 &  5.6 &  1.0 & 0.00 & 0.01 &  ... & 65.0 \\
4 & 0.00 & 0.02 &  ... & 33.0 & 0.00 & 0.20 &  ... &  2.8 & 0.00 & 0.00 &  ... &  ... \\
5 & 0.00 & 0.01 &  ... & 66.0 & 0.00 & 0.07 &  ... &  8.1 & 0.00 & 0.03 &  ... & 21.7 \\
6 & 0.00 & 0.00 &  ... &  ... & 0.00 & 0.02 &  ... & 28.5 & 0.00 & 0.21 &  ... &  3.1 \\
7 & 0.00 & 0.00 &  ... &  ... & 0.00 & 0.01 &  ... & 57.0 & 0.00 & 0.65 &  ... &  1.0 \\
8 & 0.00 & 0.00 &  ... &  ... & 0.00 & 0.00 &  ... &  ... & 0.00 & 0.00 &  ... &  ... \\
\hline
 & \multicolumn{4}{c}{$R_B^{\text{h}}$}& \multicolumn{4}{c}{$R_V^{\text{h}}$}& \multicolumn{4}{c}{$R_I^{\text{h}}$}\\
\cmidrule(rr){2-5} \cmidrule(rr){6-9} \cmidrule(rr){10-13}
$i^a$&$w_i$& P$_i$ &$\frac{w_5}{w_i}$& $\frac{\text{P}_2}{\text{P}_i}$ & $w_i$ & P$_i$ &$\frac{w_2}{w_i}$& $\frac{\text{P}_2}{\text{P}_i}$&$w_i$&P$_i$&$\frac{w_2}{w_i}$&$\frac{\text{P}_2}{\text{P}_i}$ \\
\hline
0 & 0.00 & 0.00 &  ... &  ... & 0.00 & 0.00 &  ... &  ... & 0.00 & 0.00 &  ... &  ... \\
1 & 0.00 & 0.00 &  ... &  ... & 0.00 & 0.01 &  ... & 89.0 & 0.37 & 0.84 &  1.0 &  1.0 \\
2 & 0.04 & 0.61 &  6.0 &  1.0 & 0.47 & 0.89 &  1.0 &  1.0 & 0.13 & 0.11 &  2.8 &  7.6 \\
3 & 0.06 & 0.18 &  4.0 &  3.4 & 0.21 & 0.09 &  2.2 &  9.9 & 0.05 & 0.02 &  7.4 & 42.0 \\
4 & 0.23 & 0.16 &  1.0 &  3.8 & 0.08 & 0.01 &  5.9 & 89.0 & 0.19 & 0.02 &  1.9 & 42.0 \\
5 & 0.24 & 0.04 &  1.0 & 15.2 & 0.05 & 0.00 &  9.4 &  ... & 0.15 & 0.01 &  2.5 & 84.0 \\
6 & 0.21 & 0.01 &  1.1 & 61.0 & 0.09 & 0.00 &  5.2 &  ... & 0.07 & 0.00 &  5.3 &  ... \\
7 & 0.10 & 0.00 &  2.4 &  ... & 0.04 & 0.00 & 11.7 &  ... & 0.02 & 0.00 & 18.5 &  ... \\
8 & 0.12 & 0.00 &  2.0 &  ... & 0.06 & 0.00 &  7.8 &  ... & 0.02 & 0.00 & 18.5 &  ... \\
\hline
\end{tabular}
\end{table}

\begin{table}
\renewcommand\thetable{F1} 
\contcaption{Akaike and Bayesian weights, and evidence ratios for $R_x^{\text{G}}$, $K^\text{s}_x/z$, $R_x^{\text{h}}$, and $f_{x,\Delta t}$.}
\begin{tabular}{@{}c@{\,\,\,\,}c@{\,\,}c@{\,\,\,\,}c@{\,\,}c@{\,\,\,\,}c@{\,\,}c@{\,\,\,\,}c@{\,\,}c@{\,\,\,\,}c@{\,\,}c@{\,\,\,\,}c@{\,\,}c@{}}
\hline
& \multicolumn{4}{c}{$R_J^{\text{h}}$}& \multicolumn{4}{c}{$R_H^{\text{h}}$}& \multicolumn{4}{c}{$R_K^{\text{h}}$}\\
\cmidrule(rr){2-5} \cmidrule(rr){6-9} \cmidrule(rr){10-13}
$i^a$&$w_i$&P$_i$&$\frac{w_0}{w_i}$&$\frac{\text{P}_3}{\text{P}_i}$&$w_i$&P$_i$&$\frac{w_0}{w_i}$&$\frac{\text{P}_3}{\text{P}_i}$&$w_i$&P$_i$&$\frac{w_1}{w_i}$&$\frac{\text{P}_6}{\text{P}_i}$\\
\hline
0 & 0.78 & 0.13 &  1.0 &  4.1 & 0.77 & 0.14 &  1.0 &  3.8 & 0.31 & 0.04 &  1.7 &  7.8 \\
1 & 0.12 & 0.04 &  6.5 & 13.2 & 0.14 & 0.05 &  5.5 & 10.6 & 0.53 & 0.17 &  1.0 &  1.8 \\
2 & 0.02 & 0.02 & 39.0 & 26.5 & 0.01 & 0.02 & 77.0 & 26.5 & 0.14 & 0.18 &  3.8 &  1.7 \\
3 & 0.08 & 0.53 &  9.8 &  1.0 & 0.08 & 0.53 &  9.6 &  1.0 & 0.02 & 0.31 & 26.5 &  1.0 \\
4 & 0.00 & 0.19 &  ... &  2.8 & 0.00 & 0.16 &  ... &  3.3 & 0.00 & 0.10 &  ... &  3.1 \\
5 & 0.00 & 0.06 &  ... &  8.8 & 0.00 & 0.07 &  ... &  7.6 & 0.00 & 0.07 &  ... &  4.4 \\
6 & 0.00 & 0.02 &  ... & 26.5 & 0.00 & 0.02 &  ... & 26.5 & 0.00 & 0.09 &  ... &  3.4 \\
7 & 0.00 & 0.01 &  ... & 53.0 & 0.00 & 0.01 &  ... & 53.0 & 0.00 & 0.04 &  ... &  7.8 \\
8 & 0.00 & 0.00 &  ... &  ... & 0.00 & 0.00 &  ... &  ... & 0.00 & 0.00 &  ... &  ... \\
\hline
& \multicolumn{4}{c}{$f_{B,\Delta t}$}& \multicolumn{4}{c}{$f_{V,\Delta t}$}& \multicolumn{4}{c}{$f_{I,\Delta t}$}\\
\cmidrule(rr){2-5} \cmidrule(rr){6-9} \cmidrule(rr){10-13}
$i^a$&$w_i$& P$_i$&$\frac{w_1}{w_i}$&$\frac{\text{P}_1}{\text{P}_i}$&$w_i$& P$_i$&$\frac{w_1}{w_i}$&$\frac{\text{P}_1}{\text{P}_i}$&$w_i$& P$_i$&$\frac{w_1}{w_i}$&$\frac{\text{P}_1}{\text{P}_i}$ \\
\hline
0 & 0.00 & 0.00 &  ... &  ... & 0.00 & 0.00 &  ... &  ... & 0.00 & 0.00 &  ... &  ... \\
1 & 0.63 & 0.70 &  1.0 &  1.0 & 0.74 & 0.74 &  1.0 &  1.0 & 0.59 & 0.59 &  1.0 &  1.0 \\
2 & 0.28 & 0.24 &  2.2 &  2.9 & 0.21 & 0.20 &  3.5 &  3.7 & 0.33 & 0.31 &  1.8 &  1.9 \\
3 & 0.07 & 0.05 &  9.0 & 14.0 & 0.04 & 0.05 & 18.5 & 14.8 & 0.07 & 0.07 &  8.4 &  8.4 \\
4 & 0.02 & 0.01 & 31.5 & 70.0 & 0.01 & 0.01 & 74.0 & 74.0 & 0.01 & 0.02 & 59.0 & 29.5 \\
5 & 0.00 & 0.00 &  ... &  ... & 0.00 & 0.00 &  ... &  ... & 0.00 & 0.01 &  ... & 59.0 \\
6 & 0.00 & 0.00 &  ... &  ... & 0.00 & 0.00 &  ... &  ... & 0.00 & 0.00 &  ... &  ... \\
\hline
& \multicolumn{4}{c}{$f_{J,\Delta t}$}& \multicolumn{4}{c}{$f_{H,\Delta t}$}& \multicolumn{4}{c}{}\\
\cmidrule(rr){2-5} \cmidrule(rr){6-9} 
$i^a$&$w_i$& P$_i$&$\frac{w_1}{w_i}$& $\frac{\text{P}_1}{\text{P}_i}$&$w_i$& P$_i$&$\frac{w_1}{w_i}$& $\frac{\text{P}_1}{\text{P}_i}$&&&&  \\
\cmidrule(rr){1-9} 
0 & 0.00 & 0.00 &  ... &  ... & 0.00 & 0.00 &  ... &  ... &      &      &      &      \\
1 & 0.31 & 0.31 &  1.8 &  1.7 & 0.54 & 0.54 &  1.0 &  1.0 &      &      &      &      \\
2 & 0.56 & 0.53 &  1.0 &  1.0 & 0.38 & 0.35 &  1.4 &  1.5 &      &      &      &      \\
3 & 0.11 & 0.12 &  5.1 &  4.4 & 0.07 & 0.08 &  7.7 &  6.8 &      &      &      &      \\
4 & 0.02 & 0.03 & 28.0 & 17.7 & 0.01 & 0.02 & 54.0 & 27.0 &      &      &      &      \\
5 & 0.00 & 0.01 &  ... & 53.0 & 0.00 & 0.01 &  ... & 54.0 &      &      &      &      \\
6 & 0.00 & 0.00 &  ... &  ... & 0.00 & 0.00 &  ... &  ... &      &      &      &      \\
\cmidrule(rr){1-9} 
\multicolumn{13}{l}{$^a$Polynomial order.}
\end{tabular}
\end{table}

\begin{table}
\caption{Fits parameters for $R_x^{\text{G}}$, $K^\text{s}_x/z$, and $R_x^{\text{h}}$.}
\label{table:AKAf_parameters}
\begin{tabular}{@{}c@{\quad}c@{\,}c@{\,\,\,\,}c@{\quad}c@{}c@{}}
\hline
$x$ & $c_{0,x}$                        & $c_{1,x}$                      & $c_{2,x}$                      & rms  & $p$(RM) \\
\hline
     &\multicolumn{4}{c}{$R_x^{\text{G}}=c_{0,x}+c_{1,x}(\bv)+c_{2,x}(\bv)^{2\dagger}$} &                   \\
\hline
$B$  & $4.074_{-0.003}^{+0.003}$    & $-0.107_{-0.012}^{+0.014}$ & $-0.046_{-0.011}^{+0.010}$ & 0.009 & 0.42 \\ 
$V$  & $3.089_{-0.002}^{+0.002}$    & $-0.048_{-0.002}^{+0.003}$ & ...                        & 0.006 & 0.32 \\ 
$I$  & $1.722_{-0.002}^{+0.002}$    & $-0.004_{-0.002}^{+0.002}$ & ...                        & 0.003 & 0.86 \\ 
$J$  & $0.8135_{-0.0008}^{+0.0008}$ & ...                        & ...                        & 0.0012    & 0.67 \\ 
$H$  & $0.5184_{-0.0003}^{+0.0003}$ & ...                        & ...                        & 0.0004    & 0.43 \\ 
$K$  & $0.3484_{-0.0001}^{+0.0001}$ & ...                        & ...                        & 0.0001    & 0.46 \\ 
\hline
     &\multicolumn{4}{c}{$K^\text{s}_x/z=c_{0,x}+c_{1,x}(\bv)^{\dagger,\diamond}$} & \\
\hline
$B$  & \phs$ 0.51_{-0.06}^{+0.06}$  & $6.23_{-0.09}^{+0.09}$     & ... & $0.41$ & $0.23$ \\
$V$  &     $-0.51_{-0.05}^{+0.05}$  & $2.87_{-0.07}^{+0.07}$     & ... & $0.35$ & $0.13$ \\
$I$  &     $-0.40_{-0.11}^{+0.11}$  & $1.14_{-0.17}^{+0.17}$     & ... & $0.46$ & $0.53$ \\
$J$  &     $-0.77_{-0.09}^{+0.09}$  & ...                        & ... & $0.21$ & $0.63$ \\
$H$  &     $-0.82_{-0.05}^{+0.05}$  & ...                        & ... & $0.14$ & $0.58$ \\
$K$  &     $-1.40_{-0.05}^{+0.05}$  & ...                        & ... & $0.15$ & $0.61$ \\
\hline
     &\multicolumn{4}{c}{$R_x^{\text{h}}=c_{0,x}+c_{1,x}(\bv)+c_{2,x}(\bv)^{2\dagger}$}   &              \\
\hline 
$B$  & $4.085_{-0.003}^{+0.002}$    & $-0.097_{-0.010}^{+0.011}$ & $-0.049_{-0.009}^{+0.008}$ & 0.008  & 0.97 \\ 
$V$  & $3.092_{-0.002}^{+0.002}$    & $-0.034_{-0.007}^{+0.007}$ & $-0.009_{-0.005}^{+0.006}$ & 0.006  & 0.26 \\ 
$I$  & $1.724_{-0.002}^{+0.002}$    & $-0.005_{-0.002}^{+0.002}$ & ...                        & 0.003  & 0.95 \\ 
$J$  & $0.8140_{-0.0009}^{+0.0008}$ & ...                        & ...                        & 0.0012 & 0.60 \\ 
$H$  & $0.5187_{-0.0003}^{+0.0002}$ & ...                        & ...                        & 0.0004 & 0.36 \\ 
$K$  & $0.3485_{-0.0001}^{+0.0001}$ & ...                        & ...                        & 0.0001 & 0.58 \\ 
\hline
\multicolumn{6}{m{0.99\linewidth}}{Fits are valid for $-0.2<\bv<1.5$ and $-0.05<\bv<1.35$ for $\bvi$ and $\jhk$, respectively ($\dagger$), and $z<0.032$ ($\diamond$). Errors are the 99 per cent CI.}
\end{tabular}
\end{table}

\begin{table*}
\caption{SN II parameters.}
\label{table:SN_EBV_t0}
\begin{tabular}{@{}l@{}c@{\,}c@{\quad}c@{\,\,}c@{\,}c@{\quad}c@{\quad}c@{\quad}l@{\quad}}
\hline
    & \multicolumn{2}{c}{$E_\text{h}(\bv)$ }                          &Discovery      &$t_\text{ln}^a$&$t_\text{fd}^a$ &$t_{0,\text{SNID}}^a$ &$t_{0,\text{final}}^a$  &References$^d$\\
    \cmidrule(lr){2-3}
 SN & $\{\bv,\vi\}$                    & O10                          & (JD)          &(d)                 &(d)                   &(d)                         &    (d)     &     \\
\hline
1999em & \phs$0.035_{-0.105}^{+0.105}(0.082)$ & \phs$0.106(0.052)$& 2451480.94\phn &     $-8.99$ &     $-1.43$ &    \phn$-6.37(4.16)$ & \phn$-5.57_{-2.63}^{+2.96}$ &           IAUC 7294, E03 \\
2002gd & \phs$0.154_{-0.106}^{+0.111}(0.084)$ &            ...    & 2452553.37\phn & $-4.09^{b}$ & $-2.84^{b}$ & \phn\phs$0.73(4.48)$ & \phn$-3.42_{-0.53}^{+0.46}$ &     IAUC 7986, IAUC 7990 \\
2002gw & \phs$0.202_{-0.108}^{+0.106}(0.084)$ & \phs$0.132(0.061)$& 2452560.77\phn &   $-31.218$ &    $-1.637$ &    \phn$-4.29(4.28)$ & \phn$-5.74_{-4.73}^{+3.22}$ &     IAUC 7995, IAUC 7996 \\
2002hj & \phs$0.086_{-0.106}^{+0.106}(0.083)$ & \phs$0.068(0.100)$& 2452567.98\phn &    $-11.48$ &       $0.0$ &       $-11.61(4.40)$ & \phn$-8.56_{-2.37}^{+4.18}$ &                IAUC 8006 \\
 2003B & \phs$0.064_{-0.122}^{+0.117}(0.093)$ & \phs$0.004(0.081)$& 2452645.0\phnn &     ...     &     $-8.41$ &       $-22.83(4.32)$ &    $-22.83_{-5.57}^{+5.61}$ &     IAUC 8042, IAUC 8058 \\
 2003E & \phs$0.299_{-0.148}^{+0.148}(0.115)$ & \phs$0.358(0.100)$& 2452644.8\phnn &     $-39.0$ &       $0.0$ &    \phn$-9.25(4.40)$ & \phn$-9.36_{-5.56}^{+5.54}$ &                IAUC 8044 \\
 2003T & \phs$0.144_{-0.137}^{+0.135}(0.106)$ & \phs$0.175(0.100)$& 2452664.9\phnn &     $-20.0$ &       $0.0$ &       $-11.91(5.19)$ &    $-11.57_{-5.54}^{+6.18}$ &                IAUC 8058 \\
2003bl & \phs$0.107_{-0.127}^{+0.121}(0.097)$ & \phs$0.004(0.052)$& 2452701.0\phnn &    $-262.2$ &       $0.0$ &       $-13.53(5.79)$ &    $-13.64_{-7.35}^{+7.23}$ &                IAUC 8086 \\
2003bn & \phs$0.019_{-0.114}^{+0.113}(0.089)$ & \phs$0.038(0.052)$& 2452697.98\phn &     $-6.48$ &     $-5.15$ &    \phn$-4.42(4.46)$ & \phn$-5.80_{-0.55}^{+0.52}$ &                IAUC 8088 \\
2003ci & \phs$0.098_{-0.135}^{+0.138}(0.107)$ & \phs$0.147(0.100)$& 2452719.9\phnn &     $-16.0$ &       $0.0$ &       $-21.56(4.57)$ &    $-14.28_{-1.45}^{+3.15}$ &                IAUC 8097 \\
2003cn & \phs$0.072_{-0.110}^{+0.105}(0.084)$ & \phs$0.003(0.081)$& 2452727.9\phnn &     $-22.0$ &       $0.0$ &    \phn$-6.77(4.55)$ & \phn$-7.16_{-5.67}^{+4.88}$ &                IAUC 8101 \\
2003hn & \phs$0.181_{-0.132}^{+0.134}(0.104)$ & \phs$0.192(0.081)$& 2452877.2\phnn &     $-20.7$ &       $0.0$ &       $-11.07(4.31)$ &    $-11.03_{-5.32}^{+5.37}$ &                IAUC 8186 \\
2004et & \phs$0.053_{-0.126}^{+0.124}(0.097)$ & \phs$0.048(0.081)$& 2453275.5\phnn &    $-4.983$ &    $-4.017$ &    \phn$-1.83(4.29)$ & \phn$-4.48_{-0.40}^{+0.38}$ &                IAUC 8413 \\
2005ay & \phs$0.045_{-0.121}^{+0.120}(0.094)$ &            ...    & 2453456.58\phn &    $-7.459$ &       $0.0$ &    \phn$-8.23(4.29)$ & \phn$-5.03_{-1.97}^{+3.19}$ &     IAUC 8500, IAUC 8502 \\
2005cs & \phs$0.054_{-0.116}^{+0.115}(0.090)$ & \phs$0.045(0.052)$&    2453550.407 &    $-1.977$ &    $-0.997$ &    \phn$-6.18(4.28)$ & \phn$-1.52_{-0.37}^{+0.42}$ &           IAUC 8553, P09 \\
2008in &    $-0.031_{-0.137}^{+0.142}(0.109)$ &            ...    & 2454827.29\phn &     $-2.84$ &     $-2.34$ &    \phn$-3.91(4.27)$ & \phn$-2.60_{-0.19}^{+0.21}$ &           CBET 1636, R11 \\
 2009N & \phs$0.289_{-0.165}^{+0.164}(0.128)$ &            ...    & 2454856.36\phn &    $-21.86$ &       $0.0$ &    \phn$-3.29(4.27)$ & \phn$-4.50_{-4.74}^{+3.47}$ &                CBET 1670 \\
2009ib & \phs$0.120_{-0.133}^{+0.133}(0.104)$ &            ...    & 2455049.8\phnn &   $-166.25$ &       $0.0$ &    \phn$-9.04(4.55)$ & \phn$-9.25_{-5.63}^{+5.52}$ &                CBET 1902 \\
2009md & \phs$0.145_{-0.133}^{+0.133}(0.104)$ &            ...    & 2455170.31\phn &    $-15.81$ &       $0.0$ &            ...$^c$   & \phn$-7.91_{-6.32}^{+6.32}$ &           CBET 2065, F11 \\
 2012A & \phs$0.108_{-0.126}^{+0.127}(0.099)$ &            ...    &    2455933.887 &    $-9.387$ &       $0.0$ &            ...$^c$   & \phn$-4.69_{-3.75}^{+3.75}$ &                CBET 2974 \\
2012aw & \phs$0.039_{-0.122}^{+0.124}(0.096)$ &            ...    & 2456003.36\phn &    $-1.591$ &    $-0.011$ &    \phn$-1.44(4.20)$ & \phn$-0.81_{-0.63}^{+0.63}$ &     CBET 3054, ATel 3996 \\
2012ec & \phs$0.085_{-0.119}^{+0.124}(0.095)$ &            ...    &    2456150.539 &  $-135.289$ &       $0.0$ &    \phn$-3.69(4.33)$ & \phn$-4.75_{-5.08}^{+3.62}$ &                CBET 3201 \\
2013ej &    $-0.028_{-0.127}^{+0.130}(0.100)$ &            ...    & 2456498.95\phn &     $-1.91$ &    $-1.325$ &       $-10.22(4.21)$ & \phn$-1.64_{-0.22}^{+0.25}$ & CBET 3606, ATel 5237, CF \\
 2014G & \phs$0.315_{-0.144}^{+0.145}(0.112)$ &            ...    &    2456672.074 &    $-3.724$ &    $-0.963$ &    \phn$-0.43(4.65)$ & \phn$-2.26_{-1.17}^{+1.03}$ &                CBET 3787 \\
\hline
\multicolumn{9}{m{0.97\linewidth}}{Column 1: SN names. Columns 2: host galaxy colour excesses estimated with the $\vi$ versus $\bv$ C3, where errors correspond to the 80 per cent CI, and the rms in parenthesis. Column 3: host galaxy colour excesses from \citet{Olivares_etal2010} recalibrated by R14, with rms in parenthesis. Column 4: discovery epochs. Column 5 and 6: last non-detection and first detection epochs, respectively. Column 7 and 8: explosion epochs estimated with SNID and our SN II template library, without any prior (rms error in parenthesis) and with photometric priors, respectively.Column 9: references for discovery, last non-detection, and first detection epochs.}\\
\multicolumn{9}{m{0.97\linewidth}}{$^a$ Epochs with respect to the discovery epoch.}\\
\multicolumn{9}{m{0.97\linewidth}}{$^b$ Value obtained through polynomial fits to pre-maximum $\vri$ photometry.}\\
\multicolumn{9}{m{0.97\linewidth}}{$^c$ Optical spectra not published.}\\
\multicolumn{9}{m{0.97\linewidth}}{$^d$ E03: \citet{Elmhamdi_etal2003}; P09: \citet{Pastorello_etal2009}; R11: \citet{Roy_etal2011}; F11: \citet{Fraser_etal2011}; CF: C. Feliciano report on the \textit{Bright Supernova} website (\url{http://www.rochesterastronomy.org/snimages/}).}
\end{tabular}
\end{table*}

\begin{table*}
\caption{CMB redshifts and PMM distance moduli.}
\label{table:SN_distances}
\begin{tabular}{@{}l@{}c@{\quad}l@{\quad}c@{\quad}c@{\quad}c@{\quad}c@{\quad}c@{}}
\hline
Galaxy       & $cz_{\text{CMB}}$  & SN & $\mu_B$ & $\mu_V$ & $\mu_I$ & $\mu_J$ & $\mu_H$ \\
             &         (\kms)         &    &         &         &         &         &      \\
\hline
       NGC 6946 &    \phn$-141(382)$ & 2004et & $28.40_{-0.60}^{+0.60}(0.47)$ & $28.57_{-0.45}^{+0.45}(0.35)$ & $28.67_{-0.27}^{+0.27}(0.21)$ & $28.83_{-0.15}^{+0.15}(0.12)$ & $28.75_{-0.11}^{+0.11}(0.09)$ \\
            M74 & \phs\phn$359(383)$ & 2013ej & $30.21_{-0.56}^{+0.56}(0.44)$ & $30.08_{-0.42}^{+0.42}(0.33)$ & $30.02_{-0.25}^{+0.25}(0.20)$ & $29.94_{-0.14}^{+0.14}(0.11)$ & $29.94_{-0.11}^{+0.11}(0.09)$ \\
           M51a & \phs\phn$636(382)$ & 2005cs & $29.62_{-0.51}^{+0.51}(0.40)$ & $29.57_{-0.39}^{+0.39}(0.30)$ & $29.54_{-0.24}^{+0.24}(0.19)$ & $29.61_{-0.14}^{+0.14}(0.11)$ & $29.61_{-0.11}^{+0.11}(0.09)$ \\
       NGC 1637 & \phs\phn$670(382)$ & 1999em & $30.22_{-0.47}^{+0.47}(0.37)$ & $30.26_{-0.36}^{+0.36}(0.28)$ & $30.29_{-0.23}^{+0.23}(0.18)$ & $30.30_{-0.15}^{+0.15}(0.12)$ & $30.33_{-0.13}^{+0.13}(0.10)$ \\
       NGC 3938 &    \phs$1038(382)$ & 2005ay & $31.65_{-0.53}^{+0.53}(0.41)$ & $31.66_{-0.40}^{+0.40}(0.31)$ & $31.68_{-0.25}^{+0.25}(0.20)$ & $31.75_{-0.15}^{+0.15}(0.12)$ & $31.79_{-0.13}^{+0.13}(0.10)$ \\
       NGC 3239 &    \phs$1084(383)$ &  2012A & $29.81_{-0.57}^{+0.57}(0.45)$ & $30.01_{-0.45}^{+0.45}(0.35)$ & $30.13_{-0.30}^{+0.30}(0.24)$ & $30.35_{-0.23}^{+0.23}(0.18)$ & $30.33_{-0.21}^{+0.21}(0.16)$ \\
       NGC 1448 &    \phs$1102(382)$ & 2003hn & $30.89_{-0.58}^{+0.58}(0.45)$ & $31.02_{-0.45}^{+0.45}(0.35)$ & $31.10_{-0.29}^{+0.29}(0.22)$ & $31.14_{-0.19}^{+0.19}(0.15)$ & $31.19_{-0.16}^{+0.16}(0.13)$ \\
       NGC 1097 &    \phs$1105(382)$ &  2003B & $31.60_{-0.53}^{+0.53}(0.41)$ & $31.67_{-0.42}^{+0.42}(0.32)$ & $31.75_{-0.29}^{+0.28}(0.22)$ & $31.95_{-0.22}^{+0.21}(0.17)$ & $31.91_{-0.20}^{+0.19}(0.15)$ \\
            M95 &    \phs$1127(383)$ & 2012aw & $30.00_{-0.54}^{+0.54}(0.42)$ & $30.06_{-0.41}^{+0.41}(0.32)$ & $30.07_{-0.25}^{+0.25}(0.19)$ & $30.01_{-0.14}^{+0.14}(0.11)$ & $29.98_{-0.11}^{+0.11}(0.09)$ \\
       NGC 1084 &    \phs$1196(382)$ & 2012ec & $31.16_{-0.54}^{+0.54}(0.42)$ & $31.24_{-0.43}^{+0.43}(0.33)$ & $31.32_{-0.29}^{+0.29}(0.23)$ & $31.37_{-0.21}^{+0.21}(0.16)$ & $31.36_{-0.19}^{+0.19}(0.15)$ \\
       NGC 1559 &    \phs$1304(382)$ & 2009ib & $31.81_{-0.60}^{+0.59}(0.46)$ & $31.85_{-0.47}^{+0.47}(0.37)$ & $31.87_{-0.33}^{+0.32}(0.25)$ & $31.86_{-0.27}^{+0.26}(0.20)$ & $31.73_{-0.34}^{+0.38}(0.29)$ \\
       NGC 4487 &    \phs$1385(383)$ &  2009N & $31.00_{-0.70}^{+0.70}(0.55)$ & $31.10_{-0.53}^{+0.53}(0.42)$ & $31.18_{-0.32}^{+0.32}(0.25)$ & $31.30_{-0.20}^{+0.19}(0.15)$ & $31.25_{-0.16}^{+0.16}(0.13)$ \\
       NGC 3448 &    \phs$1528(382)$ &  2014G & $31.62_{-0.62}^{+0.62}(0.49)$ & $31.92_{-0.47}^{+0.47}(0.37)$ & $32.05_{-0.29}^{+0.28}(0.22)$ & $32.05_{-0.21}^{+0.21}(0.17)$ & $32.26_{-0.19}^{+0.18}(0.14)$ \\
       NGC 3389 &    \phs$1656(383)$ & 2009md & $31.89_{-0.68}^{+0.67}(0.53)$ & $32.06_{-0.56}^{+0.55}(0.43)$ & $32.13_{-0.45}^{+0.43}(0.34)$ & $32.22_{-0.41}^{+0.39}(0.31)$ & $32.17_{-0.38}^{+0.37}(0.29)$ \\
            M61 &    \phs$1913(383)$ & 2008in & $31.81_{-0.59}^{+0.60}(0.47)$ & $31.58_{-0.45}^{+0.45}(0.35)$ & $31.53_{-0.27}^{+0.27}(0.21)$ & $31.39_{-0.15}^{+0.15}(0.12)$ & $31.35_{-0.11}^{+0.11}(0.09)$ \\
       NGC 7537 &    \phs$2303(383)$ & 2002gd & $32.61_{-0.49}^{+0.49}(0.38)$ & $32.61_{-0.37}^{+0.37}(0.29)$ & $32.63_{-0.24}^{+0.24}(0.19)$ & $32.72_{-0.16}^{+0.16}(0.13)$ & $32.75_{-0.14}^{+0.14}(0.11)$ \\
        NGC 922 &    \phs$2878(382)$ & 2002gw & $32.81_{-0.48}^{+0.48}(0.37)$ & $33.08_{-0.37}^{+0.37}(0.29)$ & $33.22_{-0.24}^{+0.24}(0.19)$ & $33.32_{-0.18}^{+0.17}(0.14)$ & $33.38_{-0.16}^{+0.16}(0.13)$ \\
    LEDA 831618 &    \phs$4172(385)$ & 2003bn & $33.79_{-0.51}^{+0.51}(0.40)$ & $33.93_{-0.39}^{+0.39}(0.30)$ & $34.00_{-0.24}^{+0.24}(0.19)$ & $33.99_{-0.20}^{+0.21}(0.16)$ & $34.01_{-0.16}^{+0.17}(0.13)$ \\
MCG --04--12--4 &    \phs$4379(382)$ &  2003E & $33.44_{-0.64}^{+0.64}(0.50)$ & $33.78_{-0.49}^{+0.49}(0.38)$ & $34.00_{-0.30}^{+0.30}(0.24)$ & $33.94_{-0.22}^{+0.22}(0.17)$ & $33.93_{-0.17}^{+0.17}(0.14)$ \\
       NGC 5374 &    \phs$4652(383)$ & 2003bl & $33.85_{-0.54}^{+0.54}(0.42)$ & $34.13_{-0.43}^{+0.43}(0.34)$ & $34.27_{-0.30}^{+0.30}(0.24)$ & $34.35_{-0.23}^{+0.23}(0.18)$ & $34.30_{-0.22}^{+0.22}(0.17)$ \\
         IC 849 &    \phs$5753(383)$ & 2003cn & $34.10_{-0.51}^{+0.51}(0.40)$ & $34.22_{-0.42}^{+0.42}(0.33)$ & $34.24_{-0.33}^{+0.32}(0.25)$ & $34.37_{-0.28}^{+0.28}(0.22)$ & $34.40_{-0.27}^{+0.26}(0.20)$ \\
 NPM1G +04.0097 &    \phs$6884(382)$ & 2002hj & $34.94_{-0.47}^{+0.47}(0.37)$ & $35.00_{-0.36}^{+0.36}(0.28)$ & $35.08_{-0.23}^{+0.23}(0.18)$ & $35.08_{-0.15}^{+0.15}(0.12)$ & $35.26_{-0.14}^{+0.13}(0.11)$ \\
       UGC 4864 &    \phs$8662(383)$ &  2003T & $35.36_{-0.59}^{+0.59}(0.46)$ & $35.33_{-0.47}^{+0.47}(0.36)$ & $35.32_{-0.31}^{+0.32}(0.25)$ & $35.45_{-0.22}^{+0.24}(0.18)$ & $35.56_{-0.19}^{+0.22}(0.16)$ \\
       UGC 6212 &    \phs$9468(384)$ & 2003ci & $36.07_{-0.63}^{+0.62}(0.49)$ & $35.81_{-0.51}^{+0.49}(0.40)$ & $35.69_{-0.35}^{+0.33}(0.29)$ & $35.61_{-0.27}^{+0.26}(0.25)$ & $35.51_{-0.25}^{+0.24}(0.24)$ \\
\hline
\multicolumn{8}{m{0.95\textwidth}}{\textit{Note:} We adopt $R_V=3.1$ for both our Galaxy and hosts. Errors correspond to the 80 per cent CI, with the rms in parenthesis, and include the TRGB zero-point systematic error.}
\end{tabular}
\end{table*}

\begin{table*}
\caption{SNII spectra library.}
\label{table:SNII_library}
\begin{tabular}{l c c c c | l c c c c }
\hline
SN      & $E_{\text{G}}(\bv)^a$ & $cz_{\text{helio}}$ & $E_{\text{h}}(\bv)^b$ & References$^d$ & SN      & $E_{\text{G}}(\bv)^a$ & $cz_{\text{helio}}$ & $E_{\text{h}}(\bv)^b$ & References$^d$ \\
        &  (mag)                   &  (\kms)             &     (mag)                &           &         &  (mag)                   &  (\kms)             &     (mag)                &           \\
\hline
 1990E     &  0.022 &  1362 &   0.616 & 1          &  2003Z     &  0.033 &  1289 &   0.065 & 5, 7       \\
1992ba     &  0.050 &  1135 &   0.049 & 2          & 2003bl     &  0.023 &  4295 &   0.103 & 2          \\
 1996W     &  0.036 &  1617 &   0.252 & 3          & 2003bn     &  0.056 &  3897 &   0.019 & 2          \\
1999br     &  0.020 &   957 &   0.151 & 4          & 2003cn     &  0.018 &  5430 &   0.078 & 2          \\
1999ca     &  0.094 &  2791 &   0.054 & 2          & 2003ef     &  0.040 &  4093 &   0.272 & 2          \\
1999cr     &  0.085 &  6134 &   0.168 & 2          & 2003gd     &  0.060 &   657 &   0.166 & 5, 2       \\
1999em$^c$ &  0.035 &   800 &   0.036 & 5, 2       & 2003hg     &  0.064 &  4186 &   0.440 & 2          \\
2000dc     &  0.068 &  3117 &   0.372 & 6          & 2003hl     &  0.062 &  2123 &   0.328 & 5, 2       \\
2000dj     &  0.062 &  4744 &   0.106 & 5          & 2003hn     &  0.012 &  1305 &   0.184 & 2          \\
 2001X     &  0.034 &  1480 &   0.061 & 5          & 2003ho     &  0.033 &  4091 &   0.764 & 2          \\
2001cm     &  0.012 &  3412 &   0.183 & 5          & 2003ib     &  0.041 &  7442 &   0.165 & 2          \\
2001cy     &  0.179 &  4478 &   0.016 & 6          & 2003ip     &  0.052 &  5398 &   0.114 & 2          \\
2001do     &  0.163 &  3124 &   0.318 & 6          & 2003iq     &  0.062 &  2331 &   0.127 & 5, 2       \\
2001fa     &  0.067 &  5182 &   0.125 & 6          & 2004du     &  0.082 &  5025 &   0.123 & 5          \\
2001hg     &  0.031 &  2387 &   0.119 & 5          & 2005ay     &  0.018 &   850 &   0.047 & 5          \\
2002an     &  0.031 &  3870 &   0.000 & 5          & 2005cs     &  0.032 &   463 &   0.054 & 5, 8, 9    \\
2002ca     &  0.020 &  3277 &   0.086 & 5          &  2009N     &  0.018 &   905 &   0.288 & 10, 2      \\
2002gd     &  0.058 &  2536 &   0.157 & 5, 2       & 2009dd     &  0.017 &   757 &   0.107 & 3          \\
2002gw     &  0.016 &  3143 &   0.203 & 2          & 2009ib     &  0.026 &  1304 &   0.119 & 11         \\
2002hj     &  0.101 &  7079 &   0.088 & 2          & 2012aw$^c$ &  0.024 &   778 &   0.041 & 12         \\
2002hx     &  0.045 &  9299 &   0.115 & 2          & 2012ec$^c$ &  0.023 &  1407 &   0.088 & 13         \\
 2003B     &  0.023 &  1141 &   0.064 & 2          & 2013ej$^c$ &  0.060 &   657 & --0.028 & 14, 15, 16 \\
 2003E     &  0.041 &  4484 &   0.297 & 2          &  2014G     &  0.010 &  1160 &   0.313 & 17         \\
\hline
\multicolumn{10}{m{0.9\linewidth}}{$^a$ Values from \citet{Schlafly_Finkbeiner2011}.}\\
\multicolumn{10}{m{0.9\linewidth}}{$^b$ Values computed in this work.}\\
\multicolumn{10}{m{0.9\linewidth}}{$^c$ SN with useful near-IR spectra.}\\
\multicolumn{10}{m{0.99\textwidth}}{$^d$
(1) \citet{Schmidt_etal1993};
(2) \citet{Gutierrez_etal2017_I};
(3) \citet{Inserra_etal2013};
(4) \citet{Pastorello_etal2004};
(5) \citet{Faran_etal2014_IIP};
(6) \citet{Faran_etal2014_IIL};
(7) \citet{Spiro_etal2014};
(8) \citet{Pastorello_etal2006};
(9) \citet{Pastorello_etal2009};
(10) \citet{Takats_etal2014};
(11) \citet{Takats_etal2015};
(12) \citet{DallOra_etal2014};
(13) \citet{Barbarino_etal2015};
(14) \citet{Valenti_etal2014};
(15) \citet{Dhungana_etal2016};
(16) \citet{Yuan_etal2016};
(17) \citet{Terreran_etal2016}.}
\end{tabular}
\end{table*}

\begin{table*}
\footnotesize
\caption{SNID templates.}
\label{table:SNID_library}
\begin{tabular}{l c c c c l}
\hline
SN         & $t_{\text{d}}$ & $t_{\text{ln}}-t_{\text{d}}$ & $t_{\text{fd}}-t_{\text{d}}$ & $t-t_{0,\text{ln}}$ & References$^c$ \\
           &  (d)        &  (d)     &  (d)     &        (d)                        &         \\
\hline
       1986L & 2446711.1\phnn &    -5.6 &     0.0 & 6,7,[27:33] & IAUC 4260, 1  \\
       1988A & 2447179.299    &  -3.099 &  -1.968 & 13 & IAUC 4533, IAUC 4540, 1  \\
       1990E & 2447937.62\phn &   -5.12 &     0.0 & 9,19 & IAUC 4965, 2  \\
      1999br & 2451280.9\phnn &    -8.0 &     0.0 & 15,18,25,33 & IAUC 7141, IAUC 7143, 3  \\
      1999em & 2451480.94\phn &   -8.99 &   -1.43 & [7:10],12,[14:16],21,26,29,31,35,39 & IAUC 7294, 4, 5, 1  \\
      1999gi & 2451522.32\phn &   -6.64 &     0.0 & 4,6,7,30,35,38 & IAUC 7329, IAUC 7334, 5  \\
      1999go & 2451535.7\phnn &    -8.0 &     0.0 & 9,10,14 & IAUC 7337, 6  \\
      2000dc & 2451765.8\phnn &    -7.0 &     0.0 & 20 & IAUC 7476, 7  \\
      2000dj & 2451795.9\phnn & -10.413 &     0.0 & 23 & IAUC 7490, IAUC 7491, 5  \\
      2000el & 2451869.53\phn &  -33.83 &  -28.93 & 39 & IAUC 7523, IAUC 7531, 6  \\
       2001X & 2451968.3\phnn &   -10.3 &     0.0 & 9,27,36 & IAUC 7591, 5  \\
      2001do & 2452135.7\phnn &    -4.0 &     0.0 & 31,39 & IAUC 7682, 7  \\
      2001fa & 2452200.9\phnn &    -5.0 &     0.0 & [4:7],30,31 & IAUC 7737, 7, 8  \\
      2002an & 2452297.02\phn &   -4.98 &     0.0 & 16,22,25 & IAUC 7805, IAUC 7808, 9, 5  \\
      2002ce & 2452375.378    &  -5.678 &     0.0 & 4 & IAUC 7875, 6  \\
  2002gd$^a$ & 2452553.37\phn &   -4.09 &   -2.84 & 6,8,12,23,27,31,35,38 & IAUC 7986, IAUC 7990, 5, 1  \\
       2003Z & 2452669.2\phnn &    -9.0 &     0.0 & 10,28,30 & IAUC 8062, 5, 10  \\
      2003bn & 2452697.98\phn &   -6.48 &   -5.15 & 14,18,37 & IAUC 8088, 1  \\
      2003ej & 2452779.8\phnn &    -9.0 &     0.0 & 6,14,19,40 & IAUC 8134, 1  \\
      2003hg & 2452869.9\phnn &    -9.0 &     0.0 & 8,32 & IAUC 8184, 9, 1  \\
      2003hl & 2452872.0\phnn &    -9.0 &     0.0 & 12,33 & IAUC 8184, 5, 1  \\
      2003iq & 2452921.458    &  -2.988 &     0.0 & 9,16,21,29 & IAUC 8219, 5, 1  \\
      2004ci & 2453173.497    &  -4.597 &  -1.697 & 6 & IAUC 8357, 6  \\
      2004er & 2453273.9\phnn &   -4.02 &     0.0 & 12,30 & IAUC 8412, IAUC 8415, 1  \\
      2004et & 2453275.5\phnn &  -4.983 &  -4.017 & 9,20,24,30,35,38 & IAUC 8413, 11, 5  \\
      2004fc & 2453299.89\phn &    -7.0 &  -4.766 & 9,33 & IAUC 8422, 1, 6  \\
      2004fx & 2453316.94\phn &  -16.02 &  -10.01 & 19,31 & IAUC 8431, 1, 6  \\
      2005ay & 2453456.58\phn &  -7.459 &     0.0 & 7,8,19,23,25 & IAUC 8500, IAUC 8502, 12, 5, 8  \\
      2005cs & 2453550.407    &  -1.977 &  -0.997 & 4,5,9,11,[13:15],17,34,36 & IAUC 8553, 13, 5, 14  \\
      2005dz & 2453623.71\phn &   -7.91 &     0.0 & 6,20 & IAUC 8598, 9, 1  \\
       2006Y & 2453770.08\phn &   -6.99 &     0.0 & 26,32 & IAUC 8668, 1  \\
      2006bc & 2453819.15\phn &  -8.063 &     0.0 & 9 & IAUC 8693, 1  \\
      2006bp & 2453835.1\phnn &  -1.423 &  -0.453 & 4,8,10,16,22,26,34 & IAUC 8700, 15, 15  \\
      2006it & 2454009.67\phn &   -4.98 &     0.0 & 11,14 & IAUC 8758, 1  \\
      2006iw & 2454011.798    &  -2.061 &     0.0 & 19 & CBET 663, 1, 16  \\
      2007hv & 2454352.87\phn &  -10.37 &     0.0 & 7 & CBET 1056, 8  \\
      2007il & 2454353.95\phn &   -8.01 &     0.0 & 12,26 & CBET 1062, 1  \\
      2007pk & 2454414.81\phn &   -4.98 &     0.0 & 3,4,6,7,28,38 & CBET 1129, 17, 8  \\
      2008bh & 2454548.66\phn &  -10.09 &     0.0 & 13,26,38 & CBET 1311, 1  \\
      2008br & 2454564.265    &  -4.942 &     0.0 & 7,21,29,36 & CBET 1332, 1  \\
      2008ho & 2454796.61\phn &   -8.84 &     0.0 & 18,23 & CBET 1587, 1  \\
      2008if & 2454812.71\phn &   -9.98 &     0.0 & 11,[13:17],22,29 & CBET 1619, 1  \\
      2008il & 2454827.64\phn &   -4.95 &     0.0 & 3 & CBET 1634, 1  \\
      2009ao & 2454894.62\phn &    -8.0 &     0.0 & 28,34 & CBET 1711, 1  \\
      2009bz & 2454919.98\phn &   -7.95 &     0.0 & 9,23,27,36 & CBET 1748, 1  \\
      2009dd & 2454935.47\phn & -11.412 &  -4.389 & [11:13],15,18,21,23 & CBET 1746, CBET 1765, 17, 8  \\
      2010id & 2455455.83\phn &   -5.01 &  -1.087 & 4,16 & CBET 2467, ATel 2862, 18, 18  \\
      2012aw & 2456003.36\phn &  -1.591 &  -0.011 & [2:15],24,29,40 & CBET 3054, ATel 3996, 19  \\
      2013am & 2456373.138    &   -1.44 &     0.0 & 2,12,16,23,29 & CBET 3440, 20, 21, 20  \\
  2013by$^a$ & 2456406.042    &   -3.17 &   -2.29 & 4,36 & CBET 3506, 22  \\
      2013ej & 2456498.95\phn &   -1.91 &  -1.325 & 4,[7:21],23,25,26,28,35,37,39 & CBET 3606, ATel 5237, CF$^b$, 23, 24, 25, 26  \\
      2013fs & 2456572.96\phn &   -2.14 &  -1.223 & [2:4],6,11,18,20,22,27,29,31,32,39 & CBET 3671, 27, 24, 27  \\
      2013hj & 2456638.8\phnn &    -3.1 &     0.0 & 9,19 & CBET 3757, 24  \\
       2014G & 2456672.074    &  -3.724 &  -0.963 & 3,4,10,14,17,26,37,39 & CBET 3787, 28  \\
     LSQ14gv & 2456674.8\phnn &    -4.1 &     0.0 & 8 & PESSTO SSDR2, 29  \\
      2014cy & 2456900.5\phnn &    -1.7 &     0.0 & 10 & CBET 3964, 29, 24  \\
\hline
\multicolumn{6}{m{0.99\linewidth}}{Column 1: SN names. Column 2: discovery epochs. Column 3 and 4: last non-detection and first detection epochs, respectively, with respect to the discovery epoch. Column 5: spectra phase values with respect to the explosion epoch, which we assume as the midpoint between the last non-detection and the first detection. Adjacent ages are listed in brackets. Column 6: references for data.}\\
\multicolumn{6}{m{0.99\linewidth}}{$^a$ Explosion time constraint obtained through polynomial fit to pre-maximum $\vri$ photometry.} \\
\multicolumn{6}{m{0.99\linewidth}}{$^b$ C. Feliciano report on the \textit{Bright Supernova} website (\url{http://www.rochesterastronomy.org/snimages/})}\\
\multicolumn{6}{m{0.99\linewidth}}{$^c$ (1) \citet{Gutierrez_etal2017_I};
(2) \citet{Schmidt_etal1993};
(3) \citet{Pastorello_etal2004};
(4) \citet{Elmhamdi_etal2003};
(5) \citet{Faran_etal2014_IIP};
(6) \citet{Shivvers_etal2017};
(7) \citet{Faran_etal2014_IIL};
(8) \citet{Hicken_etal2017};
(9) \citet{Harutyunyan_etal2008};
(10) \citet{Spiro_etal2014};
(11) \citet{Sahu_etal2006};
(12) \citet{Gal-Yam_etal2008};
(13) \citet{Pastorello_etal2009};
(14) \citet{Pastorello_etal2006};
(15) \citet{Quimby_etal2007};
(16) \citet{Sako_etal2018_SDSS-II_SN_DR};
(17) \citet{Inserra_etal2013};
(18) \citet{Gal-Yam_etal2011};
(19) \citet{DallOra_etal2014};
(20) \citet{Tomasella_etal2018};
(21) \citet{Zhang_etal2014};
(22) \citet{Valenti_etal2015};
(23) \citet{Valenti_etal2014};
(24) \citet{Childress_etal2016};
(25) \citet{Dhungana_etal2016};
(26) \citet{Yuan_etal2016};
(27) \citet{Yaron_etal2017};
(28) \citet{Terreran_etal2016};
(29) \citet{Valenti_etal2016}.}
\end{tabular}
\end{table*}




\bsp	
\label{lastpage}
\end{document}